\documentclass[aps,prb,twocolumn,floatfix]{revtex4-1}
\usepackage{amsfonts}
\usepackage{amsmath}
\usepackage{amssymb}
\usepackage[final,dvips]{graphicx}%

\newcommand{\xy}{ {\scriptscriptstyle\parallel} }
\newcommand{\lxy}{ {\!\scriptscriptstyle\parallel} }
\newcommand{\llxy}{ {\!\!\scriptscriptstyle\parallel} }

\newcommand{\scaledy}{ {\bar{y}} }
\newcommand{\scaledz}{ {\bar{z}} }
\newcommand{\scaledr}{ {\bar{r}} }
\newcommand{\scaledrvec}{ {\bar{\mathbf{r}}} }
\newcommand{\etal}{\textit{et al.}}

\newcommand{\Euler}{ \gamma_{\scriptscriptstyle\rm E} }
\begin{document}

\title{Josephson vortex lattice in layered superconductors}
\author{Alexei E. Koshelev}
\affiliation{Materials Science Division, Argonne National Laboratory, Argonne,
Illinois 60439}
\author{Matthew J. W. Dodgson}
\affiliation{Theory of Condensed Matter Group, Cavendish Laboratory, Cambridge,
\hbox{CB3 0HE}, UK} \affiliation{Institut de Physique, Universit\'e de Neuch\^atel,
Rue A.\ L.\ Breguet 1, 2000 Neuch\^atel, Switzerland} \affiliation{Department of
Physics and Astronomy, University College London, Gower Street, London \hbox{WC1E
6BT}, UK}
\date{\today }

\begin{abstract}
Many superconducting materials are composed of weakly coupled conducting layers.
Such a layered structure has a very strong influence on the properties of vortex
matter in a magnetic field. This review focuses on the properties of the Josephson
vortex lattice generated by the magnetic field applied in the layers direction. The
theoretical description is based on the Lawrence-Doniach model in the London limit
which takes into account only the phase degree of freedom of the superconducting
order parameter. In spite of its simplicity, this model leads to an amazingly rich
set of phenomena. We review in details the structure of an isolated vortex line as
well as various properties of the vortex lattice, both in dilute and dense limits.
In particular, we present an extensive consideration on the influence of the layered
structure and thermal fluctuations on the selection of lattice configurations at
different magnetic fields.
\end{abstract}

\pacs{PACS numbers...
}

\maketitle




\section{Introduction}
\label{sec:intro}

\begin{figure}[ptb]
\centerline{\includegraphics[width=3.2in,clip]{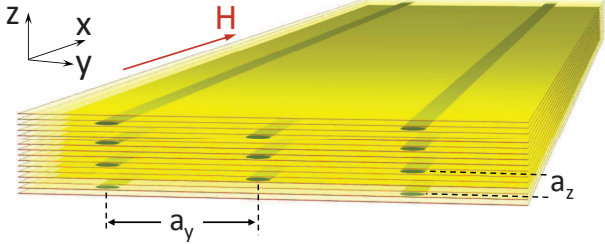}}
\caption{Illustration of a dilute lattice of Josephson vortices generated in a
layered superconductor by magnetic field applied along the layer direction.
} \label{fig:JosLatSchem}
\end{figure}

Layered superconductors are materials made from a stack of alternating thin
superconducting layers separated by non-superconducting regions. The superconducting
layers are essentially two-dimensional (2D) as long as they are so thin that there
is no variation in fields, or in the superconducting order parameter, across each
layer. Such structures frequently occur naturally in anisotropic crystals. A layered
superconductor can carry supercurrents along the layers, as well as between the
layers. This is due to the Josephson tunneling of Cooper pairs \cite{Josephson1962}
across the insulating regions that separate neighboring superconducting layers,
i.e., each pair of neighboring layers forms one Josephson junction. In general, the
$z$-axis (Josephson) supercurrents are weaker than the supercurrents along the
layers. A mere "layeredness" of atomic structure, however, does not make a material
automatically a layered superconductor. When the interlayer electrical coupling is
strong enough, this discrete system of layers approximates to a continuous
superconductor with uniaxial anisotropy. The case we are interested in here is when
the approximation to a uniaxial continuous superconductor breaks down, which happens
when the layer separation $d$ is greater than the $z$-axis superconducting coherence
length,  $d\gg \xi_{\mathrm{c}}$.

The most prominent example is the high-$T_c$ cuprate superconductors, discovered in
1986 \cite{Bednorz1986,WuATHMGHWC1987,MaedaTFA1988,ChuBGHHMSWX1988}, which led to a
huge interest in physics of layered superconductors. The two most studied cuprate
compounds, YBa$_2$Cu$_3$O$_7$ (YBCO) and Bi$_{2}$Sr$_{2} $CaCu$_{2}$O$_{x}$ (BSCCO),
have similar transition temperatures $T_c\approx 90$K and represent two important
particular cases. YBCO is moderately anisotropic, with the anisotropy factor $\gamma
\approx 5-7$, and its ``layeredness'' becomes essential at low temperatures when the
c-axis coherence length $\xi_c$ drops below the layer spacing $d$. On the other
hand, BSCCO has a huge anisotropy factor, $\gamma \approx 400-1000$, and behaves as
a layered superconductor practically in the whole temperature range below $T_c$.
Other naturally layered superconductors include the transition metal
dichalcogenides\cite{WilsonY69,Bulaevskii1975} and organic charge-transfer salts
formed with the molecule BEDT-TTF
\cite{IshiguroY1990,Singleton2000}. %
Important new family of atomically layered superconducting materials, iron pnictides
and chalcogenides, has been discovered in 2008\cite{KamiharaWHH2008} and are being
extensively explored since then, see, e.g., Reviews
\cite{Johnston2010,CanfieldB2010,Stewart2011}. Anisotropy of most compounds is
actually not very high and typically they behave as anisotropic three-dimensional
materials. There are, however, important exceptions. The most studied compound in
which the layered structure is clearly essential is
SmFeAsO$_{1-x}$F$_x$\cite{ChenWWLCF2008} with $T_c$ up to $55K$. For example,
Josephson nature of the in-plane
vortices at low temperatures has been recently demonstrated in this compound \cite{MollBGB2013}. %
Also, several iron pnictide compounds with extremely high anisotropy have been
discovered \cite{ZhuHM2009,OginoMK2009,SatoOK2010}. Properties of these compounds
remain mostly unexplored due to their rather complicated composition.

All layered superconductors share a very similar general behavior of the vortex
matter generated by external magnetic field, which is insensitive to the microscopic
nature of superconductivity inside the layers. Several excellent review articles
have been published in the past covering different aspects of the vortex matter in
type-II superconductors
\cite{BlatterFGLV1994,Feinberg1994,Brandt1995,NattermannS2000,BlatterG2003}.
Nevertheless, we feel that further progress on the understanding of the Josephson
vortices in layered superconductors warrants a specialized review, which will
provide much more details and discuss important results obtained in recent years.

This short review narrowly focuses on the vortex lattice which appears at magnetic
fields applied along the layers. In this case the flux line winds its phase around
an area between two neighboring layers and is called a Josephson vortex in analogy
with a vortex in a superconducting tunneling junction. The Josephson vortex contains
out-of-plane currents that tunnel via the Josephson effect from layer to layer. The
current distribution around vortex is anisotropic. As a consequence, the vortex
lattice is also anisotropic, it is a triangular lattice strongly stretched along the
layers, see Fig. \ref{fig:JosLatSchem}.  In addition, the restriction to lie between
the layers leads to commensurability effects and an energy barrier to tilting the
field away from the layers. There are two very different regimes depending on the
strength of the magnetic field $B_x$. The crossover field scale, $B_{\mathrm{cr}}$,
separating these two regimes is set by the anisotropy factor $\gamma$ and the layer
periodicity, $d$, $B_{\mathrm{cr}}=\Phi_{0}/(2\pi\gamma d^{2})$, where
$\Phi_{0}=hc/2e$ is the flux quantum. In the case of BSCCO this field scale is
around 0.5 tesla. In the dilute lattice regime, $B_x<B_{\mathrm{cr}}$, the nonlinear
cores of Josephson vortices are well separated and the distribution of currents and
fields is very similar to that in continuous anisotropic superconductors
\cite{CampbellDK1988}. The dense lattice regime is realized at high fields,
$B_x>B_{\mathrm{cr}}$, where the cores of the Josephson vortices overlap. In this
regime the Josephson vortices fill all layers homogeneously \cite{BulaevskiiC1991}.
This state is characterized by rapid oscillations of the Josephson current and by
very weak modulation of the in-plane current. In this review we characterize in more
detail these two lattice regimes.

Note that we do not consider in this review properties of vortices generated by
magnetic field applied perpendicular to the layers, along the c axis.\footnote{In
the literature the layer plane and the axis perpendicular to the layers are
frequently called ``ab plane'' and ``c axis''.} The structure of a c-axis vortex is
very different from the structure of an in-plane vortex. In layered superconductors
the c-axis vortex can be viewed as a stack of weakly coupled pointlike pancake
vortices. Properties of the pancake vortex lattice were also extensively explored,
see, e.g., Reviews \onlinecite{BlatterG2003,Clem04} and references therein.

Several experimental techniques have been employed to explore the Josephson vortex
lattices. The dilute stretched lattice at small fields ($< 100$G) has been directly
observed in YBCO with Bitter decoration by Dolan \etal \cite{DolanHFD1989}, where
the elliptical distribution of the flux around each Josephson vortex was also seen.
At high fields ($>1$ tesla) the commensurability between the c-axis parameter of the
Josephson vortex lattice and the interlayer separation leads to magnetic field
oscillations which have been observed experimentally in underdoped YBCO in
irreversible magnetization \cite{OussenaGGT94,ZhukovKPC1999} and nonlinear
resistivity \cite{GordeevZdG2000}.

In much more anisotropic BSCCO direct observation of Josephson vortices is not
possible. However, when the magnetic field tilted at small angles with respect to
the layers, the c-axis field component generates the pancake-vortex stacks which
preferably enter the superconductor along the Josephson vortices forming chains.
Visualizing the flux of these chains, it is possible to find locations of vertical
rows of the Josephson vortices and measure the in-plane lattice parameter $a_{y}$.
This was done using a variety of visualization techniques, such as Bitter
decorations \cite{BolleGGMBMK1991,TokunagaTFC2003}, scanning Hall probes
\cite{GrigorenkoBTOH2001}, Lorentz microscopy
\cite{MatsudaKKHYATNSKHK2001,TonomuraKKMHYASKHKMTKGBSO2002} and magnetooptical
imaging \cite{VlaskoVlasovKWCK2002,*VlaskoVlasovKWCK2003,TokunagaKTT2002} These
observations have been summarized in the Review \onlinecite{BendingD2005}.

Most extensively, properties of the Josephson vortex lattice have been explored in
BSCCO using c-axis transport in small-size mesas
\cite{LeeNH95,*LeeGHE97,HechtfischerKUM97,*HechtfischerKSW97,LatyshevGYMM01,LatyshevKB2003,KimWH2005}.
These studies revealed a very rich \emph{dynamical} behavior of the lattice, which
is beyond the scope of this review. The very important feature is that, due to low
dissipation, the Josephson vortex lattice can be accelerated up to very high
velocities. It is clear that understanding dynamics is not possible without good
understanding of static lattice properties. The dynamic phenomenon closely related
to static lattice configurations is magnetic-field oscillations of resistance for
very slow lattice motion which have been discovered and explored in small-size BSCCO
mesas \cite{OoiMK02,ZhuWK2005,LatyshevPOH2005,KatterweKrasnov2009,KakeyaKK2009}. The oscillation
period may correspond to either flux quantum or half flux quantum per junction
depending on the magnetic field and lateral size of the mesa. Interplay between the
bulk shearing interaction and the interaction with edges leads to very nontrivial
evolution of lattice structures which we consider in this review.

This review is organized as follows.  We start in Sec.\ \ref{sec:LD} in which we
present the energy functional and equilibrium equations for the phase and vector
potential. In Sec.\ \ref{sec:Jvortex} we describe the structure and energetics of a
single flux line. In Sec.\ \ref{sec:jvldilute} we discuss the dilute JVL and
consider in details the role of layered structure in selecting lattice
configurations. The properties of the dense JVL at high fields are considered in
Sec.\ \ref{sec:densejvl}.  In this regime the structure and energy of the lattice
can be evaluated analytically using expansion with respect to the Josephson
coupling. In this section we also review the magnetic field dependence of lattice
configurations and oscillations of the critical current in finite-size samples.
Elastic properties of both dilute and dense lattice are discussed in the
corresponding sections. Based on the elastic energies, in Sec. \ref{sec:Thermal} we
review effects caused by thermal fluctuations.

\section{Energy functional and equations for the superconducting phases and vector-potential}
\label{sec:LD}

Theoretical analysis of the Josephson vortex matter in layered superconductors is
based on a phenomenological model in which only phase degree of freedom of the
superconducting order parameters is taken into account and its amplitude variations
are neglected.
\begin{eqnarray} \label{eq:flld}
&&F_{\rm LLD}\left[ \phi_n({\mathbf r}_\lxy),{\mathbf A}(\mathbf {r})\right]
=\int d^3r \frac{B^2}{8\pi}\nonumber\\
&+&\sum_n \int d^2 r_{\!\xy}
\left\{
\frac{E_0}{2}
\left(\nabla_{\lxy}\phi_n +\frac{2\pi}{\Phi_0}{\mathbf A}_{\xy,n}\right)^2
\right.\nonumber\\
&+&\left.\frac{E_{\rm J}}{d^2}\left[1-\cos(\phi_{n+1}-\phi_n+\chi_{n,n+1})\right]
\right\},
\end{eqnarray}
where $E_0=\Phi_0^2 d/(16\pi^3\lambda_{\mathrm{ab}}^2)$ gives the in-plane phase
stiffness and $E_{\rm J}=E_0/\gamma^2=\Phi_0^2 d/(16\pi^3\lambda_{\mathrm{c}}^2)$ is
the phase stiffness for smooth inter-layer phase variations, $\lambda_{\mathrm{ab}}$
and $\lambda_{\mathrm{c}}$ are the components of the London penetration depth, and
$\gamma= \lambda_{\mathrm{c}}/\lambda_{\mathrm{ab}}$ is the anisotropy factor. The
$z$-component of the vector potential enters the tunneling term in the form
$\chi_{n,n+1}=(2e/\hbar c)\int_{nd}^{(n+1)d} d z\,A_z$.\footnote[4]{ Here $e$ is
chosen to be positive, $e>0$, i.e., the charge of an electron is $-e$.}
Near the transition temperature the above phase model can be obtained from the
celebrated Lawrence-Doniach model \cite{Lawrence1970} by fixing the order-parameter
amplitude (London approximation). However, the model is actually more general and
describes Josephson properties of layered material in the whole temperature range.
Starting from the phase model,  a rich variety of the lattice properties can be
derived, which we review in this article.

Subject to some given boundary conditions, the configuration of $\{\phi_n,{\mathbf
A}\}$ is determined by minimizing the free energy. This leads to a set of
differential equations, e.g. minimizing with respect to the phase gives the
current-conservation condition,
\begin{equation} \label{eq:phaseA}
\nabla_{\!\xy}^{\,2}\phi_n \!+\! \frac{2\pi}{\Phi_0} \nabla_{\!\xy}\cdot
{\mathbf A}_{\xy,n}\!=\! \frac{1}{(\gamma d)^2}
\left( \sin\varphi_{n-1,n}\!-\! \sin\varphi_{n,n+1}\right),
\end{equation}
with the gauge-invariant phase difference defined as
$\varphi_{n,n+1}\!=\!\phi_{n+1}\!-\!\phi_{n}+\chi_{n,n+1}$. In this equation the
Josephson length, $\Lambda_{\rm J}=\gamma d$, appears for the first time. This
length plays a very important role in layered superconductors, as it determines the
scale over which the phase can relax to minimize the Josephson coupling energy
without costing too much energy in the gradient term. Three more equations result
from minimizing with respect to the three components of the vector potential. We can
write these in terms of the electric current density by using the Maxwell equation
${\mathbf j}=(c/4\pi)\nabla\times(\nabla\times{\mathbf A})$, giving,
\begin{eqnarray}
{\mathbf J}_{\xy,n} &=& -  \frac{2\pi cE_0}{\Phi_0}
\left(\nabla_{\!\xy}\phi_n + \frac{2\pi}{\Phi_0}{\mathbf A}_{\xy,n}\right), \label{eq:Jinplane}\\
J_{n,n+1}&=&- j_{\rm J} \sin\varphi_{n,n+1},
\label{eq:Joutplane}
\end{eqnarray}
where ${\mathbf J}_{\xy,n}$ is the 2D current density in the $n$th layer and
$J_{n,n+1}$ is the current density in the $\hat{\mathbf z}$-direction between the
$n$th and $(n+1)$th layer which has the maximum value,
\begin{equation}
j_{\rm J}= \frac{2\pi cE_0}{\Phi_0(\gamma d)^2}.
\label{eq:jJdef}
\end{equation}

The four equations \eqref{eq:phaseA}--\eqref{eq:Joutplane}, are the starting point
for finding the structure of vortices in layered superconductors. In fact, we can
make the job of solving this set of equations slightly clearer by combining them to
find a differential equation for the gauge-invariant phase differences alone. This
is done by using the general result,
\begin{eqnarray}
\frac{4\pi d}{c}J_{n,n+1}&=&
\int_{nd}^{(n+1)d} d z\,[\nabla\times(\nabla\times{\mathbf A})]_z\nonumber\\
&=&\!\nabla_{\!\!\xy}\cdot({\mathbf A}_{\xy,n+1}\!-\!{\mathbf A}_{\xy,n})
\!-\!\frac{\Phi_0}{2\pi}\nabla_\llxy^2\chi_{n,n+1},
\end{eqnarray}
and combining this with \eqref{eq:phaseA} and \eqref{eq:Joutplane} to arrive at,
\begin{eqnarray}\label{eq:varphi}
&&\nabla_{\!\!\xy}^2\varphi_{n,n+1}
+\frac 1 {\lambda_{\mathrm{c}}^2}\sin\varphi_{n,n+1}\\
&&+\frac 1 {(\gamma d)^2}
\left[
\sin\varphi_{n+1,n+2}
\!-\!2\sin\varphi_{n,n+1}
\!+\!\sin\varphi_{n-1,n}
\right]=0.\nonumber
\end{eqnarray}
Solving this equation will give the entire solution for currents using
\eqref{eq:Joutplane} to find $J_{n,n+1}$, and the conservation law,
\begin{equation} \nabla_{\!\!\xy}\cdot{\mathbf J}_{\xy,n} = J_{n,n+1} - J_{n-1,n}
\end{equation}
to find ${\mathbf J}_{\xy,n}$.

\section{Structure of a Josephson vortex in a layered superconductor}
\label{sec:Jvortex}

If we place a flux line directed along the layers, the singularity associated with
the vortex core can be avoided by placing the center in the insulating layer between
two superconducting layers (first noticed by Bulaevskii\cite{Bulaevskii1973}). The
structure of the ``core'' is similar to the structure of the phase drop across a
flux line in a two-dimensional Josephson junction\cite{Josephson1965}. This
well-studied problem has a solution where the phase difference across the two layers
drops by $2\pi$ over a distance of the Josephson length $\Lambda_{\rm
J}$.\footnote[3]{ This characteristic length was noted soon after the discovery of
the Josephson effect \cite{Ferrell1963}.} For the 3D layered superconductor, this
length is given by $\Lambda_{\rm J}=\gamma d$, and we can think of a central region
$\gamma d$ wide and $d$ high as the core of an in-plane vortex. Beyond this core,
the flux density and currents are quite similar to that for a continuous anisotropic
superconductor\cite{CampbellDK1988}. The screening by $z$-axis currents is much
weaker than that by in-plane currents, and the flux line is stretched into an
ellipsoidal shape with a large width $\sim\lambda_{\mathrm{c}}$ along the layers.
Even though only the ``core'' resembles the vortex in a 2D Josephson junction, it
has become common usage in the literature to label the entire flux line with this
orientation a Josephson vortex.

Consider now a flux line directed along the $x$-axis. The general structure of this
Josephson vortex was first described by Bulaevskii\cite{Bulaevskii1973}: The center
of the vortex lies between two layers, so that there is no core with suppressed
amplitude of the order parameter, while at large distances from the center the
structure is similar to a conventional flux line. The phase around the vortex is not
given trivially by symmetry, but is a solution to the non-linear equations
\eqref{eq:phaseA}. The most convenient path to a quantitative solution is to
separate the problem into two different scales: At large scales we can ignore the
non-linearity and there is an analytical solution. At small scales the numerical
solution is simplified by ignoring the screening contribution of the vector
potential. Fortunately, for $\lambda_{\mathrm{ab}}/d\gg 1$ there is a large region
of intermediate scales where both approximations work well, allowing us to match the
small-scale and long-scale solutions.

We consider a vortex centered between layers $0$ and $1$, and $y=0$, which is
defined by the limiting values,
\begin{eqnarray}\label{eq:Jvbc}
\phi_{n}(y)  &  =&0,\hbox{ for }y\rightarrow+\infty,\label{eq:jvbound}\\
\phi_{n}(y)  &  =&\left\{
\begin{array}
[c]{l}
-\pi,\hbox{ for }n\geq1,\\
\pi,\hbox{ for }n\leq0,
\end{array}
\right.  \hbox{ for }y\rightarrow-\infty.\nonumber
\end{eqnarray}
This corresponds to the following conditions for the interlayer phase difference
\begin{eqnarray}
\varphi_{n,n+1}  &  =&0,\hbox{ for }y\rightarrow\pm\infty\hbox{ and }n\neq
0,\label{eq:JV_phaseDif}\\
\varphi_{0,1}  &  =&\left\{
\begin{array}
[c]{l}
0,\hbox{ for }y\rightarrow+\infty,\\
-2\pi,\hbox{ for }y\rightarrow-\infty.
\end{array}
\right.  \nonumber
\end{eqnarray}

To obtain the distributions of current and field, we first derive a useful exact
equation for the magnetic field. The current components \eqref{eq:Jinplane} and
\eqref{eq:Joutplane} can be represented as
\begin{eqnarray}
&&J_{n,n+1}   =\!-\!\frac{c}{4\pi}\nabla_{\!\! y}B_x^{n, n+1}\!=\!-\frac{c\Phi_{0}}{8\pi
^{2}\lambda_{\mathrm{c}}^{2}d}\sin\varphi_{n,n+1},\\
&&J_{y,n}  \!=\!\frac{c}{4\pi}\nabla_{n}B_x^{n-1, n}
\!=\!-\!\frac{c\Phi_{0}d}{8\pi
^{2}\lambda_{\mathrm{ab}}^{2}}\!\left(  \!\nabla_{\!\! y}\phi
_{n}\!+\!\frac{2\pi}{\Phi_{0}}A_{y}\right)\!,
\end{eqnarray}
where $B_x^{n, n+1}$ is the average magnetic field between the layers $n$ and $n+1$
and $\nabla_{n}$ is a difference operator, $\nabla_{n}A_{n}\equiv
A_{n+1}\!-\!A_{n}$. Collecting the combination $\left(  4\pi/c\right)  \left(
-\lambda_{\mathrm{c}}^{2} \nabla_{\!\!
y}J_{n,n+1}\!+\!(\lambda_{\mathrm{ab}}^{2}/d)\nabla_{n}J_{y,n}\right)  $, we obtain
\begin{eqnarray}\label{ed:JV-Beq}
&&\left[ 1-\lambda_{\mathrm{c}}^{2}\nabla_{\!\! y}^{2}
-(\lambda_{\mathrm{ab}}^{2}/d^{2})\nabla_{n}^{2}\right]
B_x^{n, n+1}\nonumber\\
&&=\frac{\Phi_{0}
}{2\pi d}\nabla_{\!\! y}\left(  \varphi_{n+1,n}-\sin\varphi_{n+1,n}
\right)
\end{eqnarray}
with $\nabla_{n}^{2}A_{n}\!\equiv A_{n+1}\!+A_{n-1}\!-2A_{n}$. The difference of
$\varphi_{n+1,n}$ and $\sin\varphi_{n+1,n}$ decays outside the nonlinear core and
satisfies the relation
\begin{equation}
\sum_{n}\int^\infty_{-\infty}
 \!\mathrm{d} y \nabla_{\!\! y}\left(
\varphi_{n+1,n}\!-\!\sin\varphi_{n+1,n}\right) \! =\! \sum_{n}
\varphi_{n+1,n}{\Big|}_{-\infty}^{\infty}\!=\!2\pi.
\end{equation}
Therefore, in the continuum limit the right-hand side of \eqref{ed:JV-Beq} converts
into $\Phi_{0}\delta(y)\delta(z)$ and (\ref{ed:JV-Beq}) transforms into the usual
equation for the vortex magnetic field\cite{ClemCoffey1990}
\begin{equation}
B_{x}-\lambda_{\mathrm{c}}^{2}\nabla_{\!\! y}^{2}B_{x}-
\lambda_{\mathrm{ab}}
^{2}\nabla_{\!\! z}^{2}B_{x}=\Phi_{0}\delta(y)\delta(z), \label{eq:JV-B-contin}
\end{equation}
which gives
\begin{equation}
B_{x}=\frac{\Phi_{0}}{2\pi\lambda_{\mathrm{c}}\lambda_{\mathrm{ab}}} K_{0}\left(  \sqrt{
\left(y/\lambda_{\mathrm{c}}\right)  ^{2}+
\left(z/\lambda_{\mathrm{ab}}\right)  ^{2}}\right)  .
\label{eq:JVfield_out}
\end{equation}
The current densities outside the core region are also given by standard formulas
for anisotropic superconductors,
\begin{eqnarray}
\label{eq:Jvjylin}
&\hspace{-0.35in}j_{y} &\!
=\!-\frac{c\Phi_{0}}{8\pi^{2}\lambda_{\mathrm{c}}
\lambda_{\mathrm{ab}}^{2}}\frac{z/\lambda_{\mathrm{ab}}}
{\sqrt{ y^{2}\!/\!\lambda_{\mathrm{c}}^{2}\!+\!
z^{2}\!/\!\lambda_{\mathrm{ab}}^{2}} }
K_{1}\!\left(\!\sqrt{
\frac{y^2}{\lambda_{\mathrm{c}}^2}\!+\!
\frac{z^2}{\lambda_{\mathrm{ab}}^2}}\right)\!,\label{eqJVjy_out}\\
&\hspace{-0.35in}j_{z}&\!=\!\frac{c\Phi_{0}}{8\pi^{2}\lambda_{\mathrm{c}}^{2}\lambda
_{\mathrm{ab}}}\frac{y/\lambda_{\mathrm{c}}}{\sqrt{
y^{2}\!/\!\lambda _{\mathrm{c}}^{2}\!+\! z^{2}\!/\!\lambda_{\mathrm{ab}}^{2}} }
K_{1}\!\left(\!\sqrt{
\frac{y^2}{\lambda_{\mathrm{c}}^2}\!+\!
\frac{z^2}{\lambda_{\mathrm{ab}}^2}}\right)\!.
\label{eq:JVjz_out}
\end{eqnarray}
These results should be valid as long as the linear approximation for the sine of
the phase difference is good. To find the range of applicability for this
approximation we compare the last equation to \eqref{eq:Joutplane}, which gives near
the vortex center,
\begin{equation}
\sin\varphi_{n,n+1}\!=\!-\frac{y/\gamma d}{\left(  y/\gamma d\right)  ^{2}\!+\!
n  ^{2}},\ \hbox{for }y^2/\lambda_{\mathrm{c}}  ^{2}\!+\!
z^2/\lambda_{\mathrm{ab}}^{2}\ll1,
\end{equation}
indicating that the linear theory breaks down at $\left( y/\gamma d\right)  ^{2}+n
^{2}\sim 1$. This condition therefore sets the boundary of the nonlinear core.

The above analysis shows that the Josephson vortex is characterized by two sets of
length scales. A region where the interlayer phase difference is large defines the
nonlinear core of the vortex. In the $z$ direction this region is essentially
localized within the central junction and in the $y$ direction it spreads over the
Josephson length $\gamma d$. At scales $\left\vert z\right\vert ,\left\vert
y\right\vert /\gamma\gg d$ the vortex structure is described by the anisotropic
London theory. In addition, in a wide region one can neglect screening effects where
the currents around the vortex decay as $1/r$, (though the current pattern is
strongly stretched along the layers). Screening of the currents and magnetic field
becomes important at the length scales $\left\vert z\right\vert
\approx\lambda_{\mathrm{ab}}$ and $\left\vert y\right\vert
\approx\lambda_{\mathrm{c}}$, which are much larger than the corresponding
boundaries of the nonlinear core.

Due to this vortex structure, a quantitative analysis may be obtained with more ease
by introducing an intermediate scale $R_{\rm int}$, with $d<R_{\rm
int}<\lambda_{\mathrm{ab}}$, such that at a distance from the vortex center
$\sqrt{z^{2}+(y/\gamma)^{2}}=R_{\rm int}$ both non-linearity and screening may be
ignored. One then considers separately the small-distance region
$\sqrt{z^{2}+(y/\gamma)^{2}}<R_{\rm int}$ (containing the nonlinear core) and the
large-distance region $\sqrt{z^{2}+(y/\gamma)^{2}}>R_{\rm int}$ (where screening
will become important). At small distances one can neglect screening. In the London
gauge, $ \mathbf{\nabla}
\cdot
\mathbf{A} =0$, this means that the vector potential $
\mathbf{A} $ can be dropped and the vortex is described in terms of in-plane phases
$\phi _{n}(y)$ only, which satisfy the following equation (from \eqref{eq:phaseA})
\begin{equation}
(\gamma d)^{2}\frac{d^{2}\phi_{n}}{dy^{2}}+\sin\left(  \phi_{n+1}-\phi
_{n}\right)  -\sin\left(  \phi_{n}-\phi_{n-1}\right)
=0\label{eq:phase_no_screen}
\end{equation}
and the boundary conditions (\ref{eq:jvbound}). These conditions are satisfied by
our knowledge that outside the nonlinear core, $(n-1/2)^{2}+(y/\gamma d)^{2} \gg1$,
the phase has to approach the scaled version of the usual form relating to the angle
around a vortex,
\begin{equation}
\phi^{\rm Jv}_{n}(y)=-\tan^{-1}\left(  \frac{\gamma d(n-1/2)}{y}\right).
\label{eq:JVphase_outside}
\end{equation}
Multiplying \eqref{eq:phase_no_screen} by $d\phi_n/dy$,
 summing over $n$, and performing an indefinite integral over $y$,
 we derive the following exact relation for all $y$,
\begin{equation}
\sum_n
\left[ (\gamma d)^{2}\left( \frac{d\phi_{n}}{dy}\right) ^{2}
\!-\!2\left( 1\!-\!\cos \left( \phi _{n+1}\!-\!\phi_{n}\right) \right)
\right] \!=\!\mathrm{const},
\label{eq:phase_exact}
\end{equation}
which is analogous to the first integral of a second-order differential equation
with one variable.
For the case of an isolated Josephson vortex the constant is zero. In contrast to
the single-variable case, this relation does not help us to find the exact solution
of the  coupled non-linear equations \eqref{eq:phase_no_screen}, and one has either
to use some approximate solution or to solve it numerically. Relation
\eqref{eq:phase_exact} can, however,  be used to test the accuracy of the
approximate and numerical solutions.

A simple approximate solution has been proposed by Clem and Coffey
\cite{ClemCoffey1990} (the CC solution) who used the following ansatz for the
magnetic field
\begin{equation}
\label{eq:jvbx} B_{x} \approx
\frac{\Phi_{0}}{2\pi\lambda_{\mathrm{c}}\lambda_{\mathrm{ab}}}
K_{0}\left( \frac{\sqrt{y^{2} +\gamma^{2}z^{2} +y_{\rm cc}^{2}}}{\lambda_{\mathrm{c}}}
\right).
\end{equation}
and found that the best approximation for the core structure is achieved by
selecting the cut off $y_{\rm cc}=\gamma d/2$. This field distribution allows one to
obtain the distribution of the phase difference
\begin{equation}
\varphi_{n,n+1}\approx -\sin^{-1} \left\{  \frac{d}{
\lambda_{\mathrm{ab} }} \frac{y}{R_{n}(y)} K_{1}\left(
\frac{R_{n}(y)}{\lambda_{\mathrm{c}}}\right) \right\}  ,
\end{equation}
where $R_{n}(y) =\sqrt{y^{2} +(\gamma d n)^{2} + y_{\rm cc}^{2}}$. In particular, at
$\gamma d\ll y \ll \lambda_{\mathrm{c}}$, this corresponds to $\phi_1(y)\approx
-\tan^{-1}(\gamma d/2y)$.

\begin{figure}[ptb]
\centerline{\includegraphics[width=3.0in,clip]{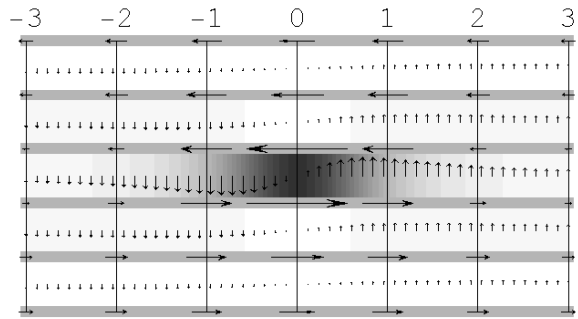}}
\caption{Visualization of the numerically computed structure of an
isolated Josephson vortex. The arrows represent the current distribution
(half interlayer distance corresponds to maximum Josephson
current). The greylevel codes for the cosine of the interlayer phase
difference.
The scale in the $y$-direction is in units of the Josephson length $\Lambda_J=\gamma d$.
} \label{fig:JosVor}
\end{figure}

\begin{figure}[ptb]
\centerline{\includegraphics[width=3.0in,clip]{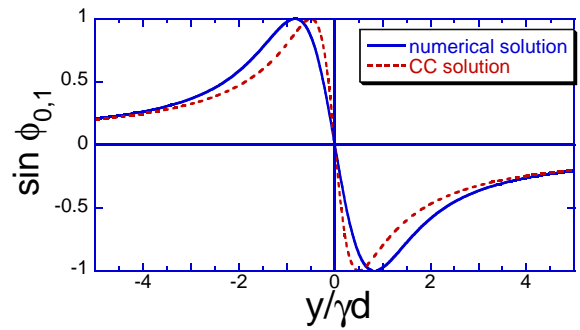}}
\caption{Sine of the phase difference between the central layers of
the Josephson vortex. For comparison, the approximate solution of
Clem and Coffey \cite{ClemCoffey1990} is also  shown.}
\label{fig:sin_phi01}
\end{figure}

The accurate numerical structure for the  core was obtained in Ref.\
\onlinecite{Koshelev1993}.
Figure \ref{fig:JosVor} presents a visualization of this numerical solution, and we
compare the phase difference in the central junction to that from the CC solution in
figure~\ref{fig:sin_phi01}. The numerical solution is characterized by the following
properties. The maximum in-plane phase gradient is given by
\begin{equation}\label{eq:maxdphidy}
\gamma d \left.\frac{\mathrm{d} \phi_1}{\mathrm{d} y}\right|_{y=0}=1.10
\end{equation}
(the CC solution gives $\gamma d~(\mathrm{d} \phi_1/\mathrm{d} y)_{y=0}=2$) and the
maximum Josephson current flows at a distance $y_{\rm max}=0.84\gamma d$ from the
vortex center (the CC solution gives $y_{\rm max}=y_{\rm cc}=0.5\gamma d$). The
maximum magnetic field in the vortex core is given by
\begin{equation}\label{eq:JVfield0}
    B_x^{0, 1}(y\!=\!0)\approx \frac{\Phi_{0}}
    {2\pi\lambda_{\mathrm{c}}\lambda_{\mathrm{ab}}}
    \left[\ln\left(\lambda_{\mathrm{ab}}/d \right)+1.03 \right].
\end{equation}
The asymptotic limits for the phase difference in the central junction are,
\begin{equation}
\varphi_{0,1}= \left\{
\begin{array}
[c]{l}
\displaystyle -\pi+\frac{2.20y}{\gamma d}, \hbox{ for } |y| \ll\gamma d,\\
\displaystyle -\frac{\gamma d }{ y }, \hbox{ for } \gamma d
\ll y \ll
\lambda_{\mathrm{c}},\\
\displaystyle -\frac{d }{\lambda_{\mathrm{ab}}}
\sqrt{\frac {\pi\lambda_{\mathrm{c}}}{2y}}\mathrm{e}^{-y/\lambda_{\mathrm{c}}}, \hbox{ for }
\lambda_{\mathrm{c}} \ll y .
\end{array}
\right.
\end{equation}


Outside the core  one can calculate the correction $\delta \phi_n(y)$ to the
continuum-limit phase asymptotics (\ref{eq:JVphase_outside}) by treating the
discreteness and nonlinearity of the Josephson current perturbatively (see
appendix~\ref{App:JVphaseAsymp}). This gives%
\begin{equation} \label{eq:JVcorr}
\hspace{-0.01in}\delta \phi_n(y) \!=\!\frac{\sin [2\,\phi^{\rm Jv}_n(y)
]}{16\,R^{2}}\left( \ln R\!+\!C_{\delta\phi}\right)\! +\!\frac{ 5\sin
[4\,\phi^{\rm Jv}_n(y)]}{96\,R^{2}},
\end{equation}%
where $R = \sqrt{(n-1/2)^2+(y/\gamma d)^2}$, and the constant $C_{\delta\phi}\approx
4.362$ is found from comparison with the numerical solution.

We can find the energy per unit length of the Josephson vortex by inserting this
solution into \eqref{eq:flld}. The simplest method\cite{Koshelev1993} is again to
split the energy into two contributions: one from the region at large distances
where the linear approximation is valid, and one from small distances where we need
the numerical solution, but can ignore the contributions of $ \mathbf{A} $ to the
current (i.e. ignore screening). The first is found analytically, while the second
needs a numerical integration. The final result is (see also \cite{ClemCH1991}),
\begin{equation}\label{eq:Jvself}
\varepsilon_{\mathrm{Jv}}=\frac{\varepsilon_{0}}{\gamma} \left[
\ln(\lambda_{\mathrm{ab}}/d) +1.55\right]  .
\end{equation}
with $\varepsilon_{0}=\Phi_0^2/(4\pi \lambda_{\mathrm{ab}})^2$. This energy
determines the lower critical field $H_{c1,x}$ above which Josephson vortices are
generated,
\begin{equation}
H_{c1,x}=4\pi\varepsilon_{\mathrm{Jv}}/\Phi_0 =\frac{\Phi_0}{4
\pi \lambda_{\mathrm{c}} \lambda_{\mathrm{ab}}}
 \ln(0.44  \lambda_{\mathrm{ab}}/d).
\end{equation}

To summarize, the solution for a Josephson vortex presented here is very similar to
the usual flux lines in isotropic superconductors, but stretched by the factor
$\gamma$ in the $y$-direction. The reason for this similarity is that the linear
approximation to the Josephson relation works well away from the vortex center. The
important feature, however, is that at the center of the vortex there is no normal
core, but rather a phase drop of nearly $2\pi$ across the central junction over a
distance of $\gamma d$.

\subsection{Line-tension energy of Josephson vortex}
\label{sec:LineTensJV}

In this section we consider the line-tension energy of a distorted Josephson vortex,
an important parameter which determines thermal wandering of the vortex line and its
response to pinning centers. We consider a kink-free vortex located in between the
layers $0$ and $1$ and defined by the planar displacement field $u(x)$. As the
energy of the straight vortex does not depend on its orientation inside the layer's
plane, for very smooth distortions with the wavelength larger that
$\lambda_{\mathrm{c}}$ the line-tension energy is simply determined by the line
energy \eqref{eq:Jvself},
\[
\delta F=\int dx\frac{\varepsilon_{\mathrm{Jv}}}{2}\left(  \frac{du}
{dx}\right)  ^{2}\hbox{ for }|du/dx|<|u/\lambda_{\mathrm{c}}|
\]
This simple result, however, has a limited interest, because most properties of the
vortex are determined by deformations with smaller wavelength,
$|du/dx|/|u|\sim|k_{x}|\gg1/\lambda_{\mathrm{c}}$. In this range, the line-tension
energy acquires nonlocality, a typical feature of vortex lines. An accurate
calculation of the line tension for this regime presented in Appendix
\ref{App:LineTension} leads to the following result
\begin{equation}\label{eq:LineTensEnJV}
\delta F=\frac{\pi}{2}\varepsilon_{\mathrm{J}}\int\frac{dk_{x}}{2\pi}k_{x}
^{2}\ln\frac{C_{t}}{\gamma dk_{x}}u^{2}
\end{equation}
with $\varepsilon_{\mathrm{J}}\equiv E_{0}/\gamma d$ and $C_{t}\approx 2.86$. The
important feature is the logarithmic dependence of the efficient line tension on the
deformation wave vector, which is a consequence of nonlocality.

\section{Dilute lattice, $B_x<\Phi_{0}/2\pi\gamma d^{2}$}
\label{sec:jvldilute}

When the Josephson vortices are well separated, the linear and continuous
approximation can be applied to the energy functional \eqref{eq:flld} everywhere
except in the core regions,  which reduces it to the anisotropic London model
\begin{eqnarray} \label{eq:flon}
F_{\rm L}\left[ \phi({\mathbf r}),{\mathbf A}(\mathbf {r})\right]
&\approx& \int d^3r \left\{
\frac{B_{x}^2}{8\pi}
\!+\!\frac{E_0}{2}\left[
\left(\nabla_{\lxy}\phi \!+\!\frac{2\pi}{\Phi_0}{\mathbf A}_{\xy}\right)^2
\right.\right.\nonumber\\
&+&\left.\left.
\frac{1}{\gamma^2}\left(\nabla_z\phi \!+\!\frac{2\pi}{\Phi_0} A_{z}\right)^2
\right]\right\}.
\end{eqnarray}
This means that the lattice solution is just a linear addition of single flux-line
solutions and the lattice energy is determined by this London model. To understand
the nature of the ground state it is useful to apply the rescaling trick
\cite{KlemmClem1980,BlatterGL1992},
\begin{equation} \label{eq:scal}
\tilde{\mathbf r}= (y,\gamma z) \hbox{ and }
\tilde{\mathbf A}=(A_y,
A_z/\gamma),
\end{equation}
which in the case of zero $z$-component of the magnetic field precisely reduces the system to the isotropic state
\cite{CampbellDK1988}. Therefore the ground state configuration in scaled coordinates is given by a regular
triangular lattice. In real coordinates this state corresponds to the triangular
lattice strongly stretched along the direction of the layers.

\begin{figure}[ptb]
\centerline{\includegraphics[width=7cm,clip]{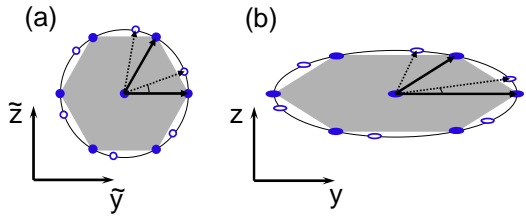}}
\caption{
Ground-state lattice configuration for an in-plane field and its
rotational degeneracy within the anisotropic London model in (a)
scaled coordinates and (b) real coordinates. The ellipse aspect
ratio corresponds to the anisotropy factor $\approx 3$, much
smaller than the anisotropy of, e.g., BSCCO.}
\label{fig:RotatDegen}
\end{figure}

\begin{figure}[ptb]
\centerline{\includegraphics[width=7cm,clip]{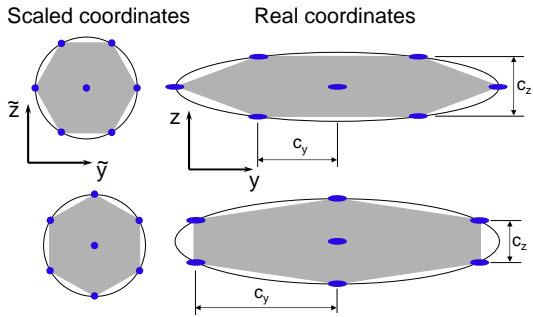}}\caption{
The two alternative lattice configurations that are aligned with
the layers, in scaled and real coordinates. } \label{fig:JVLalign}
\end{figure}

Within the anisotropic London model the lattice is degenerate with respect to
rotation in scaled coordinates. In real coordinates this corresponds to an
``elliptic rotation'' illustrated in figure~\ref{fig:RotatDegen}. In particular,
there are two aligned configurations, in which Josephson vortices form vertical
stacks along the $z$ axis (see figure~\ref{fig:JVLalign}).  For these configurations
the vertical distance between the Josephson vortices in the stacks, $a_z$, and the
separation between the stacks, $a_y$, are given by
\begin{equation}
a_{z}=\sqrt{\beta \Phi_0/(\gamma B_x)},\ a_{y}=\sqrt{
\gamma\Phi_0/(\beta B_x)}
\end{equation}
where the constant $\beta$ is equal to $2\sqrt{3}$ and $2/\sqrt{3}$ for the upper
and lower configuration in figure~\ref{fig:JVLalign} respectively.

The interaction energy of the Josephson vortex lattice can be reduced to interaction
energy of Abrikosov vortex lattice using scaling trick. This energy must be added to
the self energy of each Josephson vortex \eqref{eq:Jvself} which, in the
intermediate field regime $H_{c1,x}\ll \overline{B}_x\ll B_{\gamma d^2}$, gives the
result
\begin{equation} \label{eq:jvlenergy}%
f_{\rm Jl} \approx
\frac{\overline{B}_x^2}{8\pi} + \frac{\overline{B}_x}{\Phi_0}
\frac{\varepsilon_0}{2\gamma} \ln\left(
\frac{1.23 \Phi_{0}}{\gamma d^2 \overline{B}_x}
\right).%
\end{equation}

The ``elliptic rotation'' degeneracy is eliminated by the layered structure of
superconductor. There are several different mechanisms of this elimination. First,
due to the strong intrinsic pinning, the vortex centers must be located in between
the layers. This limits the possible lattice orientations. A second, less trivial,
mechanism is from the corrections due to the discrete lattice structure to the
vortex interactions. The degeneracy is also eliminated by thermal fluctuations,
because the Josephson vortices mostly fluctuate along the layers directions and this
selects preferential lattice orientations. All these mechanisms will be considered
in details below.


\subsection{Selection of ground-state configurations by layered structure}

\begin{figure}[ptb]
\centerline{\includegraphics[width=7cm,clip]{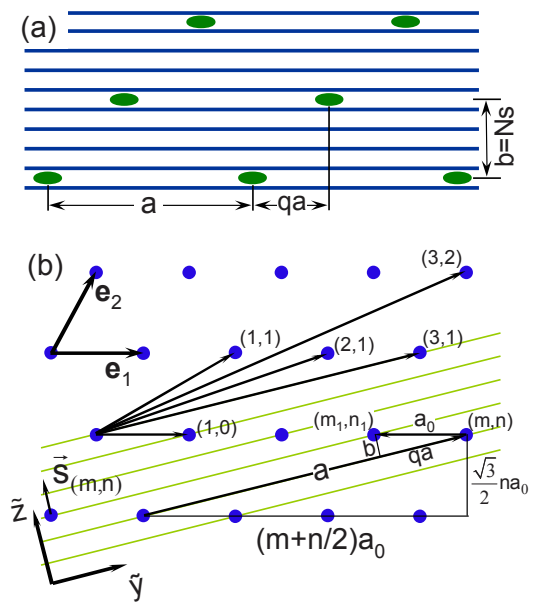}}\caption{
(a) General Josephson vortex lattice and its parameters. (b)
Orientation of layered structure with respect to ideal lattice (in
scaled coordinates). The layered structure fits the ideal lattice only
if it is oriented along one of the crystallographic directions,
which is characterized by two numbers $(m,n)$, corresponding to
expansion of the direction vector over  the two basic lattice
vectors, ${\mathbf e}_{1}$ and ${\mathbf e}_{2}$.
 Several  possible
directions are shown with the corresponding indices $(m,n)$.
The layers, together with the lattice parameters, $a$,
$b$, and $q$, are drawn here for the (3,1) orientation.} \label{fig:JVLcommens}
\end{figure}
As the centers of the Josephson vortices must be located between the layers, the
layered structure plays a crucial role in the selection
 of the ground-state lattice configurations. The Josephson-vortex
lattice is commensurate with the layered structure only at a discrete set of
magnetic fields. Due to the ``elliptic rotation'' degeneracy of the lattice within
the London approximation, the family of commensurate lattices includes lattices
aligned with the layers (see Fig.\ \ref{fig:JVLalign}), as well as misaligned ones.
To make a full classification of commensurate lattices we consider a general lattice
as shown in Fig.\ \ref{fig:JVLcommens}a \cite{IvlevKP1990,Levitov1991}. The lattice
is characterized by three parameters: the in-plane period $a$, the distance between
vortex rows in the $z$ direction $b=Nd$, and the relative shift between the
neighboring vortex rows in $qa$. The lattice shape is characterized by the two
dimensionless parameters, $q$ and the ratio $r=b/a$. The lattice parameters are
related to the in-plane magnetic field, $B_x$, as $B_x=\Phi_0/(ab)$. The two aligned
structures in Fig.\ \ref{fig:JVLalign} correspond to $q=1/2$.  As the replacement
$q\rightarrow 1-q$ corresponds to a mirror reflection with respect to the $x$-$z$
plane, every structure with $q\ne 1/2$ is doubly degenerate. In addition to giving
the general ground states, these lattices describe multiple metastable states with
unique properties studied in Refs.\ \onlinecite{IvlevKP1990,Levitov1991}, which we
will review below.

We now classify the exactly commensurate lattices to give the set of commensurate
fields. An equivalent geometrical analysis has been done in Ref.\
\onlinecite{LagunaDB2000} following a somewhat different line of reasoning, but with
the same final result for the commensurate fields. The analysis of commensurability
conditions can be done most conveniently in scaled coordinates \eqref{eq:scal}. In
these coordinates the ground-state configuration corresponds to a regular triangular
lattice with period $\tilde{a}_\Delta=\sqrt{2\gamma\Phi_0/\sqrt{3}B_x}$. It is
convenient to consider the orientation of the layered structure with respect to this
lattice rather than the other way round. The layered structure fits this lattice
only if it runs along one of the crystallographic directions, see Fig.\
\ref{fig:JVLcommens}b. This direction, $(m,n)$, is defined by the lattice vector,
${\mathbf e}_{m,n}$, which can be expanded over the two basic lattice vectors,
${\mathbf e}_{(m,n)}=m{\mathbf e}_{1}+n{\mathbf e}_{2}$. For nonequivalent
directions $m$ and $n$ must be relatively prime numbers (i.e., there is no integer
other than one that divides into both $m$ and $n$). Any such direction corresponds
to a set of matching fields which we notate as $B_{(m,n)}(N)$. We also notate the
lattice parameters corresponding to such an orientation as $a_{(m,n)}$, $b_{(m,n)}$,
and $q_{(m,n)}$. Immediately, we obtain
\begin{equation} \label{eq:anm}
a_{(m,n)}=e_{(m,n)}=\tilde{a}_\Delta\sqrt{m^2+mn+n^2}.
\end{equation}

It is useful to write the unit vector perpendicular to the layers, $\hat{{\mathbf
z}}$, in terms of ${\mathbf e}_{(m,n)}$. This vector is labelled  ${\mathbf
s}_{(m,n)}$ in Fig.\ \ref{fig:JVLcommens}b, and is given by,
\begin{equation}
{\mathbf s}_{(m,n)}\equiv \hat{{\mathbf z}} =\frac{{\mathbf e}_{(m,n)}\times \hat{\mathbf x}} {e_{(m,n)}}.\label{eq:snm} \end{equation}
Commensurability means that the projections of the two basic lattice vectors on
${\mathbf s}_{(m,n)}$ should be an integer number of layers, i.e.,
\begin{equation}
{\mathbf e}_1 \cdot {\mathbf s}_{(m,n)} =\tilde{n} \gamma
d,\hspace{1cm} {\mathbf e}_2\cdot {\mathbf s}_{(m,n)} =\tilde{m} \gamma d,
\end{equation}
 (in scaled coordinates the interlayer distance is $\gamma d$).
Using \eqref{eq:anm} and \eqref{eq:snm}, we rewrite these conditions as
\begin{eqnarray} \frac{\sqrt{3}}{2}\tilde{a}_\Delta
\frac{n}{\sqrt{m^2+mn+n^2}}&=&\tilde{n}\gamma d,\\
\frac{\sqrt{3}}{2}\tilde{a}_\Delta \frac{m}{\sqrt{m^2+mn+n^2}}&=&\tilde{m}\gamma d.
\end{eqnarray} These equations mean that $\tilde{m}/\tilde{n}=m/n$. As $m$ and $n$
are by definition relatively prime numbers, the set of allowed $\tilde{m}$ and
$\tilde{n}$ is simply given by $\tilde{m}=N m$ and $\tilde{n}=N n$. Therefore we can
represent the commensurability condition as
\begin{eqnarray} \frac{\sqrt{3}}{2}\tilde{a}_\Delta=N \sqrt{m^2+mn+n^2}\gamma d,
\label{eq:CommensCond} \end{eqnarray} which gives the following set of
commensurate fields, distances between neighboring rows $b=Nd$ and ratios
$r_{(m,n)}$
\begin{eqnarray}
B_{(m,n)}(N)&=&\frac{\sqrt{3}}{2}\frac{\Phi_0}{N^2\gamma
d^2(m^2+mn+n^2)}, \label{eq:CommensFields}\\
b_{(m,n)}&=&\frac{\frac{\sqrt{3}}{2}\tilde{a}_\Delta}{\sqrt{m^2+mn+n^2}},
\label{eq:bnm}\\
r_{(m,n)}&=&\frac{\sqrt{3}/2}{m^2+mn+n^2}.\label{eq:rnm}
\end{eqnarray}

Finding the parameter $q_{(m,n)}$ for a general orientation is a more complicated
problem. Defining the direction to the nearest-row site $(m_1,n_1)$ (see
figure~\ref{fig:JVLcommens}), we have
\begin{equation}
q_{(m,n)}\!=\!\frac{{\mathbf e}_{(m,n)}\!\cdot \!{\mathbf e}_{(m_1,n_1)}}{|{\mathbf e}_{(m,n)}|^2} =\frac{|mm_1\!+\!(m_1n\!+\!mn_1)/2\!+\!nn_1 |}{m^2+mn+n^2}.
\label{eq:qnm-gen} \end{equation} Expressing the neighboring-row separation
via $(m_1,n_1)$,
$$
b_{(m,n)}=\frac{|[{\mathbf e}_{(m,n)}\times {\mathbf e}_{(m_1,n_1)}]|}{e_{(m,n)}}=\frac{\frac{\sqrt{3}}{2}
\tilde{a}_\Delta|m_1n-mn_1|}{\sqrt{m^2+mn+n^2}}
$$
and comparing it with Eq.\ \eqref{eq:bnm},  we can see that pair $(m_1,n_1)$ must
satisfy the condition
\begin{equation}
|m_1n-mn_1|=1.
\label{eq:n1m1_cond}
\end{equation}
It is well-known from the theory of numbers that for any relatively prime pair
$(m,n)$ one can find a complimentary pair $(m_1, n_1)$ satisfying this condition,
and there is a general recipe to find complimentary pairs based on the Euclid
algorithm (see, e.g., Ref.\ \onlinecite{Stewart64}).  Moreover, as the combination
$m_1n-mn_1$ does not change with the substitution $m_1\rightarrow m_1+m$,
$n_1\rightarrow n_1+n$, there is an infinite set of pairs which satisfy condition
\eqref{eq:n1m1_cond} (physically, this corresponds to different lattice sites in the
neighboring row). Therefore, the problem to find $q_{(m,n)}$ can be formulated as
follows: among all pairs $(m_1, n_1)$ satisfying condition \eqref{eq:n1m1_cond} one
must find the pair which minimizes $|mm_1+(m_1n+mn_1)/2+nn_1 |$ and use this pair in
Eq.\ \eqref{eq:qnm-gen}. (Practically, we need not search to very high-order
directions.) In the case $n=1$ and arbitrary $m$ the choice of $(m_1,n_1)$ is
obvious, $(m_1,n_1)=(-1,0)$, and we obtain
\begin{equation}
q_{(m,1)}=\frac{m+1/2}{m^2+m+1}.
\end{equation}

We should stress that these results essentially rely on the linear London
approximation, which implies a very strong inequality $\tilde{a}_\Delta\gg \gamma
d$, or equivalently, $N\sqrt{m^2+mn+n^2}\gg 1$. The number of vortex-free layers per
unit cell is given by $N-1$. The case $N=1$ represents a special situation when all
the layers are filled with vortices and are equivalent. It is interesting to note
that even for a dilute lattice one can have Josephson vortices in every layer
($N=1$) in the case of high-order commensurability ($m,n\gg 1$). In an ideal
situation,  the lattice transfers with changing magnetic field between different
commensurate configurations via a series of first-order phase transitions. The
number of competing states rapidly increases as the field decreases.

A full analysis of the structural evolution requires consideration of the energy. In
the London limit a very useful expression for the energy of the general lattice in
Fig.\ \ref{fig:JVLcommens}a has been derived in Ref.\ \onlinecite{IvlevKP1990}. We
outline this derivation and present the final result in a somewhat different form.
For the lattice in Fig.\ \ref{fig:JVLcommens}a the interaction energy in the London
limit is given by,
\begin{eqnarray*}
f_{\mathrm{Jl}}^{\mathrm{int}} &=&\frac{B_{x}^{2}}{8\pi }\left[
\sum_{l,k} \frac{1}{1\!+\!\lambda_{\mathrm{ab}} ^{2}\left[ \gamma ^{2}\left( 2\pi
l\right)
^{2}\!/a^{2}\!+\!\left[ 2\pi (k\!-\!ql)\right]^{2}\!/b^{2}\right] }\right.  \\
&&\left. -\int dydz\frac{1}{1+\lambda_{\mathrm{ab}} ^{2}\left[ \gamma
^{2}\left( 2\pi z\right) ^{2}/a^{2}+\left( 2\pi y\right)
^{2}/b^{2}\right] }\right]
\end{eqnarray*}
Using the  formula
\begin{equation*}
\sum_{k=-\infty }^{\infty }\frac{1}{(k+v)^{2}+u^{2}}=\frac{\pi
}{u}\frac{ \sinh \left( 2\pi u\right) }{\cosh \left( 2\pi u\right)
-\cos \left( 2\pi v\right) }
\end{equation*}
we can sum over $k$ and integrate over $y$ leading to
\begin{eqnarray*}
&&f_{\mathrm{Jl}}^{\mathrm{int}} =\frac{B_{x}^{2}}{8\pi
}\frac{b^{2}}{2\pi \lambda_{\mathrm{ab}} ^{2}}\left[ \frac{\pi \lambda_{\mathrm{ab}}
}{b}\frac{\sinh \left( b/\lambda_{\mathrm{ab}}
\right) }{\cosh \left( b/\lambda_{\mathrm{ab}} \right) -1}\right.  \\
&&\left. +\sum_{l=1}^{\infty }\frac{1}{g_b(l)}\frac{\sinh \left[
2\pi g_b(l)\right] }{\cosh \left[ 2\pi g_b(l) \right] -\cos \left(
2\pi ql\right) } -\int_{0}^{\infty }dz\frac{1}{g_b(z)}\right]
\end{eqnarray*}
with $g_b(z)=\sqrt{\left(b/ 2\pi \lambda_{\mathrm{ab}} \right)^{2}+r^{2}z^{2}}$ and
$r=b\gamma /a$. This expression significantly simplifies in the intermediate region
$ b\ll 2\pi \lambda_{\mathrm{ab}} $, where we can use the expansion
\begin{equation*}
\frac{\pi \lambda_{\mathrm{ab}} }{b}\frac{\sinh \left( b/\lambda_{\mathrm{ab}} \right)
}{\cosh \left( b/\lambda_{\mathrm{ab}} \right) -1}\approx \frac{2\pi
\lambda_{\mathrm{ab}} ^{2}}{b^{2}}+\frac{\pi }{6}
\end{equation*}
and drop $b^{2}/\left( 2\pi \lambda_{\mathrm{ab}} \right) ^{2}$ in $g_b(z)$ meaning that
$g_b(z)\rightarrow rz$. This allows us to represent the interaction energy in this
regime as \cite{IvlevKP1990}
\begin{eqnarray}
&\hspace{-0.2in}&f_{\mathrm{Jl}}^{\mathrm{int}}=\frac{B_{x}^{2}}{8\pi
}+\frac{B_{x}\Phi _{0}}{ \left( 4\pi \right) ^{2}\lambda_{\mathrm{ab}}
\lambda_{\mathrm{c}}}\nonumber\\
&&\times\left[ \frac{1}{2}\ln \left( \frac{\Phi _{0}}{2\pi
\lambda_{\mathrm{ab}} \lambda_{\mathrm{c}}B_{x}}\right)\! +\!\Euler\!-\!\ln 2+G_{\rm
L}(r,q)\right] \label{eq:EnLonGenInter}
\end{eqnarray}
with $\Euler=0.5772$ being the Euler constant and
\begin{equation}
G_{\rm L}(r,q)\!=\!\frac{\pi r}{6}+\sum_{l=1}^{\infty
}\frac{
\cos \left( 2\pi ql\right)\!-\!\exp \left( -2\pi rl\right) }{l[\cosh \left( 2\pi rl\right) \!-\!\cos \left( 2\pi
ql\right)] }-\frac{1}{2}\ln (2\pi r). \label{eq:gfun}
\end{equation}
The dimensionless function $G_{\rm L}(r,q)$ depends only on the lattice shape. Its
absolute minimum corresponding to the triangular lattice is given by $G_{\rm
L}(\sqrt{3}/2,1/2)=-0.4022$. A peculiar property of the function $G_{\rm L}(r,q)$,
following from the rotational degeneracy, is that it also has this value for the
whole set of pairs $(r,q)=(r_{(m,n)},q_{(m,n)})$ corresponding to the different
lattice orientations. In particular, for $(m,n)=(m,1)$ we have $G_{\rm L}\left(
\frac{\sqrt{3}/2}{m^{2}+m+1},\frac{m+1/2}{m^{2}+m+1}\right)=G_{\rm
L}(\sqrt{3}/2,1/2)$. This function also has very peculiar behavior at small $r$
which is important for the statistics of metastable states \cite{Levitov1991}: at
$r\rightarrow 0$ it acquires peaks at all rational values of $q=k/l$. Large-order
peaks with denominator $l$ develop as $r$ drops below $1/(2\pi l)$.

For layered superconductors we have
\begin{equation*}
b=Nd,\ a=\frac{\Phi _{0}}{B_{x}dN},\ r=N^{2}\frac{B_{x}}
{B_{\gamma d^{2}}}
\end{equation*}
with $B_{\gamma d^2}=\Phi_{0}/(\gamma d^2)$ and, adding the energy of isolated
Josephson vortices, we can write the total energy of the lattice as
\begin{eqnarray}
&&f_{\mathrm{Jl}}(N,q,h)=\frac{B_x^{2}}{8\pi }\nonumber\\
&&+\frac{B_x\Phi
_{0}}{(4\pi)^{2}\lambda_{\mathrm{ab}} \lambda_{\mathrm{c}}}\left[ \frac{1}{2}\ln \left(
\frac{1}{h}\right) +1.432+G_{\rm L}(r,q)\right]
\label{eq:EnLonGenTotal}
\end{eqnarray}
with $h\equiv 2\pi B_{x}/B_{\gamma d^{2}}$ and $r=N^{2}h/2\pi $. For given $h$ the
ground state configuration is determined by the minimum of $G_{\rm L}(N^2h/2\pi,q)$
with respect to discrete $N$ and continuous $q$. As follows from Eq.\
\eqref{eq:CommensFields}, perfect fits where $G_{\rm L}$ reaches its absolute
minimum occur at the set of reduced fields $h=h_{(m,n)}(N)$ with
\begin{equation} \label{eq:RedCommensFields}
h_{(m,n)}(N)=\frac{\sqrt{3}\pi}{N^{2}(m^{2}+mn+n^{2})}.
\end{equation}
 At these
fields this energy reproduces the result \eqref{eq:jvlenergy}.
The field dependence of $G_{\rm L}$ for the ground state is shown in Fig.\
\ref{fig:JVL-GrSt-Lon}. The continuous London model does not not accurately describe
layered superconductors at high fields. To obtain lattice structures in this region
one has to consider the more general Lawrence-Doniach model. The transition between
the \emph{aligned} lattices have been studied within this model by Ichioka
\cite{Ichioka95}. However, our analysis in the next section shows that at many
fields the true ground state is not given by an aligned lattice.

\subsection{Evolution of ground-state configurations within
Lawrence-Doniach model} \label{sec:grst-num}

The accurate analysis of the lattice configurations within the Lawrence-Doniach
model which we report in this section was only published in short Proceeding
\footnote{A. E. Koshelev, Proceedings of FIMS/ITS-NS/CTC/PLASMA 2004, Tsukuba,
Japan, Nov. 24–28, 2004; arXiv:cond-mat/0602341.}. Independently, such numerical
analysis was done by Nonomura and Hu \cite{NonomuraHu2006} with fully consistent
results.

At high in-plane magnetic fields the spatial variations of the field are very small
and in the first approximation can be neglected. In this limit the only relevant
degrees of freedom are the superconducting phases and the relevant part of the LLD
energy \eqref{eq:flld} per unit volume, $f_{\phi}\equiv F_{\rm
LLD}/(L_xL_yL_z)-B_x^2/(8\pi)$, can be written as
\begin{eqnarray}
&&f_{\phi}[\phi_{n}({\mathbf r})]\!=\frac{E_0}{L_yL_z}\sum_{n}\!\int\!dy\!\left[
\frac{1}{2}\left(  \nabla_{\!\!y}\phi_{n}\right)  ^{2}\right.\nonumber\\
&&\!+\!
\left.\frac{1}{\left(  \gamma d\right)  ^{2}}\left( 1\!-\!\cos\left(
\phi_{n+1}\!-\!\phi_{n}\!+\!\frac{2\pi dB_{x}y}{\Phi_{0}}\right)
\right)  \right]. \label{eq:PhaseEnInPlane}
\end{eqnarray}
To simplify the analysis we will introduce the reduced in-plane length
$\scaledy\equiv y/\gamma d$ and the reduced magnetic field $h\equiv2\pi\gamma
d^{2}B_{x}/\Phi_{0}$ leading to
\begin{eqnarray}
&&f_{\phi}[\phi_{n}({\mathbf r})]=\frac{\varepsilon_{\rm J}}{L_yL_z} \sum_{n}\int d\scaledy\left[  \frac {1}{2}\left(
\nabla_{\!\scaledy}\phi_{n}\right)  ^{2}\right.\nonumber\\
&&+\left.1\!-\!\cos\left(  \phi_{n+1}
\!-\!\phi_{n}+h\scaledy\right)  \right] \label{eq:PhaseEnInPlaneR}
\end{eqnarray}
with $\varepsilon_{\rm J}\equiv E_0/\gamma d$. Varying this energy with respect to
the phases $\phi_{n}(\scaledy)$ we obtain an equation for the equilibrium phase
distribution [equivalent to \eqref{eq:phaseA} when we ignore the spatial dependence
in $B_x$]
\begin{equation}\label{eq:PhaseInplane}
\hspace{-0.07in}\nabla_{\!\scaledy}^{2}\phi_{n}\!+\!\sin\left(
\phi_{n+1}\!-\!\phi_{n}+h\scaledy\right) \! -\!\sin\left(
\phi_{n}\!-\!\phi_{n-1}+h\scaledy\right) \! =\!0.
\end{equation}

We again consider a general lattice shown in Fig.\ \ref{fig:JVLcommens}a with
in-plane period $a$, $N$ layers between neighboring rows, and relative shift $qa$
between relative rows with $a$ and $N$ being related to the reduced field as
$h=2\pi\gamma d/Na$. It is sufficient to find the solution for the phase in one unit
cell, $0<y<a$, $1\leq n\leq N$, using appropriate quasi-periodicity conditions for
the phase. The total lattice energy  per unit volume can be represented as
\begin{equation}
f_{\phi}=\frac{B_{x}\Phi_{0}}{(4\pi)^{2}
\lambda_{\mathrm{ab}} \lambda_{\mathrm{c}}}u(N,q,h), \label{eq:LDLatEnTotal}
\end{equation}
where the reduced energy $u(N,q,h)$ per unit cell is given by
\begin{eqnarray}
&&u(N,q,h)=\frac{1}{\pi}\sum_{n=1}^{N}\int_{0}^{a} d\scaledy\left[
\frac{1}{2}\left( \frac{d\phi_{n}}{d\scaledy}\right)
^{2}\right.\nonumber\\
&&\left.+1-\cos\left( \phi_{n+1}-\phi_{n}+h\scaledy\right)
\right] . \label{eq:LDEner}
\end{eqnarray}
Using a relaxation method to solve \eqref{eq:PhaseInplane} numerically within one
unit cell, we can find the energy $u$ for any given values of $N$, $q$ and $h$. To
match the London representation \eqref{eq:EnLonGenInter}, we write $u(N,q,h)$ in the
form
\begin{equation}
u(N,q,h)=\frac{1}{2}\ln\frac{1}{h}+1.4323+G(N,q,h) \label{eq:LDEner-pres}
\end{equation}
where the function $G(N,q,h)$ defined by this equation approaches the London limit
$G_{\rm L}(r\!=\!N^{2}h/2\pi,q)$ for $h\rightarrow0$.

\begin{figure*}[ptb]
\begin{center}
\includegraphics[width=0.8\textwidth]{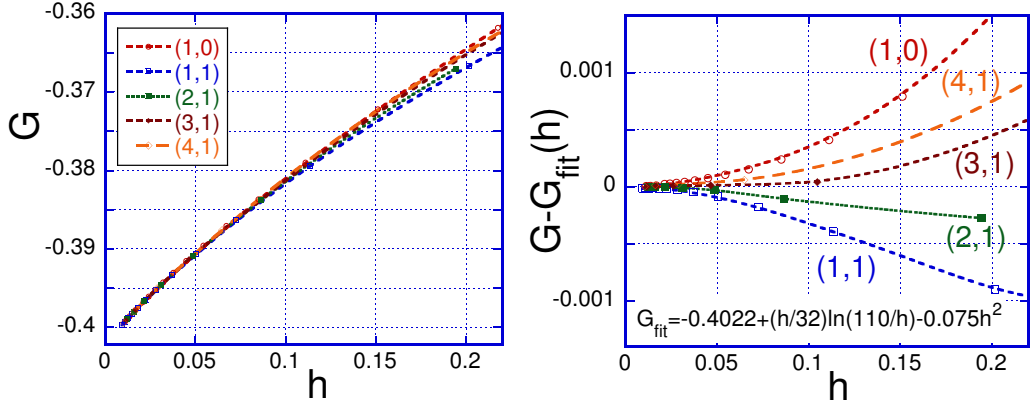}
\caption{{\emph Left panel}: Field dependence of the reduced-energy function
$G(N,h,q)$ for several lattice orientations $(m,n)$ at the commensurate field
$h_{(m,n)}(N)$. To enlarge small differences, we plot in the right panel the
difference between $G$ and its fit obtained using all data for $h<0.05$.}
\label{fig:G-h-orient}
\end{center}
\end{figure*}
We consider first influence of the layered structure at small fields. As shown in
the appendix \ref{App:JVphaseAsymp}, in the lowest order with respect to $h$ the
layered structure gives orientation-independent correction to energy, $G\approx
(h/32)\ln(C_h/h)$. In higher (quadratic) order the layered structure generates
orientation-dependent correction to the lattice energy leading to a break down of
the ``elliptic-rotation'' degeneracy of the lattice. To study this effect
quantitatively we plotted in Fig.\ \ref{fig:G-h-orient} the computed field
dependencies of $G(N,q,h)$ for several lattice orientations at the corresponding
reduced commensurate fields $h_{(m,n)}(N)$ given by \eqref{eq:RedCommensFields}. At
small $h$, $h<0.05$, neglecting a very weak dependence on orientation, one can
accurately fit the correction from the layeredness as
\begin{equation}\label{eq:GhGLFit}
G(h)-G_{\rm L}\approx \frac{h}{32}\ln\frac{110}{h}-0.075h^2.
\end{equation}
One can see that among the two aligned structures shown in Fig.\ \ref{fig:JVLalign}
the layers favor the lower structure with indices $(1,1)$. However, for $h<0.1$ the
energy difference between the two structures is tiny and in real samples external
factors may select the lattice orientation. On the other hand, one can expect that
at sufficiently large fields the ground-state configuration will be selected by the
layered structure even in real samples.

\begin{figure}[ptb]
\begin{center}
\includegraphics[width=3.4in]{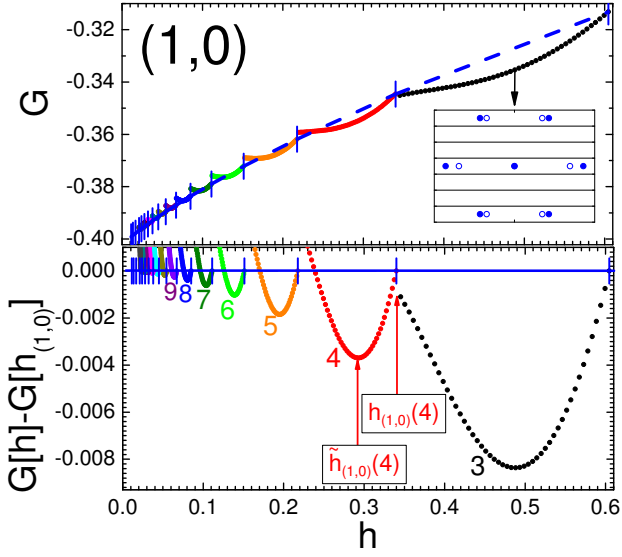}
\caption{\emph{Upper panel}: The field dependences of the
reduced-energy function $G=G(N,h,0.5)$ for the aligned lattice (1,0)
and different $N$. Vertical bars mark locations of the London-model
commensurate fields $h_{(1,0)}(N)$. \emph{Lower panel} shows
difference between $G(N,h,0.5)$ and smooth curve through the points
($h_{(1,0)}(N)$, $G(N,h_{(1,0)}(N),0.5)$) (dashed line in the upper
panel). One can see that the matching fields systematically
displaced to the lower values $\tilde{h}_{(1,0)}(N)$, as illustrated
for $N=4$. Inset in the upper panel shows lattice structure at the
displaced matching field for $N=3$ (solid symbols) in comparison
with the regular-hexagon structure at the London matching field.}
\label{fig:dh10}
\end{center}
\end{figure}
Energy corrections due to the layered structure favor lattice stretching along the
layer direction and shift down the matching fields. This effect is strongest for the
aligned lattice (1,0) and is illustrated in Fig.\ \ref{fig:dh10}. In this figure we
show the field dependences of $G(N,h,0.5)$ for different $N$. When a smooth function
is subtracted from these dependences, local minima are realized at fields
$\tilde{h}_{(1,0)}(N)$ which are smaller then the London matching field
$h_{(1,0)}(N)$. The shift $\tilde{h}_{(1,0)}(N)-h_{(1,0)}(N)$ rapidly decreases with
increasing magnetic field. We found that this shift is described by the following
equation
\[
\tilde{h}_{(1,0)}(N)\approx
\frac{h_{(1,0)}(N)}{1+(0.63/N^2)\ln(19/\tilde{h}_{(1,0)}(N))}.
\]
For other lattice orientations the shift is smaller but still noticeable. To
quantify the energy difference between the aligned lattices due to the layered
structure, we fit their energies at the \emph{shifted} matching field for $h<0.1$ to
smooth curves and subtract these curves. This procedure gives
$G_{(1,1)}-G_{(1,0)}\approx -0.011 h^2$.

\begin{figure}[ptb]
\begin{center}
\includegraphics[width=3.4in]{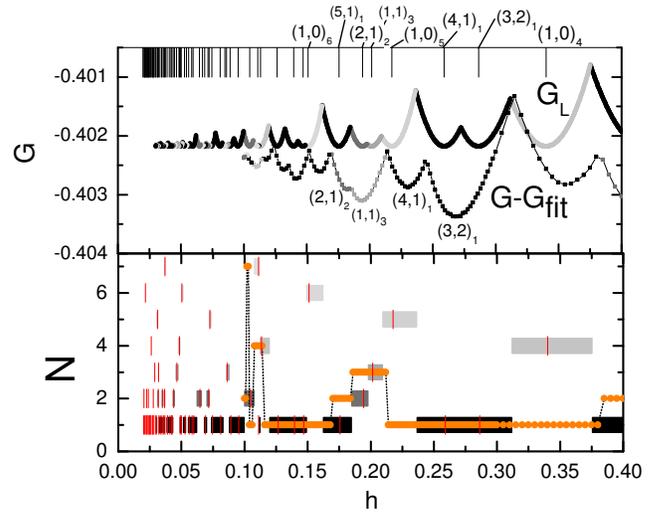}
\caption{\emph{Upper panel:} The field dependence of the reduced
energy functions for London model ($G_{\rm L}$, upper curve) and
full Lawrence-Doniach model ($G-G_{\rm fit}$, lower curve). For
clearer comparison we subtracted from $G(h)$ its fit at small $h$
given by Eq.\ \eqref{eq:GhGLFit} . Values of commensurate fields
$h_{(m,n)}(N)$ are shown in the top axis and the corresponding
indices for several of them are written in the format $(m,n)_{N}$.
As expected, $G_{\rm L}$ reaches its absolute minimum for every
$h_{(m,n)}(N)$. The \emph{lower panel} shows field dependence of $N$
for ground state for both models (stripes for the London model and
circles for the Lawrence-Doniach model). The same graylevel codes
the value of $N$ in the upper panel and the London-model plot in the
lower panel.}
\label{fig:JVL-GrSt-Lon}
\end{center}
\end{figure}
\begin{figure}[ptb]
\begin{center}
\includegraphics[width=3.4in]{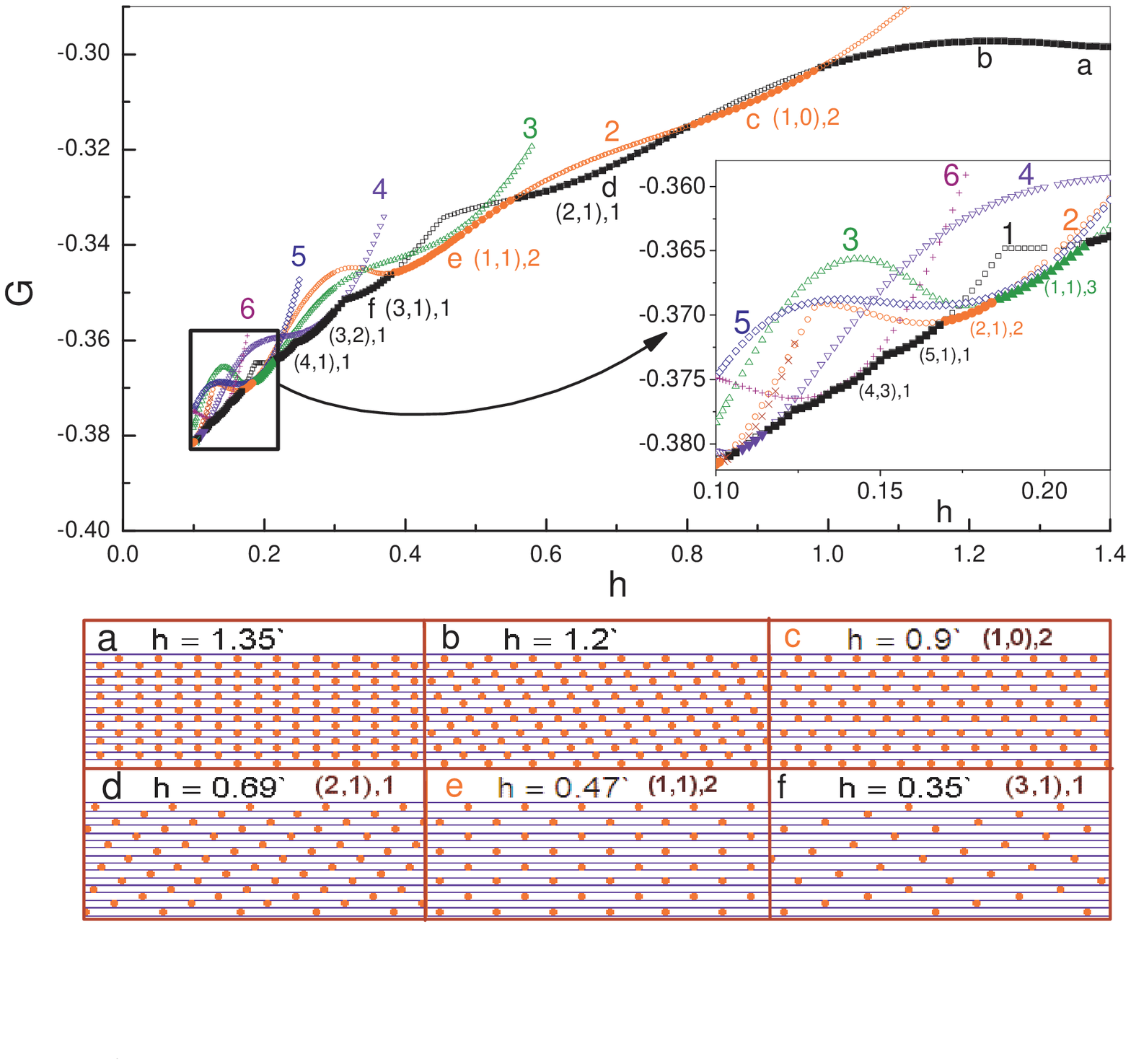}
\caption{Field dependence of the energy function $G(N,q,h)$ for the
ground state and competing states. Each curve corresponds to the
minimum of $G(N,q,h)$ with respect to $q$ at fixed $h$ and $N$. The
curves are marked by the value of $N$. Lattice configurations in
scaled coordinates are shown at six marked fields. Inset illustrates
competition between different configuration at
smaller fields }
\label{fig:G-h-Configs}
\end{center}
\end{figure}

We can now explore the evolution of the ground-state configuration by direct
minimization of the energy with respect to the lattice parameters $N$ and $q$ as
defined in Fig.\ \ref{fig:JVLcommens}. To this end we have computed the reduced
ground-state energy defined as $G(h)\equiv\min_{N,q}[G(N,q,h)]$. We checked that if
we consider only aligned lattices, the results of Ichioka \cite{Ichioka95} are
reproduced for the transition fields between lattices with different periods $N$ in
the case of large anisotropy. For comparison, we also made a similar calculation for
the London model and computed the field dependence of the function $G_{\rm
L}(h)=\min_{N,q}[G_{\rm L}(r\!=\!N^{2}h/(2\pi),q)]$ where $G_{\rm L}(r,q)$ is
defined in Eqs.\ \eqref{eq:EnLonGenInter} and \eqref{eq:gfun}. In figure
\ref{fig:JVL-GrSt-Lon} we compare field evolutions of these ground-state reduced
energies and the corresponding c-axis period $N$. For clearer comparison we
subtracted from $G(h)$ its fitted correction from
 $G_{\rm L}(\sqrt{3}/2,1/2)$ at small $h$ given in \eqref{eq:GhGLFit}.
Values of the London commensurate fields $h_{(m,n)}(N)$ are shown on the top axis
with several low-order fields being marked by corresponding indices using the format
$(m,n)_{N}$. As expected, $G_{\rm L}(h)$ reaches its absolute minimum for every
$h_{(m,n)}(N)$. We can observe several interesting properties. As the lattice
orientation with indices $(m,n)=(1,0)$ is not favored by the layered structure,
several low-$N$ configurations, $3\leq N\leq 6$, expected at $h=h_{(1,0)}(N)$, are
skipped. However, as one can see from the inset in Fig.\ \ref{fig:G-h-Configs}, for
$N=5$ and $6$ the ground-state energy is smaller than the energies of these states
at $h=h_{(1,0)}(N)$ only by a tiny value. For $h<0.2$ the actual evolution of
lattice structure starts to follow roughly the London route (except for skipped
state $(1,0)_{6}$ near $h=0.16$) but with small negative offset, i.e., we again see
that the matching fields systematically shifted down in comparison with their London
values.

The field dependence of the energy function $G(N,q,h)$ in an extended field range is
shown in figure~\ref{fig:G-h-Configs} for the ground state and competing states.
Each curve corresponds to the minimum of $G(N,q,h)$ with respect to $q$ at fixed $h$
and $N$ and is marked by its value of $N$. We also show the first six lattice
configurations which are realized with decreasing field. The inset of the figure
blows up the low-field region. One can see that many lattice configurations compete
for the ground state at small fields and at several fields (e.g., at $h\approx,
0.19, 0.137, 0.105 \ldots$) one or more lattice configurations have energies very
close to the ground-state energy. We also note that there are several extended field
ranges where in the ground state all layers are homogeneously filled with vortices
($N=1$) even in the region of the dilute vortex lattice, e.g., $0.115<h<0.17$,
$0.21<h<0.38$.

We see that an accurate consideration within both London and Lawrence-Doniach models
shows that the ground state of the Josephson vortex lattice at low temperatures does
not give any preference to the lattices aligned with the layers. Therefore for
equilibrium field dependencies one can not expect to observe any strong features at
the matching fields of these lattices, $B_{(1,0)}(N)$ and $B_{(1,1)}(N)$ given by
Eq.\ \eqref{eq:bnm}. Nevertheless, clear commensurability oscillations have been
observed experimentally in underdoped YBCO in irreversible magnetization
\cite{OussenaGGT94,ZhukovKPC1999} and nonlinear resistivity \cite{GordeevZdG2000}.
The period of these oscillations corresponds to the fields $B_{(1,0)}(N)$ indicating
that for some reason in this material the aligned lattice $(1,0)$ occurs to be
preferable. We note that, due to small differences between the energies of different
configurations, in real materials aligned lattices can be selected by external
factors, such as interaction with correlated disorder (twin boundaries or
dislocations) or sample surface. We will also see that the aligned lattice with
indices $(1,0)$ is favored by thermal fluctuations. Finally, we mention the work of
Ikeda and Isotani \cite{IkedaI1999} who performed similar analysis of the ground
state configurations for field applied along the layers within \emph{the lowest
Landau level approximation}.

\subsection{Properties of metastable states in London model}

Josephson vortices can slide easily along the layers but there is a huge barrier for
the motion across the layers. This property makes it hard to equilibrate the
lattice. It also leads to the appearance of a very large number of metastable
states. The properties of these states have been considered in Refs.\
\onlinecite{IvlevKP1990,Levitov1991}. Systematically, metastable states at fixed
c-axis period can be sampled by first slowly cooling down the superconductor at
fixed magnetic field and then in a second step decreasing the magnetic field at a
low temperature \cite{Levitov1991}. We assume that the prepared starting
configuration is the \emph{aligned} lattice. As the c-axis period, $N$, is locked by
the layers, the lattice stretches along the layers with lowering the field, i.e, the
ratio $r=b/a$ decreases. During stretching, these fixed-$N$ metastable states go
through a sequence of nontrivial structural transformations. In the London regime,
the aligned configuration becomes unstable at $r_0\approx 1.51/(2\pi) \approx 0.24$
\cite{IvlevKP1990}. This instability is driven by the repulsion between neighboring
vortices in the vertical stack. At low $r<r_0$ the parameter $q$ continuously
decreases starting from $1/2$ to lower values. We found that the layeredness
stabilizes the aligned structures: the critical ratio decreases to $0.231$ at $N=3$
and to $0.224$ at $N=2$. It is important to note that the shear instability occurs
\emph{in the ground state} only for $N=1$ (we consider in detail this structural phase transition below). At higher values of $N$ this instability
only occurs when the state for this given $N$ is metastable with respect to other
values of $N$. This instability is considered in detail below.

\begin{figure}[ptb]
\centerline{\includegraphics[width=3.4in]{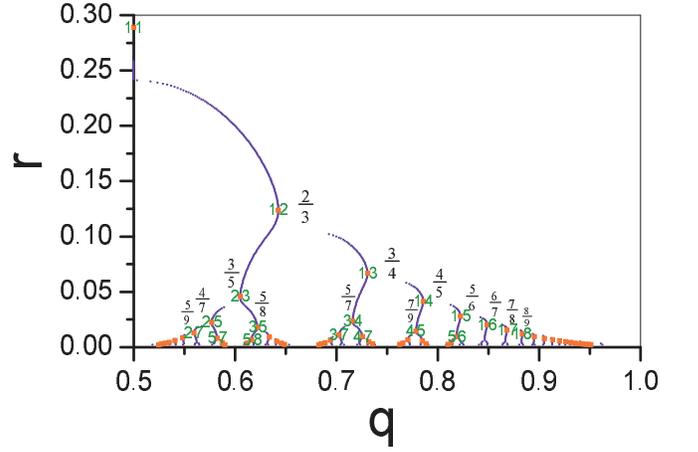}}\caption{Levitov's
hierarchical plot of metastable states in $q$-$r$ plane
\cite{Levitov1991} (in this plot $q$ is selected within the interval
[0.5,1]). Each dotted curve is obtained from the local minima of
function $G_{L}(r,q)$ with respect to $q$ at fixed $r$. New branches
appear as a result of ``quasi-bifurcations''. Each
``quasi-bifurcation'' is associated with a rational number. The
branches turn at points $(q_{(m,n)},r_{(m,n)})$ corresponding to
ground states (marked by squares and labelled by the indices $m n$
in the plot). } \label{fig:Levitov}
\end{figure}
The statistics of metastable states has been explored in detail in Ref.\
\onlinecite{Levitov1991} where the similarity to the phyllotaxis phenomenon in
biological systems has been pointed out. For every $r$ one can find all local
minima, $q_i(r)$ of the energy function $G(r,q)$ with respect to $q$ and plot all
these minima in the $q$-$r$ plane (see Fig.\ \ref{fig:Levitov}). The obtained
pattern is quite peculiar. At $r>r_0$ the only minimum is at $q_0(r)=1/2$. Below
$r=r_0$ this trajectory symmetrically splits into two. With further decrease of $r$
many more minima appear forming a complex hierarchical structure. The pattern can be
viewed as a series of ``quasi-bifurcations'' occurring near rational values of $q$.
``Quasi-bifurcation'' corresponds to the appearance of a new branch below a certain
value of $r$ in the vicinity of the old branch. The branches turn at the points
$(q_{(m,n)},r_{(m,n)})$ corresponding to ground states. The evolution of the initial
state is described by the two main trajectories symmetrically split from $q=1/2$.
The trajectory with $q>1/2$ ``quasi-bifurcates'' at $q=F_j/F_{j+1}$ where $F_j$ are
the Fibonacci numbers and approaches the ``golden ratio'' $(\sqrt{5}-1)/2\approx
0.618$ as $r\rightarrow 0$. It goes through ground states with the indices also
described by the Fibonacci sequence, $(m,n)=(F_{j+1},F_j)$. Unfortunately, these
exciting predictions have never been verified experimentally because there is no
direct way to probe the structure of the Josephson vortex lattice.

\subsection{Elasticity of dilute Josephson vortex lattice}

Josephson vortices easily slide along the layers but motion across the layers is
strongly suppressed by intrinsic pinning from the layers. Due to the intrinsic
pinning, $z$ axis fluctuations of the vortex lines occur via kink formation. In
moderately anisotropic layered superconductors, such as YBCO, in which the $c$-axis
coherence length is larger than or comparable with the interlayer spacing $d$, the
intrinsic pinning potential $V(u_{z})$ can be described as a cosine function of $z$
axis vortex displacements $V(u_{z} )=-V_{0}\cos(2\pi u_{z}(x)/d)$. However such
description becomes inadequate in strongly layered materials, where the structure of
a kinks is very similar to the structure of a pancake vortex.

In strongly layered materials at low temperatures one can neglect kink formation and
take into account only in-plane lattice deformations $u(\mathbf{r})\equiv
u_{y}(\mathbf{r})$ (planar-fluctuations model). In this case, one can derive the
following nonlocal elastic energy in the $k$-space
\begin{equation}
F_{el}=\frac{1}{2}\int\frac{d^{3}\mathbf{k}}{(2\pi)^{3}}\left[
c_{11}
(\mathbf{k})k_{y}^{2}+c_{44}(\mathbf{k})k_{x}^{2}+c_{66}k_{z}^{2}\right]
\left\vert u(\mathbf{k})\right\vert ^{2} \label{eq:Fel-JVL-dilute}
\end{equation}
with elastic moduli
\begin{eqnarray} c_{66}  &
=&\frac{B_{x}\Phi_{0}}{(8\pi)^{2}\lambda_{\mathrm{c}}^{2}\gamma
},\label{eq:C66-JVL-dil}\\
c_{11}(\mathbf{k})  &  =&\frac{B_{x}^{2}/4\pi}{1\!+\!\lambda_{\mathrm{ab}}^{2}k_{z}
^{2}\!+\!\lambda_{\mathrm{c}}^{2}\left(  k_{y}^{2}+k_{x}^{2}\right)}\!-\!\frac{B_{x}\Phi_{0}
}{(8\pi)^{2}\lambda_{\mathrm{ab}}\lambda_{\mathrm{c}}},\label{eq:C11-JVL-dil}\\
c_{44}(\mathbf{k})  & =&\frac{B_{x}^{2}/4\pi}{1\!+\!\lambda_{\mathrm{ab}}^{2}k_{z}
^{2}\!+\!\lambda_{\mathrm{c}}^{2}\left(  k_{y}^{2}+k_{x}^{2}\right) }\nonumber\\&\!+\!&\frac{B_{x}\Phi_{0}
}{(4\pi)^{2}\lambda_{\mathrm{ab}}\lambda_{\mathrm{c}}}\ln\frac{1}{s\sqrt{c_{z}^{-2}\!+\!\left(
\gamma k_{x}/\pi\right)  ^{2}}}. \label{eq:C44-JVL-dil}
\end{eqnarray}
While the tilt [$c_{44}(\mathbf{k})$] and compression [$c_{11}(\mathbf{k})$] moduli
are not sensitive to exact lattice structure, the formula for the shear modulus
$c_{66}$ is valid only for perfect matching between the Josephson vortex lattice and
layered structure, which is achieved at matching fields \eqref{eq:CommensFields}.
For a general lattice shown in Fig. \ref{fig:JVLcommens}a one can derive a more
general expression for $c_{66}$ using representation
\eqref{eq:EnLonGenInter}--\eqref{eq:gfun} for the lattice energy \cite{IvlevKP1990}
and relation between lattice deformation and change of parameter $q$, $\delta
q=rdu/dz$,
\begin{equation}
c_{66}=\frac{B_{x}\Phi_{0}}{\left(  8\pi\right)
^{2}\lambda_{\mathrm{c}}^{2}\gamma
}g_{66}(r,q) \label{eq:C66-JVL-dil-gener}
\end{equation}
with
\begin{eqnarray}
&&g_{66}(r,q)  =4r^{2}\frac{\partial^{2}}{\partial q^{2}}G_{L}(r,q)=\!-\!\left(4\pi\right)^{2}r^{2}\hspace{1.4in}\nonumber \\
&&
\times\sum_{l=1}^{\infty}
\!\frac{\cos(2\pi ql) \cosh(2\pi rl) \!-\!\sin^{2}\!(2\pi ql) \!-\!1}
{\left(\cosh(2\pi rl)  -\cos(2\pi ql) \right)
^{3}}l\sinh(  2\pi rl).\nonumber
\end{eqnarray}
This formula reproduces the result \eqref{eq:C66-JVL-dil} for the commensurate
configurations $(r,q)=(r_{(n,m)},q_{(n,m)})$. It also describes instability of
aligned configuration ($q=1/2$) at $r\approx0.24$ \cite{IvlevKP1990}.

Softest mode in the planar model corresponds to shearing between neighboring planar
arrays of Josephson vortices. The harmonic approximation breaks for this mode first.
The simplest extension of the linear elastic energy which describes strong
interplanar fluctuation is amounts to replacement of the continuous displacement
field $u(\mathbf{r})$ by displacement of the planar arrays $u_{j}(x,y)\equiv
u(x,y,jb)$ and replacement the shear term in the energy by the nonlinear interaction
term
\begin{eqnarray*}
&&\int d^{3}\mathbf{r}\frac{c_{66}}{2}\left(  \frac{du}{dz}\right)
^{2}\\
&&\rightarrow\int d^{2}\mathbf{r}\frac{a^{2}c_{66}}{\left(
2\pi\right) ^{2}b}\sum_{j}\left[  1\!-\!\cos\left(
2\pi\frac{u_{j+1}\!-\!u_{j}}{a}\right) \right]
\end{eqnarray*}
Such extension has been used to study strong-fluctuations region
\cite{Korshunov1991}.

\section{Dense lattice, $B_x>\Phi_{0}/2\pi\gamma d^{2}$}
\label{sec:densejvl}


With increasing magnetic field distance between Josephson vortices decreases and at
field $B\sim B_{\mathrm{cr}}=\Phi_{0}/2\pi\gamma d^{2}=B_{\gamma d^2}/2\pi$ this
distance becomes of the order of the vortex-core size. In contrast to the Abrikosov
vortex lattice, for which overlap of the vortex cores marks disappearance of
superconductivity, for the Josephson vortex lattice this field just marks a
crossover to a new regime, the dense Josephson vortex lattice. Existence of this
regime was pointed out by Bulaevskii and Clem \cite{BulaevskiiC1991}. In the dense
Josephson vortex lattice the gauge invariant phase difference is a smoothly
increasing function of distance and the Josephson coupling energy can be treated as
a small perturbation. This allows for the following quantitative description.

\subsection{Very high fields: Quantitative description using expansion
in Josephson coupling}

\begin{figure}[ptb]
\centerline{\includegraphics[width=7cm,clip]{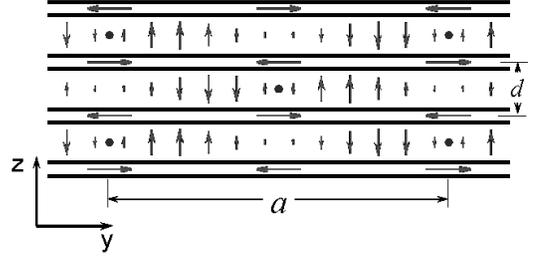}}\caption{
Schematic distribution of currents in the dense Josephson vortex lattice.
The circles mark the centers of the Josephson vortices. }
\label{fig:DenseJosLat}
\end{figure}

At high fields $B_x>B_{\mathrm{cr}}$, vortices homogeneously fill all of
the layers. This means that all layers are equivalent and the in-plane lattice
period is $\tilde{a}=2\pi/h$ (see figure~\ref{fig:DenseJosLat}). When the strong
inequality $B_x\gg B_{\mathrm{cr}}$ ($h\gg1$) is satisfied Eq.\
\eqref{eq:PhaseInplane} for the phases can be solved using an expansion with respect
to the Josephson currents. In the zeroth order we can construct a regular lattice
with an arbitrary translation from layer to layer by using the form,
\[
\phi_{n}^{(0)}=\kappa\frac{n(n-1)}{2}.
\]
This corresponds to the gauge-invariant phase difference
\[
\varphi_{n,n+1}^{(0)}=\kappa n+h\scaledy,
\]
i.e., the planar lattices in the neighboring layers are shifted by the fraction
$q=\kappa/2\pi$ of the in-plane lattice spacing $\tilde{a}$. In the first order we
obtain
\[
\nabla_{\!\scaledy}^{2}\phi_{n}^{(1)}+
\sin\left(  \kappa n+h\scaledy\right)  -\sin\left(
\kappa(n-1)+h\scaledy\right)  =0
\]
which gives
\[
\phi_{n}^{(1)}(\scaledy)=\frac{1}{h^{2}}\left[  \sin\left(  \kappa n+h\scaledy\right)
-\sin\left(  \kappa(n-1)+h\scaledy\right)  \right].
\]
Substituting this solution into \eqref{eq:PhaseEnInPlaneR}, we obtain the energy per
unit volume up to second order with respect to the Josephson coupling,
\begin{equation}
f_{\phi}(\kappa,h)= \frac{\varepsilon _{\rm J}}{\gamma d^{2}}
\left(1-\frac{1-\cos\kappa}{2h^{2}}\right).\label{eq:EnInplane_kappa}
\end{equation}
We can immediately see that the minimum energy $f_{\min}(h)= (\varepsilon _{\rm
J}/\gamma d^{2})(1-1/h^{2})$ is achieved at $\kappa=\pi$, corresponding to the
triangular lattice shown in figure \ref{fig:DenseJosLat}. The phase distribution in
the ground state is given by
\begin{equation}
\phi_{n}(\scaledy)\approx\pi\frac{n(n-1)}{2}+\frac{2(-1)^{n}}{h^{2}}\sin\left(
h\scaledy\right) \label{eq:DenseLatPhase}
\end{equation}
From this solution we can recover the distributions of the in-plane and Josephson
current
\begin{eqnarray*}
j_{y,n}(y)  \!&\approx&-\frac{2(-1)^{n}}{h}\gamma j_{\rm J}\cos\left(  \frac{2\pi
dB_xy}{\Phi_{0}}\right)  ,\\
j_{z,n}(y)  \!& \approx&\!-\!(-1)^{n}j_{\rm J}\sin\left[
\!-\!\frac{4(-1)^{n}}{h^{2}} \sin\left(\!\frac{2\pi
dB_{x}y}{\Phi_{0}}\!\right)  \!+\!\frac{2\pi dB_{x}y} {\Phi_{0}}\right]\!,
\end{eqnarray*}
and a weak modulation of the in-plane field
\[
B_x(y)\approx B_x - \frac{(-1)^{n}\Phi_{0}^{2}}{B_x(2\pi
d\lambda_{\mathrm{c}})^{2} }\cos\left(  \frac{2\pi dB_xy}{\Phi_{0}}\right)
.
\]
A schematic distribution of the currents is shown in figure~\ref{fig:DenseJosLat}.

\subsection{Dense lattice close to the crossover region. Structural phase
transition}

When the magnetic field approaches the crossover field $\Phi_{0}/(2\pi\gamma
d^{2})$, the perturbative approach of the previous section becomes insufficient and
one has to obtain a full solution of the nonlinear equation \eqref{eq:PhaseInplane}.
The general solution for the lattice with the arbitrary phase shift $\kappa$ can be
written as
\begin{equation}
\phi_{n}(\scaledy)=\kappa\frac{n(n-1)}{2}+g\left(  \scaledy+\frac{\kappa n}{h}\right)
,\label{eq:PhDenGen}
\end{equation}
where $g(\scaledy)$ is a periodic function, $\;g\left(  \scaledy+2\pi/h\right)
=g(\scaledy)$, which obeys the following equation
\begin{eqnarray}
\frac{d^{2}g}{d\scaledy^{2}}&+&\sin\left(  g\left(
\scaledy+\frac{\kappa}{h}\right) -g\left(  \scaledy\right)  +h\scaledy\right)\nonumber\\
&-&\sin\left(  g\left(  \scaledy\right)  -g\left(
\scaledy-\frac{\kappa}{h}\right) +h\scaledy-\kappa\right)  =0.\label{eq:DenEq_g}
\end{eqnarray}
The reduced energy $\bar{f}\equiv f_{\phi}\gamma d^2/\varepsilon _{\rm J}$ can also
be written in terms of the function $g(\scaledy)$,
\begin{eqnarray}
\bar{f}&=&\int_{0}^{2\pi/h}\frac{hd\scaledy}{2\pi}
\left\{
\frac{1}{2}\left( \frac{dg} {d\scaledy}\right)  ^{2}
\right.
\nonumber\\
&+&
\left.
1-\cos\left[
g\left( \scaledy+\frac{\kappa}{h}\right)  -g\left( \scaledy\right)
+h\scaledy\right]
\right\}. \label{eq:DenEn_g}
\end{eqnarray}

\begin{figure}[ptb]
\centerline{\includegraphics[width=7cm,clip]{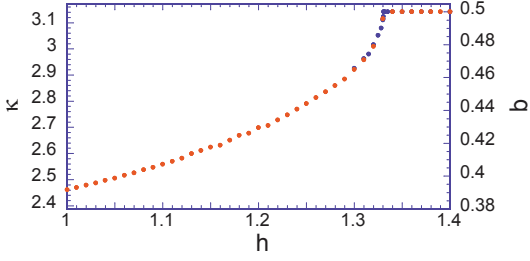}}\caption{
Field dependence of the phase shift $\kappa$ and corresponding
lattice shift $q$ for the dense Josephson vortex lattice. At
$h\approx1.332$ the lattice experiences a continuous structural
phase transition.} \label{fig:kappa_h}
\end{figure}

Equation \eqref{eq:DenEq_g} does not have an analytical solution and has to be
solved numerically. Lattice configurations of the dense lattice also has been
investigated using the code developed for the lattice with general period $N$. Both
approaches give identical results. Numerical investigation shows that the triangular
lattice with $\kappa=\pi$ gives the ground state for $h>1.332$. At $h\approx1.332$
the system has a second-order phase transition to a lattice of lower symmetry, see
lattice structures for $h=1.35$ (a) and $h=1.2$ (b) in figure \ref{fig:G-h-Configs}.
The field dependence of $\kappa$ and corresponding lattice shift $q$ are shown in
figure~\ref{fig:kappa_h}. Ikeda and Isotani \cite{IkedaI1999} found that within the
lowest Landau level approximation this structural phase transition occurs at
somewhat higher value, $h\approx 1.4$.

\begin{figure}[ptb]
\centerline{\includegraphics[width=7cm,clip]{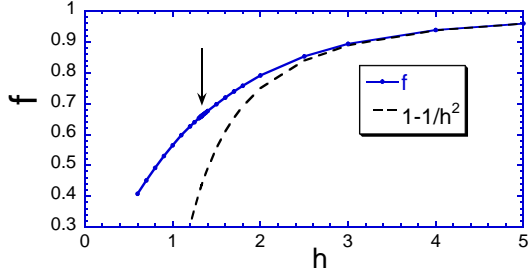}}\caption{
Field dependence of the reduced energy for the dense Josephson
vortex lattice. The dashed line shows the high-field asymptotics.
The arrow marks the position of the structural phase transition at
$h\approx 1.332$.} \label{fig:En_h_dense}
\end{figure}

To study validity range of the high-$h$ approximation of the previous section we
plot in figure~\ref{fig:En_h_dense} the computed field dependence of the reduced
energy together with its high-field asymptotics, derived in the previous section. As
one can see, the perturbative approach gives a good approximation for the energy
down to $h\sim 2$.

\subsection{Elasticity of dense lattice}

In this section we consider the deformation energy of the dense Josephson vortex
lattice in the limit $h=2\pi\gamma d^{2}B_{x}/\Phi_{0}\gg1$. In particular, this
energy serves as a starting point for analysis of fluctuations. We will follow the
approach used by Korshunov and Larkin \cite{KorshunovL1992}. The starting point of
the analysis is again reduced LLD energy in the phase approximation \eqref{eq:PhaseEnInPlaneR} which we
rewrite now for the general case of the phase depending on both reduced coordinates
$\bar{\mathbf{r}}\equiv(\bar {x},\bar{y})=\mathbf{r}/\gamma d$,
\begin{equation}
\hspace{-0.08in}F_{\phi}\!=\!E_{0}\!\sum_{n}\!\int\!d^2\bar{\mathbf{r}}\!\left[
\frac {1}{2}\left( \! \frac{d\phi_{n}}{d\bar{\mathbf{r}}}\!\right)
^{2}\!-\!\cos\left(
\phi_{n+1}\!-\!\phi_{n}\!+\!h\bar{y}\right)  \right]\! .  \label{eq:PhaseEn3D}
\end{equation}
The ground-state phase distribution is given by Eq.\ \eqref{eq:DenseLatPhase}. Now
we consider small deformations of the lattice and split the total phase into the
smooth ($v_{n}$) and rapidly oscillating in the $y$ direction ($\tilde{\phi}_{n}$)
parts
\begin{equation}
\phi_{n}(\bar{\mathbf{r}})=\pi \frac{n(n+1)}{2}+v_{n}(\bar{\mathbf{r}})+\tilde{\phi}_{n}
(\bar{\mathbf{r}}), \label{eq:DensePhaseSplit}
\end{equation}
where we assume $dv_{n}/d\bar{y}\ll v_{n}$ and $\tilde{\phi}_{n}\ll1$. As the smooth
part of the gauge-invariant phase difference is given by $h(\bar {y}+\left(
v_{n+1}-v_{n}\right) /h)+\pi n$, the quantity $u_{n}=-\left(
v_{n+1}-v_{n}\right)  /h$ represents a local lattice displacement. Substituting
representation \eqref{eq:DensePhaseSplit} in the energy \eqref{eq:PhaseEn3D},
expanding with respect to $\tilde{\phi}_{n}$, and dropping rapidly-oscillating
terms, we obtain
\begin{eqnarray}
&&F_{\phi}\!\approx \!E_{0}\sum_{n}\int d\bar{\mathbf{r}}\left[
\frac{1}{2}\left( \frac{d\tilde{\phi}_{n}}{d\bar{y}}\right)
^{2}\!+\!\frac{1}{2}\left( \frac{dv_{n}}{d\bar{\mathbf{r}}}\right)
^{2} \right.\nonumber\\
&&\left.\!+\! \left(  \tilde{\phi} _{n+1}\!-\!\tilde{\phi}_{n}\right)
\sin\left( v_{n+1}\!-\!v_{n}\!+\!h\bar{y}\!+\!\pi n\right)  \right].
\label{eq:Ejvl-split}
\end{eqnarray}
As $\tilde{\phi}_{n}$ rapidly oscillates
only in $y$ direction, we kept only its $\bar{y}$ derivative. Minimization of this
energy with respect to $\tilde{\phi}_{n}$ gives
\[
\tilde{\phi}_{n}\!\approx\!(-1)^n\frac{\sin\left(v_{n+1}\!-\!v_{n}\!+\!h\bar{y}
\right)\! +\!\sin\left(v_{n}\!-\!v_{n-1}\!+\!h\bar{y}\right)  }{h^{2}}.
\]
Substituting this solution into Eq.\ \eqref{eq:Ejvl-split} and averaging with
respect to $\bar{y}$, we finally obtain the coarse-grained energy of the deformed
dense Josephson vortex lattice \cite{KorshunovL1992}, which we write in real units
\begin{equation}
F_{\phi}\!\approx\!\frac{E_{0}}{2}\sum_{n}\!\int\! d\mathbf{r}\left[
\left(\!\frac{dv_{n}}{d\mathbf{r}}\!\right)^{2}\!-\!\frac{\cos\left(  v_{n-1}
\!+\!v_{n+1}\!-\!2v_{n}\right) \! +\!1}{(\Lambda_{J}h)^{2}}\right]  .
\label{eq:Ejvl-dense-real}
\end{equation}
This energy describes the phase fluctuations in large in-plane magnetic field. The
first term is just usual in-plane phase stiffness energy. In the elasticity-theory
language this term represents the compression ($dv_{n}/dy$) and tilt ($dv_{n}/dx$)
contributions. The second term represents the shearing interactions between the
Josephson vortex arrays in neighboring junctions. It originates from the Josephson
coupling energy and can be viewed as the effective Josephson coupling renormalized
by the in-plane magnetic field. Roughly, we can state that with increasing magnetic
field the effective Josephson energy decreases as $1/h^{2}$ and the effective
Josephson length, $\Lambda_{Jh}$, increases linearly with $h$,
\begin{equation}
\Lambda_{Jh}=\Lambda_{J}h=\frac{2\pi\gamma^{2}d^{3}B_{x}}{\Phi_{0}}.
\label{eq:Ejvl-Length}
\end{equation}

In the case of slowly-changing from layer to layer deformation, we can expand cosine
in Eq. \eqref{eq:Ejvl-dense-real} and obtain the harmonic elastic energy of the
dense Josephson vortex lattice in terms of smooth phase deformations
\begin{eqnarray}
&&\hspace{-0.15in}F_{\phi\!-\!el} \! \approx \!\frac{E_{0}}{2}\!\sum_{n}\!\int
\!d\mathbf{r}\!\left[ \left( \frac{dv_{n}}{d\mathbf{r}}\right)^{2}\!+\!\frac{\left(
v_{n-1}\!+\!v_{n+1}\!-\!2v_{n}\right)^{2}}{2(\Lambda_{J}h)^{2}}\right] \label{eq:Ejvl-elast}\\
&&\hspace{-0.15in} \!=\!\frac{E_{0}}{2d}\int\frac{d^{2}\!\mathbf{k}_{\xy}}{(2\pi)^{2}}\int
_{-\pi\!/\!d}^{\pi\!/\!d}\frac{dk_{z}}{2\pi}\left[ k_{\xy}^{2}\!+\!\frac{2\left( 1\!-\!\cos
k_{z}d\right) ^{2}}{(\Lambda_{J}h)^{2}}\right]\! |v^{k}|^{2} \label{eq:Ejvl-elast-k}
\end{eqnarray}

Using relation between the phase perturbation and lattice displacements
\[
v^{k}=-\frac{hu^{k}}{\Lambda_{J}\left(  \exp(ik_{z}d)-1\right)  },
\]
we can rewrite the elastic energy in a more traditional way, via lattice
deformations
\begin{equation}
F_{\phi-el}=\frac{1}{2}\int\frac{d^{2}\mathbf{k}_{\xy}}{(2\pi)^{2}}
\int_{-\pi/d}^{\pi/d}\frac{dk_{z}}{2\pi}\left[  c_{11}(k_{z})k_{\xy}
^{2}+c_{66}\tilde{k}_{z}^{2}\right]  |u^{k}|^{2} \label{eq:Ejvl-elast-u}
\end{equation}
with elastic constants
\[
c_{11}(k_{z})=\frac{B_{x}^{2}}{4\pi}\frac{1}{\tilde{k}_{z}^{2}\lambda_{\mathrm{ab}}^{2}};\ c_{66}=\frac{\Phi_{0}^{2}}{32\pi^{3}d^{2}\gamma^{4}\lambda_{\mathrm{ab}}^{2}}
\]
where we used notation $\tilde{k}_{z}\equiv2\sin\left( k_{z}d/2\right)/d$.
Note that in our case the nonlocal tilt modulus $c_{44}(k_{z})$ is identical to
compression modulus $c_{11}(k_{z})$ and they coincide with elastic moduli within the
anisotropic London model \eqref{eq:C11-JVL-dil} and \eqref{eq:C44-JVL-dil} in the
limit $\tilde{k}_{z}\lambda_{\mathrm{ab}}\gg 1,k_{x}\lambda _{z}$. This elastic
energies \eqref{eq:Ejvl-elast}, \eqref{eq:Ejvl-elast-k}, and \eqref{eq:Ejvl-elast-u}
can be used to study weak fluctuations and weak pinning of the dense Josephson
vortex lattice. The shear modulus is field independent in the dense-lattice regime.
One can check that it matches the dilute-lattice result \eqref{eq:C66-JVL-dil} at
the crossover field.

\subsection{Lattice configurations and magnetic oscillations in finite-size
samples} \label{sec:FinSize}

\begin{figure}[ptb]
\begin{center}
\includegraphics[width=3.4in]{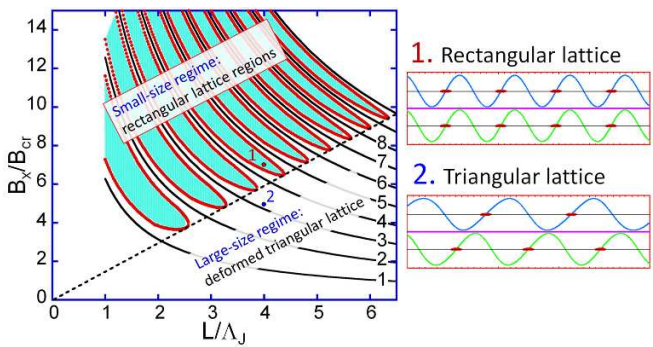}
\end{center}
\caption{Size-magnetic field phase diagram of the confined Josephson-junction
stack. Dashed line separates the large-size and small-size regimes. Black
lines correspond to integer flux quanta per junction. Shaded areas mark
regions of rectangular-lattice ground state. Representative lattice
configurations in two points are illustrated by plots of oscillating Josephson
currents in two neighboring layers. Small ellipses mark the centers of the
Josephson vortices.}%
\label{fig:h-Ldiag}%
\end{figure}
\begin{figure}[ptb]
\begin{center}
\includegraphics[width=3.4in]{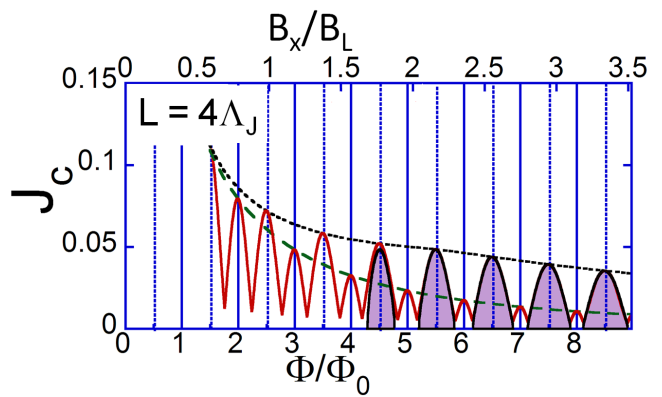}
\end{center}
\caption{Illustration of the oscillating magnetic field dependence of the critical current for $L=4\Lambda_{J}$. Crossover between $\Phi_0/2$ and $\Phi_0$ periodicity is seen at $h\bar{L}=B_{x}/B_{L}\sim 1$.
Shaded areas show the regions of stable rectangular lattice.}
\label{fig:J_PhiRect}%
\end{figure}

In this section we consider dense-lattice configurations in finite-size samples.
This study is actually motivated by experimental observations of magnetic
oscillations in small-size BSCCO mesas with lateral sizes 2-20 $\mu$m
\cite{OoiMK02,ZhuWK2005,LatyshevPOH2005,KatterweKrasnov2009,KakeyaKK2009}. Such small-size mesas
behave as stacks of intrinsic Josephson junctions with strong inductive coupling
between the neighboring junctions. The detailed analytical theory describing the
magnetic field dependences of lattice configurations and critical current has been
developed in Refs. \onlinecite{Koshelev2002,*Koshelev2007}. Lattice structures also
have been extensively explored numerically
\cite{Machida2003,*Machida2006,ZhuWK2005,IrieOya2007,KatterweKrasnov2009} and both
approaches give identical results. In a small-size sample the lattice structure is
determined by two competing interactions: the interaction with boundaries that
favors an aligned rectangular configuration and the bulk shearing interaction
between neighboring layers which favors a triangular configuration. Depending on the
mesa width $L$ and magnetic field, two very different regimes realize. In the
large-size regime the vortex lattice is triangular and it is only deformed near the
edges. In the small-size regime the lattice structure experiences a periodic series
of phase transitions between rectangular and triangular configurations. The
triangular configurations in this regime are realized only in narrow regions near
magnetic-field values corresponding to an integer number of flux quanta per junction
where the interaction with edges vanishes. The typical width of the mesa which
separates these two regimes is given by the length $\Lambda_{Jh}$, Eq.\
\eqref{eq:Ejvl-Length}, which is proportional to the applied magnetic field. Hence,
the crossover from one regime to another is driven by the
magnetic field and the corresponding crossover field scale is
$B_{L}\!=\!B_{\mathrm{cr}}L/\Lambda_{J}\!=\!L\Phi_{0}/(2\pi\gamma^{2}d^{3})$, for $B_{x}>B_{L}$ the small-size regime
realizes. The size-field phase diagram is shown in Fig.\ \ref{fig:h-Ldiag}. The
regimes are characterized by distinctly different oscillating behavior of the
critical current as function of the magnetic field. In the small-size regime, the
critical current oscillates with the period of one flux quantum per junction,
similar to a single junction. In the large-size regime, due to the triangular
lattice ground state, the oscillation period is half flux quantum per junction.

The quantitative study of the described behavior is based on the reduced energy
\eqref{eq:PhaseEn3D}, which have to be rewritten for the finite-size case,
$0<\bar{y}<\bar{L}\equiv\bar{L}_{y}$, and also assuming that the system
is uniform along the field direction, i.e., $\int\!d\bar{\mathbf{r}%
}\rightarrow\bar{L}_{x}\int_{0}^{\bar{L}}d\bar{y}$. This energy has to be
supplemented with the boundary conditions at the edges, $d\phi_{n}/d\bar{y}=0$ for
$\bar{y}=0,\bar{L}$. The important parameter in the case of finite-size sample is
the total magnetic flux through one junction, $\Phi=B_{x}dL$, which is connected
with the reduced magnetic field by relation $h\bar{L}=2\pi \Phi/\Phi_{0}$. In the
dense-lattice limit, we again use presentation of Eq.\ \eqref{eq:DensePhaseSplit}
containing the smooth phase $v_{n}$, and rapidly oscillating component
$\tilde{\phi}_{n}$. It is natural to assume that the interactions with the
boundaries preserve the alternating nature of the vortex lattice. In this case
symmetry allows us to take the smooth phase in the form
\begin{equation}
v_{n}(\bar{y})=\alpha n+(-1)^{n}v(\bar{y}) \label{SmoothPhaseFinSize}%
\end{equation}
where $\alpha$ describes the translational displacement of the lattice and $v$
describes lattice deformations with respect to the triangular lattice. In
particular, one can show that the maximum value of $v(\bar{y})$,
$v_{\mathrm{max}}=\pi/4$, describes the rectangular lattice, i.e., identical $\phi_n$ in all layers up to $2\pi$ phase shift. The corresponding rapid
phase becomes $\tilde{\phi}_{n}(\bar{y})\approx\!(-1)^{n}2\cos (2v)\sin\left(
\alpha+\!h\bar{y} \right)  /h^{2}$. Averaging with respect to the rapid oscillations
for such $v_{n}(\bar{y})$, gives the reduced energy per
layer and per unit length along $x$, $f_{\phi}=F_{\phi}\Lambda_{J}%
/(NL_{x}E_{0})$,
\begin{align}
f_{\phi}\!  &  \approx-\frac{1}{h}\left[  \sin\left(  2v_{0}\right)
\cos\alpha\!-\!\sin\left(  2v_{L}\right)  \cos\left(  h\bar{L}+\alpha\right)
\right] \nonumber\label{eq:Ejvl-dense-FinSize}\\
&  +\!\frac{1}{2}\!\int_{0}^{\bar{L}}\!d\bar{y}\left[  \left(  \frac{dv}%
{d\bar{y}}\right)  ^{2}\!-\!\frac{1+\cos(4v)}{h^{2}}\right] ,
\end{align}
in which the bulk part directly follows from Eq. \eqref{eq:Ejvl-dense-real} for
general $v_{n}(\bar{y})$. Variating this energy with respect to $v(\bar{y})$, we obtain that it obeys the static sine-Gordon equation
\begin{equation}
\frac{d^{2}v}{d\bar{y}^{2}}-\frac{2}{h^{2}}\sin\left(  4v\right)  =0
\label{eq:SmoothFin}%
\end{equation}
with the boundary conditions
\begin{eqnarray}\label{eq:SmoothBoundCond}
\frac{dv}{d\bar{y}}(0)  &  =&-\frac{2}{h}\cos(2v_{0})\cos\alpha
,\nonumber\\
\frac{dv}{d\bar{y}}(L)  &  =&-\frac{2}{h}\cos(2v_{L})\cos\left(  h\bar
{L}+\alpha\!\right)  .
\end{eqnarray}
Substituting solution of these equations into the energy functional
\eqref{eq:Ejvl-dense-FinSize} gives the energy as a function of the lattice shift
$\alpha$, $f_{\phi}(\alpha)$. Minimum of the energy with respect to $\alpha$ gives
the ground state for given $h$ and $\bar{L}$. Higher-energy states at other values
of $\alpha$ typically carry a finite current. The total Josephson current flowing
through the stack is proportional to $df_{\phi }/d\alpha$. Taking derivative of the
functional \eqref{eq:Ejvl-dense-real} with respect to $\alpha$, assuming that at
every $\alpha$ it is minimized with
respect to $v(u)$, we obtain the total current in units of $j_{\mathrm{J}%
}\Lambda_{J}L_{x}$
\begin{equation}
J(\alpha)\!=\!\frac{1}{h}\left[  \sin\!\left(  2v_{0}\right)  \sin
\!\alpha\!-\!\sin\!\left(  2v_{L}\right)  \sin\!\left(  h\bar{L}%
\!+\!\alpha\right)  \right]  . \label{eq:JosCurrSmooth}%
\end{equation}
An important consequence of this equation is that nonzero current exists only if the
surface deformations $v_{0}$ and $v_{L}$ are finite.

The general solution of equations \eqref{eq:SmoothFin} and
\eqref{eq:SmoothBoundCond} can be written in terms of the elliptic integrals and elaborated analytical analysis is possible \cite{Koshelev2002,*Koshelev2007}. Here we summarize the most important results of this analysis for two limiting cases.

In the \emph{large-size regime}, $L\gg \Lambda_{Jh}$ or $B_{x}\ll B_{L}$, the smooth
alternating deformation $v(\bar{y})$ has solution in the form of two isolated surface solitons \cite{Koshelev2002}. For example, near the edge $\bar{y}=0$ such soliton solution decaying from the surface into the balk is given by the
well-known formula for the sine-Gordon kink%
\begin{equation}
\tan v=\tan v_{0}\exp\left(  -2\sqrt{2}\bar{y}/h\right) ,
\label{eq:vSolution}%
\end{equation}
where the boundary value $v_{0}$ can be found from the boundary condition
(\ref{eq:SmoothBoundCond}) leading to $\tan\left(  2v_{0}\right)  =\sqrt
{2}\cos\alpha$. Using this solution, one can find the surface energy and surface
current for the edge $\bar{y}=0$ as functions of the lattice displacement $\alpha$,
\begin{align}
f_{s}(\alpha)  &  =\frac{1}{\sqrt{2}h}\left(  1-\sqrt{2+\cos2\alpha}\right),
\label{eq:SurfEn}\\
j_{s}(\alpha)  &  =-\frac{1}{\sqrt{2}h}\frac{\sin2\alpha}{\sqrt{2+\cos2\alpha
}}. \label{eq:SurfCur}%
\end{align}
The $2\alpha$ periodicity of these results is a consequence of the triangular lattice
structure: the change of $\alpha$ by $\pi$ corresponds to the vertical lattice
displacement by one layer. Similar solution is realized at the opposite edge
$\bar{y}=\bar{L}$. Its energy and current can be obtained from the above results using
the substitution $\alpha\rightarrow\alpha+h\bar{L}$. For a wide stack one can neglect
the interaction between the solitons and the total
Josephson current is given by the sum of two independent surface currents,%
\[
J(\alpha)=j_{s}(\alpha)+j_{s}(\alpha+h\bar{L}).
\]
The critical current $J_{c}$ can be found as maximum of $J(\alpha)$ with respect to
$\alpha$ giving the following result in real units
\begin{equation}
J_{c}(B)=J_{J}\frac{\Phi_{0}}{2\pi dLB_{x}}\mathcal{F}\left(  \frac{2\pi
dLB_{x}}{\Phi_{0}}\right)  , \label{I_B}%
\end{equation}
where $J_{J}=j_{J}LL_{x}$ is the maximum Josephson current through the sample at
zero field, and oscillating function $\mathcal{F}(\chi)$ has period $\pi$ and in the
range $0<\chi<\pi/2$ can be approximated by the following formula,
$\mathcal{F}(\chi)\approx0.128+0.888\,\cos (\chi)+0.021\,\cos(3\,\chi)$ . We can see
that, in this regime the product $B_{x}J_{c}$ has periodicity of half flux-quantum
per junction and reaches maxima at points $\Phi\!=dLB_{x}=\!j\Phi_{0}/2$ with
$B_{x}J_{c,\max}\approx1.035J_{J}\Phi_{0}/(2\pi dL)$. This corresponds to the
low-field part of the plot in Fig.\ \ref{fig:J_PhiRect}. All other properties of the
sample should also oscillate with the period of half flux quantum. Such oscillations
of the flux-flow resistivity in BSCCO micro-mesas have been first detected
experimentally in Ref. \onlinecite{OoiMK02} and later confirmed by several
experimental groups.

In the \emph{small-size regime} $L<\Lambda_{Jh}$ or $B_{x}>B_{L}$ the interaction with edges dominates. As a
consequence, extended regions of the rectangular lattice appear in the phase
diagram, see Fig.\ \ref{fig:h-Ldiag}. The energy of the rectangular lattice,
$v=\pm\pi/4$, coincides with the well-known result for a single junction
\begin{equation}
f_{\mathrm{rect}}(\alpha)=-\frac{2}{h}\sin\left(  \frac{hL}{2}\right)
\sin\left(  \alpha+\frac{hL}{2}\right)  .\label{eq:En_rect}%
\end{equation}
and has minimum $f_{\mathrm{rect}}=-2\left\vert \sin\left(  hL/2\right)  \right\vert
/h$ at $\alpha=-hL/2+\delta\pi/2$ with $\delta=\mathrm{sign}\left[  \sin\left(
hL/2\right) \right]$. An accurate analysis \cite{Koshelev2007} shows that the
rectangular lattice is stable with respect to small deformations at
$\alpha=-h\bar{L}/2+\pi/2$ in the regions $|h\bar{L}/2\pi -(k+1/2)|<1/4$ only if the
inequality
\begin{equation}
\left\vert \sin\left(  h\bar{L}/2\right)  \right\vert <\tan\left(  \sqrt
{2}\bar{L}/h\right) / \sqrt{2}\label{eq:Cond-Rect-Ground}%
\end{equation}
is satisfied. These regions are plotted in the phase diagram, Fig.\
\ref{fig:h-Ldiag}. This means that the rectangular lattices first appear in the
ground state at points $h\bar{L}=(k+1/2)2\pi$ for $\bar{L}/h\leq l_{1}=\arctan\left(
\sqrt{2}\right) /\sqrt{2}\approx0.675$. This corresponds to the dashed line shown in
the phase diagram of Fig.\ \ref{fig:h-Ldiag}. If, however, $L/h$ is only slightly
smaller than this value, the rectangular lattice becomes unstable with increasing
current and the configuration at the critical current still corresponds to the
deformed lattice. The accurate analysis shows that there is another typical value of
the ratio $L/h$, $L/h=l_{2}\approx0.484$, below which \emph{the rectangular lattice
remains stable up to the critical current}.

In the region $h\gg \bar{L}$ the rectangular lattice is realized in the most part of
the phase diagram except narrow regions in the vicinity of the integer-flux quanta
lines, $h\bar{L}/2\pi=\Phi/\Phi_{0}=k$ where the interaction with edges vanishes.
Switching between the rectangular and triangular lattices in the ground state occurs
via a first-order phase transition \cite{Koshelev2007} at the transition fields
which are determined
by equation%
\begin{equation}
\left\vert \sin\left(  \frac{h_{t}\bar{L}}{2}\right)  \right\vert =\frac{3}{2}%
\frac{\bar{L}}{h_{t}}.\label{eq:first-order-tran}%
\end{equation}
At high fields the critical current approaches the classical Fraunhofer dependence
for a single small junction, $J_{F}(\Phi)=J_{J}|\sin(\pi\Phi
/\Phi_{0})|/|\pi\Phi/\Phi_{0}|$. Two important deviations persist at all fields and
sizes: (i) Near the points $\Phi=k\Phi_{0}$, due the phase transitions to the
triangular lattice, the critical current never drops to zero and actually always has
small local maxima; (ii) Away from the points $\Phi=k\Phi_{0}$ the critical current
is reached at the instability point of the rectangular vortex lattice and it is
always somewhat smaller than the ``Fraunhofer'' value $J_{F}(\Phi)$.

In the region $B\sim B_{L}$ the crossover between the two described regimes takes
place. In the oscillations of the critical current this crossover manifests itself
by breaking the $\Phi_{0}/2$ periodicity, the maxima at the half-integer
flux-quantum points $\Phi=(k+1/2)\Phi_{0}$ progressively become larger while the maxima at the integer flux-quantum points $\Phi=k\Phi_{0}$ become smaller. This crossover
behavior of the critical current is illustrated in Fig.\ \ref{fig:J_PhiRect}. Such
behavior was indeed observed experimentally in very narrow BSCCO mesas
\cite{ZhuWK2005,KatterweKrasnov2009,KakeyaKK2009}.

\section{Thermal fluctuations \label{sec:Thermal}}

In this section we consider thermal fluctuations effects for the Josephson vortex
lattice. Confinement of the vortex cores in between the layers leads to strong
suppression of the vortex motion across the layers, which can only occur via
formation of kinks. Therefore, as a first step, one can neglect these energy-costly
displacements and consider only planar fluctuations of vortices along the layers.
This simple model describes fluctuation behavior in the most part of the
field-temperature phase diagram but it occurs to be insufficient for description of
the melting transition of the lattice. In general, thermal effects for Josephson
vortices are much weaker then for pancake-vortex lattice and phase transformations
are expected only in the vicinity of the transition temperature. On the other hand,
due to the intrinsic pinning potential and involvement of the kink excitations, the
overall behavior near the melting line is rather complicated and, in spite of quite
extensive theoretical effort
\cite{Efetov1979,Korshunov1991,KorshunovL1992,Horovitz1993II,BalentsN1995,BalentsR1996,HuLM2005}
and numerical simulations \cite{HuT2000,HuT2004}, there is no clear consensus on the
nature of melting transition and structure of the phase diagram for magnetic field
aligned with direction of the layers.

\subsection{Thermal effects for dilute Josephson vortex lattice: problem of
intermediate phase}

A standard first step to study thermal fluctuation effects is to evaluate the
mean-squared local fluctuation displacement from the elastic energy
\eqref{eq:Fel-JVL-dilute} \footnote[5]{As in most theoretical papers, the temperature is measured in energy units.}
\begin{equation}
\left\langle u^{2}\right\rangle =\int\frac{d^{3}\mathbf{k}}{(2\pi)^{3}}
\frac{T}{c_{11}(\mathbf{k})k_{y}^{2}+c_{44}(\mathbf{k})k_{x}^{2}+c_{66}
k_{z}^{2}}.\label{eq:u2-JVL-dil-gen}
\end{equation}
Introducing reduced wave vector $\mathbf{\tilde{k}}$ as
\begin{equation}
k_{x}=k_{\mathrm{BZ} }\tilde{k}_{x}/\sqrt{\gamma},\
k_{y}=k_{\mathrm{BZ}}\tilde{k}_{y}/\sqrt {\gamma},\
k_{z}=k_{\mathrm{BZ}}\sqrt{\gamma}\tilde{k}_{z},
\label{eq:JVL-red-k}
\end{equation}
where $k_{\mathrm{BZ}}=\sqrt{4\pi B_{x}/\Phi_{0}}$ is the average wave vector of the
Brillouin zone, we rewrite this integral in a more explicit form.
\begin{eqnarray*}
&&\left\langle u^{2}\right\rangle =\frac{\left(  4\pi\right)  ^{2}
k_{\mathrm{BZ}}\lambda_{\mathrm{c}}^{2}T}{\sqrt{\gamma}\Phi_{0}B_{x}}\int\frac
{d^{3}\mathbf{\tilde{k}}}{(2\pi)^{3}}\left[  \left(  \frac{1}{\tilde{k}^{2}
}-\frac{1}{4}\right)  \tilde{k}_{y}^{2}\right.\\
&&\left.+\left(  \frac{1}{\tilde{k}^{2}}
+\ln\frac{b/s}{\sqrt{1+\beta^{2}\tilde{k}_{x}^{2}}}\right)  \tilde{k}_{x}
^{2}+\frac{\tilde{k}_{z}^{2}}{4}\right]  ^{-1}
\end{eqnarray*}
with $\beta
\sim1$. Evaluation of this integral leads to the following result
\begin{equation}
\frac{\left\langle u^{2}\right\rangle
}{a_{0}^{2}}=\frac{0.12T}{b_{0}\sqrt
{\ln\left(  b_{0}/d\right)  }\varepsilon_{0}},\label{eq:u2-JVL-dil-res}
\end{equation}
where $a_{0}=\sqrt{\gamma\Phi_{0}/B_{x}}\ $and $b_{0}=\sqrt{\Phi_{0}/\gamma B_{x}}$
are the typical lattice constant in the $y$ and $z$ directions. From this result one
can obtain estimate for the typical temperature at which fluctuations become strong
\cite{Korshunov1991}
\begin{equation}
T_{f}\sim b_{0}\sqrt{\ln\left(  b_{0}/d\right)  }\varepsilon_{0}
(T_{f}).\label{eq:Tf-JVL-dil}
\end{equation}
Unfortunately, this temperature is located very close to $T_{c}$ where one can not
use approximations behind Eq.\ \eqref{eq:Fel-JVL-dilute}, e.g., neglect thermal
activation of kinks and antikinks. We can conclude that \emph{the planar-fluctuations model given by the elastic energy} \eqref{eq:Fel-JVL-dilute} \emph{is not sufficient to describe the melting of the Josephson-vortex lattice }\cite{Korshunov1991}
The temperature scale \eqref{eq:Tf-JVL-dil} is much higher than the corresponding
temperature scale for the pancake vortex lattice \cite{BlatterFGLV1994}
meaning that thermal-fluctuation effects for the
Josephson vortex lattice are much weaker than for the pancake vortex lattice.

We can estimate the typical temperature above which kink formation strongly
influences the fluctuation displacements of the vortex lines. In the isolated line
the typical distance between thermally excited kinks is given by
\begin{equation}
L_{\mathrm{kink}}=\xi_{\mathrm{kink}}\exp(E_{\mathrm{kink}}/T),
\label{eq:kink-sep}
\end{equation}
where $E_{\mathrm{kink}}\approx d\varepsilon_{0}\ln(\gamma d/\xi_{\mathrm{ab}})$ is
the kink energy. Usually, it is assumed that the preexponential factor
$\xi_{\mathrm{kink}}$ is of the order of the in-plane coherence length
$\xi_{\mathrm{ab}}$ \cite{BalentsN1995}. Analysis of fluctuations of the order
parameter near the core \cite{Koshelev1994} gives somewhat more accurate estimate
$\xi_{\mathrm{kink} }\sim$ $\xi_{\mathrm{ab}}\sqrt{T/d\varepsilon_{0}}$. Typical
$k_{x}$ contributing to the fluctuation displacement \eqref{eq:u2-JVL-dil-gen} can
be estimated as $k_{x}\sim\pi/b_{0}$. Therefore, the kinks start to contribute to
thermal wandering if $L_{\mathrm{kink}}<b_{0}$. This gives estimate for typical
temperature
\begin{equation}
T_{\mathrm{kink}}=E_{\mathrm{kink}}/\ln(b_{0}/\xi_{\mathrm{kink}}).
\label{eq:T-kink-gen}
\end{equation}
In the limit $\gamma>\xi_{\mathrm{ab}}/d$, we obtain
\begin{equation}
T_{\mathrm{kink}}=d\varepsilon_{0}(T_{\mathrm{kink}})
\frac{\ln(\gamma d/\xi_{\mathrm{ab}})}
{\ln (b_{0}/\xi_{\mathrm{kink}})}.
\label{eq:T-kink-anis}
\end{equation}
One can see that even though this temperature is smaller then $T_{f}$
\eqref{eq:Tf-JVL-dil}, it is also located close to the fluctuation region near
$T_{c}$ and very slowly decreases with increasing magnetic field.

The planar-fluctuations model belongs to universality class of the three-dimensional
XY model meaning that the phase transition described by this model has to be
continuous. In spite of insufficiency of this model, this suggests that the melting
transition for the magnetic field applied along the layers may become continuous for
sufficiently high anisotropy. It was indeed observed experimentally by Kwok \emph{et
al.} \cite{KwokFWFDC1994} and by Gordeev \emph{et al.} \cite{GordeevZdG2000} that
the melting transition in YBCO becomes continuous when magnetic field is aligned
with the layers. Continuous melting of the Josephson vortex lattice also has been
observed in numerical simulations by Hu and Tachiki \cite{HuT2000}. The simulation
parameters in this work, however, correspond to the regime of dense lattice, which
will be considered below.

\begin{figure}[ptb]
\centerline{\includegraphics[width=7cm,clip]{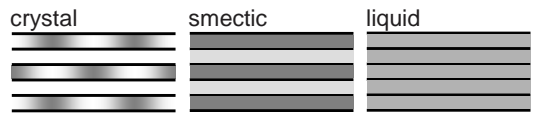}}
\caption{Possible phases for field applied along the layers. Grey
level illustrates average vortex density. In the intermediate
smectic phase suggested in Ref.\ \onlinecite{BalentsN1995} density is
modulated only in the direction perpendicular to the layers.}
\label{fig:JV-phases}
\end{figure}

A description of the fluctuating Josephson vortices taking into account
kink-antikink formation is much more complicated problem and possibilities for
analytical progress are quite limited. General scenarios of Josephson-vortex-lattice
melting have been discussed by Balents and Nelson \cite{BalentsN1995}.
They argued that an aligned lattice may melt via an intermediate smectic phase, in
which the average vortex density is modulated only in the direction perpendicular to
the layers but no order is preserved in the direction of the layers, as illustrated
in Fig.\ \ref{fig:JV-phases}. The period of density modulation has to be equal to
the integer number of layers. The developed Landau theory of the liquid-to-smectic
transition suggests that this transition has to be of the second order. Static and
dynamic properties of the intermediate smectic phase have been described in detail.
In particular, Balents and Nelson argued that this phase is characterized by finite
but very large tilt modulus, corresponding to very small transversal susceptibility
$\mu_z=B_z/H_z$, and by very small in-plane resistivity. Both these properties
appear due to the thermally-activated ``superkink'' excitations, in which one vortex
is moved across the layers by one smectic period. While the density modulation
remains static and oriented parallel to the layers, these excitations may facilitate
tilting of the magnetic induction with respect to the layers and flux motion in the
$z$-axis direction. In spite of its physical appeal, the theory of Balents and
Nelson is not quantitative. It does not predict locations of the transitions in the
field-temperature plane, their thermodynamic signatures, and the width of the
intermediate-phase region. The very existence of the intermediate smectic phase has
been not rigorously proven. Alternatively, the crystal may melt directly into the
liquid via a first-order phase transition.

More quantitative study based on the density-functional theory has been performed
recently by Hu, Luo, and Ma \cite{HuLM2005}. The intrinsic pinning potential in this
study has been modeled by the cosine function and its strength has been used as an
adjusting parameter. It was found that the smectic phase exists for sufficiently
strong periodic potential only for one type of aligned lattice, which in our
notations corresponds to $(m,n)=(1,0)$ and with one empty layer between the layers
filled with Josephson vortices, i.e., with $N=2$. According to analysis of Sec.\
\ref{sec:grst-num} such lattice is realized in ground state within field interval
$[0.8-0.98]\Phi_{0}/(2\pi\gamma d^{2})$. The melting scenario via the intermediate
smectic phase is most probable in this field range.

\subsection{Elimination of the lattice rotational degeneracy by thermal fluctuations}

The dilute lattice at small field is approximately degenerate with respect to
elliptic rotations, as it was discussed in Sec. \ref{sec:jvldilute}. This degeneracy
is partially eliminated by intrinsic pinning potential and by the corrections to the
intervortex interactions due discreteness of the layered structure. The latter
effect becomes noticeable only at high magnetic field approaching the crossover
field. As the Josephson vortices mainly fluctuate along the layer direction, the
fluctuation correction to the free energy depends on the lattice orientation with
respect to the layers and also eliminates the elliptic degeneracy. Therefore the
Josephson vortex lattice at small fields gives a physical realization of a system in
which the ground state is highly degenerate at zero temperature and this degeneracy
is eliminated by thermal fluctuations. Similar behavior is realized in some
frustrated magnetics and is known as ``order as an effect of disorder''
\cite{VillainBCC1980}. As a natural way to prepare the ground state is to cool
system in fixed field, it is important to understand how the ground-state
configuration evolves with the temperature.

In this section we consider the orientation-dependent entropy correction to the free
energy. This will allow us to trace evolution of the ground-state configurations
with increasing field at finite temperature. Qualitatively, fluctuations favor soft
lattices, with smaller elastic constants. One can expect then that the entropy
correction favors the aligned lattice $(1,0)$, because for this lattice the shear
deformations take place along the closed-packed direction.

The orientation-dependent entropy correction is determined by the short-wavelength
lattice deformations and the long-wavelength elastic approximation of the previous
section is not sufficient. The elastic energy for planar deformations in the whole
Brillouin zone is given by
\begin{equation}
F_{\mathrm{el}}=\int\frac{d^{3}\mathbf{k}}{(2\pi)^{3}}\frac{\Phi
_{\mathrm{JVL}}(\mathbf{k})}{2}\left\vert u(\mathbf{k})\right\vert^{2}
\label{eq:PlanElEner}
\end{equation}
with
\begin{eqnarray}
&&\Phi_{\mathrm{JVL}}(\mathbf{k})\!=\!\frac{B_{x}^{2}}{4\pi}\sum_{\mathbf{Q}
}\!\left(\!\frac{\left(  k_{y}\!-\!Q_{y}\right)^{2}+k_{x}^{2}}
{1\!+\!\lambda_{\mathrm{ab}}^{2}\!(k_{z}\!-\!Q_{z}) ^{2}
\!+\!\lambda_{\mathrm{c}}^{2}\!(k_{y}\!-\!Q_{y})^{2}
\!+\!\lambda_{\mathrm{c}}^{2}k_{x}^{2}}\right.\nonumber\\
&&\left.-\frac{Q_{y}^{2}
}{1\!+\!\lambda_{\mathrm{ab}}^{2}Q_{z}^{2}\!+\!\lambda_{\mathrm{c}}^{2}Q_{y}^{2}}\right),
\label{eq:JVLel_matr}
\end{eqnarray}
where $\mathbf{Q}=(Q_{y},Q_{z})$ are the reciprocal-lattice vectors. The fluctuation
correction to the free energy is given by
\begin{equation}
\delta f_{T} =-\frac{T}{2}\int_{-\infty}^{\infty}\frac{dk_{x}}{2\pi}
\int_{\mathrm{BZ}}\frac{dk_{y}dk_{z}}{(2\pi)^{2}}\ln\frac{C}{\Phi
_{\mathrm{JVL}}(\mathbf{k})}.\label{eq:EnCorrDef}
\end{equation}
Calculation of this correction is described in detail in Appendix
\ref{App:EntCorrCalc}. Combining result of this calculation with the London-limit
presentation of the lattice interaction energy \eqref{eq:EnLonGenInter}, we
represent the orientation-dependent part of the total free energy at finite
temperature in the form
\begin{equation}
\delta
f_{a}=\frac{B_{x}}{\Phi_{0}}\frac{\varepsilon_{0}}{\gamma}\left(
G_{L}-\frac {T}{\varepsilon_{0}}\sqrt{\frac{\gamma
B_{x}}{\pi\Phi_{0}}}g_{a}\right). \label{eq:JVL-AnisEn-EntrCorr}
\end{equation}
Numerically computed orientation-dependent correction $g_{a}(\theta,h)$ in the range
$0.001<h<0.1$ is well described by
\begin{equation}
\hspace{-0.02in}g_{a}(\theta,h)\!\approx \!g_{6}(h)\cos(6\theta) \hbox{ with
}g_{6}(h)\!\approx \!\frac{0.01}{\sqrt{\ln(514/h)}}. \label{eq:ga-num}
\end{equation}
The fluctuations give the largest negative contribution for $\theta=0$, meaning that
they indeed favor the aligned lattice (1,0).

Let us compare the orientation-dependent entropy correction with the correction due
to the layered structure considered in Sec. \ref{sec:grst-num}. We can see that
these corrections compete: the first one favors the $(1,0)$ orientation while the
second one favors the (1,1) orientation. The entropy correction decays with
decreasing fields as $\sqrt{B_{x}}$ and always exceeds at small fields the
``layeredness'' correction, which decays as $B_{x}^{2}$. We estimate that the
``layeredness'' correction exceeds the fluctuation correction when $B_{x}$ exceeds
the temperature-dependent field scale
\[
B_{x,T}=\frac{\Phi_{0}}{2\pi\gamma d^{2}}\left[  \frac{T/d\varepsilon_{0}
}{\sqrt{\ln(C_{T}d\varepsilon_{0}/T)}}\right]  ^{2/3}
\]
with $C_{T}\approx 2.6\cdot10^{4}$.

\subsection{Fluctuations and melting of dense Josephson vortex lattice}

Using elastic energy \eqref{eq:Ejvl-elast-k} we can evaluate a mean-squared
fluctuation of the in-plane phase
\[
\left\langle \phi_{n}^{2}\right\rangle \!\approx \!\left\langle v_{n}
^{2}\right\rangle \!=\!\frac{dT}{E_{0}}\int\frac{d^{2}\mathbf{k}_{\xy}}
{(2\pi)^{2}}\int_{-\pi\!/\! d}^{\pi\! /\! d}\!\frac{dk_{z}}{2\pi}\frac{1}{k_{\xy}
^{2}+\frac{8 }{(\Lambda_{J}h)^{2}}s_z^4}
\]
with $s_z(k_z)=\sin(k_{z}d/2)$ and the lattice displacement $u_{n}=-\Lambda_{J}\left( v_{n+1}-v_{n}\right) /h$
\[
\left\langle u^{2}\right\rangle =\frac{dT\Lambda_{J}^{2}}{E_{0}}\int
\frac{d^{2}\mathbf{k}_{\xy}}{(2\pi)^{2}}\int_{-\pi/d}^{\pi/d}
\frac{dk_{z}}{2\pi}\frac{4 s_z^2 }{k_{\xy}^2
\!+\!\frac{8 }{(\Lambda_{J}h)^{2}}s_z^4}.
\]
Renormalization of the effective coupling is determined by the following average
\[
\left\langle \!\left(  v_{n\!-\!1}\!+\!v_{n\!+\!1}\!-\!2v_{n}\right)
^{2}\!\right\rangle
\!=\!\frac{dT}{E_{0}}\hspace{-0.05in}\int\!\frac{d^{2}\mathbf{k}_{\xy}}{(2\pi)^{2}}\hspace{-0.08in}\int\limits_{-\pi
/d}^{\pi/d}\hspace{-0.07in}\frac{dk_{z}}{2\pi}\frac{16 s_z^4}{k_{\xy}
^{2}\!+\!\frac{8 }{(\Lambda_{J}h)^{2}}s_z^4}.
\]
All above integrals are logarithmically diverging at large $k_{\xy}$. This
divergency has to be cut off at $k_{\xy}\sim1/\xi_{\mathrm{ab}}$. As usual for
quasi-two-dimensional systems, the weak interlayer coupling cuts off logarithmic
divergency at small $k_{\xy}$. Evaluating the integrals, we obtain
\begin{eqnarray*} &  \left\langle \phi_{n}^{2}\right\rangle \approx\frac{T}{2\pi
E_{0}} \ln\left(  \Lambda_{J}h/\xi_{\mathrm{ab}}\right)  ;\ \left\langle
u^{2}\right\rangle
\approx\frac{T\Lambda_{J}^{2}}{\pi h^{2}E_{0}}\ln\left(  \Lambda_{J}
h/\xi_{\mathrm{ab}}\right)  ;\\
&  \left\langle \left(  v_{n-1}+v_{n+1}-2v_{n}\right)
^{2}\right\rangle \approx6\left\langle \phi_{n}^{2}\right\rangle
\approx\frac{3T}{\pi E_{0}} \ln\left(\Lambda_{J}h/\xi_{\mathrm{ab}}\right).
\end{eqnarray*}
Fluctuations become strong and harmonic approximation breaks down when $\left\langle
\left( v_{n-1}\!+\!v_{n+1}\!-\!2v_{n}\right) ^{2}\right\rangle \sim1$, corresponding
to $\left\langle \phi_{n}^{2}\right\rangle \sim1/6$ and $\left\langle
u^{2}\right\rangle \sim$ $a^{2}/3$ with $a=\Lambda_{J}/h$ being the in-plane lattice
constant. This gives the temperature scale
\begin{equation}
T_{f}=\frac{E_{0}(T_{f})}{\ln\left(  \Lambda_{J}h/\xi_{\mathrm{ab}}\right)  }
=\frac{\varepsilon_0 (T_{f}) d}{\pi \ln\left(
\Lambda_{J}h/\xi_{\mathrm{ab}}\right) }. \label{eq:Tf-JVL-dense} \end{equation}
As $E_{0}(0)\gg T_{c}$ (typically for BSCCO $E_{0}(0)\sim 250-300K$), this temperature scale usually
corresponds to temperatures close to $T_{c}$. It is somewhat lower than the
corresponding temperature scale for the dilute lattice \eqref{eq:Tf-JVL-dil} and
even smaller than the temperature scale for kink formation in the dilute lattice
\eqref{eq:T-kink-anis}.

We turn now to discussion of the melting transition of the dense lattice based on
the energy \eqref{eq:Ejvl-dense-real} describing weakly coupled two-dimensional
systems. Behavior of such system has to be similar to layered XY model
\cite{PokrovskiiU1973} and to a layered superconductor in zero magnetic field
\cite{GlazmanKoshelev1990}. In the ordered phase of such systems, below the
Kosterlitz-Thouless temperature for a single layer,  a weak interlayer coupling is
always relevant, can not be treated as a small perturbation, and it restores
three-dimensional long-range order. The transition in such system is expected to be
continuous and to occur slightly above the Berezinskii-Kosterlitz-Thouless
transition of an isolated layer given by equation $T_{KT}=\pi E_{0}(T_{KT})/2$. This is in spite of the fact that the interplane fluctuations actually become strong at the temperature \eqref{eq:Tf-JVL-dense} which is significantly smaller than the transition temperature in the isolated layer
$T_{KT}$.

Hu and Tachiki studied numerically melting transition of the dense lattice using the
frustrated XY model \cite{HuT2000}. They claimed that the melting transition is
continuous at high field and it changes to a first-order transitions when the field
drops below $B=\Phi_{0}/2\sqrt{3}\gamma d^{2}\approx 1.8\Phi_{0}/2\pi\gamma d^{2}$.
It is not clear how universal is this field. In principle, it may be sensitive to
the kink energy, which depends on the ratio $\gamma d/\xi_{\mathrm{ab}}$.

Experimentally, indication of the melting transition in the dense-lattice regime was found in small-size BSCCO mesas by Latyshev \etal \cite{LatyshevPOH2005} exploring the temperature dependence of magnetic oscillations discussed in Sec. \ref{sec:FinSize}. It was found that in the field range 0.6-0.8 tesla the magnetic oscillations of the flux-flow voltage rapidly decrease with increasing temperature and are completely suppressed by thermal fluctuations at temperatures $\sim 4$K below the transition temperature.


\section{Summary}

In this review we considered in detail the static properties of the Josephson vortex
lattice following from the Lawrence-Doniach model in London approximation, which
mostly describes properties of superconductors in terms of the distribution of the
order-parameter phase. We reviewed the properties of an isolated vortex as well as
the structure and energetics of the vortex lattice both in dilute and dense regimes.
In addition to standard properties, our consideration includes quite subtle
nontrivial effects, such as the influence of thermal fluctuations on the orientation
of the vortex lattice. Note that we did not touch on dynamic properties of the
lattice which has became a separate large field.

\begin{acknowledgements}

AEK would like to thank L.\ N.\ Bulaevskii, M.\ Tachiki and X.\ Hu for many useful
discussions of theoretical issues and Yu.\ I.\ Latyshev, I.\ Kakeya, T.\ Hatano,
S.\ Bending, V.\ K.\ Vlasko-Vlasov, A.\ Tonomura, and A.\ A.\ Zhukov for discussions of relevant experimental data. AEK is supported by UChicago Argonne, LLC, operator of
Argonne National Laboratory, a U.S. Department of Energy Office of Science
laboratory, operated under contract No. DE-AC02-06CH11357.
\end{acknowledgements}




\appendix

\section{Calculation of the nonlocal line-tension energy of a single line\label{App:LineTension}}

For deformations with wave vectors $|k_{x}|\gg1/\lambda_{\mathrm{c}}$ screening
effects can be neglected and the energy variation is determined by the phase part of
energy, which we write using scaled in-plane coordinates, $(\bar{x},\bar
{y})=(x/\gamma d,y/\gamma d)$,
\begin{eqnarray} \label{eq:LineEnerDef}
\delta F  &  \approx E_{0}\sum_{n}\int d\bar{x}\int d\bar{y}\left[
\frac
{1}{2}\left(  \nabla_{\Vert}\phi_{n}\right)  ^{2}-\cos\left(  \phi_{n+1}
-\phi_{n}\right)  \right.\nonumber\\
&  \left.  -\frac{1}{2}\left(  \nabla_{\bar{y}}\phi_{n}^{(0)}\right)
^{2}+\cos\left(  \phi_{n+1}^{(0)}-\phi_{n}^{(0)}\right)  \right],
\end{eqnarray}
where $\phi_{n}^{(0)}(\bar{y})$ is the straight-vortex solution. The phase of the
deformed vortex obeys the following equation
\begin{equation}
\nabla_{\Vert}^{2}\phi_{n}+\sin\left(  \phi_{n+1}-\phi_{n}\right)
-\sin\left(  \phi_{n}-\phi_{n-1}\right)  =0 \label{eq:PhaseEqDefJV}
\end{equation}
with the condition $\phi_{1}(\bar{x},\bar{u}(\bar{x}))-\phi_{0}(\bar{x}
,\bar{u}(\bar{x}))=\pi$ defining the vortex core and $\bar{u}(\bar {x})=u(x)/\gamma
d$. In the elastic limit, $|du/dx|\ll1$, at distances, smaller that the typical
wavelength of deformation, the phase approximately can be represented as
\[
\phi_{n}(\bar{x},\bar{y})\approx\phi_{n}^{(0)}\left[
\bar{y}-\bar{u}(\bar {x})\right].
\]
On the other hand, at large distances we can use London approximation of Eq.
\eqref{eq:PhaseEqDefJV} and find the phase using Fourier transform. This gives for
the phase perturbation $\phi^{(1)}(\bar{\mathbf{k}})=\phi
(\bar{\mathbf{k}})-\phi^{(0)}(\bar{\mathbf{k}})$
\begin{equation}
\phi^{(1)}(\bar{\mathbf{k}})\approx\frac{2\pi i\bar{k}_{z}\bar{u}(\bar{k}
_{x})}{\bar{k}^{2}}, \label{eq:JVPhaseFar}
\end{equation}
with $(\bar{k}_{x},\bar{k}_{y},\bar{k}_{z})=(\gamma dk_{x},\gamma dk_{y},dk_{z})$
and $\bar{k}^{2}=\bar{k}_{x}^{2}+\bar{k}_{y}^{2}+\bar{k} _{z}^{2}$. We will use
this result in mixed $(\bar{k}_{x},\bar{y},\bar{z})$-representation, which is
obtained by the reverse Fourier transform of the above equations with respect to
$\bar{y}$ and $\bar{z}$
\begin{equation}
\phi^{(1)}(\bar{\mathbf{r}},\bar{k}_{x})\approx\bar{u}(\bar{k}_{x}
)\nabla_{\bar{z}}K_{0}(\bar{k}_{x}\bar{r}). \label{eq:phiLargeR}
\end{equation}
with $\bar{\mathbf{r}}=(\bar{y},\bar{z})$.

We split the total energy loss given by Eq.\ \eqref{eq:LineEnerDef} into the
$x$-gradient and transverse parts, $\delta F=F_{x}+F_{zy}$. The $x$-gradient part,
\[
F_{x}=\frac{E_{0}}{2}\sum_{n}\int d\bar{x}\int d\bar{y}\left(
\nabla_{\bar {x}}\phi_{n}\right)  ^{2},
\]
can be computed by introducing intermediate scale, $1\ll R\ll1/\bar{k}_{x}$, which
splits the integral into the two contributions, from small and at large distances.
The contribution from $\bar{r}=\sqrt{\bar{y}^{2}+\bar{z}^{2}}<R$, with
$\bar{z}=n-1/2$ is given by
\[
F_{x,<} \approx\frac{E_{0}}{2}\int d\bar{x}\left(
\frac{du}{d\bar{x} }\right)
^{2}\sum_{n}\int\limits_{-y_{n}}^{y_{n}}d\bar{y}\left( \nabla
_{\bar{y}}\phi_{n}^{(0)}\right)  ^{2}
\]
with $y_{n} =\sqrt{R^{2}-(n-1/2)^{2}}$. The quantity
\[
\sum_{n}\int\limits_{-y_{n}}^{y_{n}}d\bar{y}\left(
\nabla_{\bar{y}}\phi _{n}^{(0)}\right)
^{2}\approx\frac{\pi}{2}\left(  \ln R+C_{y}\right)
\]
is determined by exact phase distribution in the core. Using the accurate numerical
solution, we estimate $C_{y}\approx0.93$. The contribution from the region $r>R$ is
computed using Eq.\ \eqref{eq:phiLargeR},
\begin{eqnarray*}
F_{x,>}&\approx&\frac{E_{0}}{2}\int
d\bar{x}\int_{\bar{r}>R}d^{2}\bar {\mathbf{r}}\left(
\nabla_{\bar{x}}\phi_{n}\right) ^{2}\\
&=&\frac{E_{0}}{2}
\int\frac{d\bar{k}_{x}}{2\pi}\bar{k}_{x}^{2}|\bar{u}(\bar{k}_{x})|^{2}
\int_{\bar{r}>R}d^{2}\bar{\mathbf{r}}\left[
\nabla_{\bar{z}}K_{0}(\bar{k}_{x} \bar{r})\right]  ^{2}.
\end{eqnarray*}
Computing integral
\[
\int_{\bar{r}>R}d^{2}\bar{\mathbf{r}}\left[
\nabla_{\bar{z}}K_{0}(\bar{k} _{x}\bar{r})\right]
^{2}\approx\pi\left(  \ln\frac{2}{\bar{k}_{x}\bar{R}
}-\Euler-\frac{1}{2}\right) ,
\]
with $\Euler\approx 0.5772$ being the Euler constant, we obtain
\[
F_{x,>}\approx\frac{\pi}{2}E_{0}\int\frac{d\bar{k}_{x}}{2\pi}\bar{k}_{x}
^{2}\left(  \ln\frac{2}{\bar{k}_{x}\bar{R}}-\Euler-\frac{1}
{2}\right)  |\bar{u}(\bar{k}_{x})|^{2}.
\]
Combining the parts $F_{x,<}$ and $F_{x,>}$, we obtain
\begin{equation}
F_{x}=\frac{\pi}{2}E_{0}\int\frac{d\bar{k}_{x}}{2\pi}\bar{k}_{x}^{2}\left(
\ln\frac{2}{\bar{k}_{x}}-\Euler-\frac{1}{2}+C_{y}\right)
|\bar{u}(\bar{k}_{x})|^{2}\label{eq:JVLineFx}
\end{equation}
In the transverse part,
\begin{eqnarray*}
F_{xy} &  \approx E_{0}\sum_{n}\int d\bar{x}\int d\bar{y}\left[  \frac{1}
{2}\left(  \nabla_{\bar{y}}\phi_{n}\right)  ^{2}-\cos\left(  \phi_{n+1}
-\phi_{n}\right)  \right.  \\
&  \left.  -\frac{1}{2}\left(  \nabla_{\bar{y}}\phi_{n}^{(0)}\right)
^{2}+\cos\left(  \phi_{n+1}^{(0)}-\phi_{n}^{(0)}\right)  \right],
\end{eqnarray*}
we replace $\phi_{n}^{(0)}(\bar{y},\bar{z})$ with $\phi_{n}^{(0)}(\bar{y}
-\bar{u}(\bar{x}),\bar{z})$ and represent $\phi_{n}(\bar{x},\bar{y})$ as
$\phi_{n}(\bar{x},\bar{y})=\phi_{n}^{(0)}(\bar{y}-\bar{u}(\bar{x}
))+\tilde{\phi}_{n}(\bar{x},\bar{y})$ where the Fourier transform of
$\tilde{\phi}_{n}(\bar{x},\bar{y})$ at small wave vectors is given by
\begin{eqnarray*}
\tilde{\phi}(\bar{\mathbf{k}})&=&2\pi i\left(
\frac{1}{\bar{k}^{2}}-\frac
{1}{\bar{k}_{y}^{2}+\bar{k}_{z}^{2}}\right)
\bar{k}_{z}\bar{u}(\bar{k} _{x})\\
&=&-\frac{2\pi
i\bar{k}_{x}^{2}}{\left(  \bar{k}_{y}^{2}+\bar{k}_{z} ^{2}\right)
\bar{k}^{2}}\bar{k}_{z}\bar{u}(\bar{k}_{x}).
\end{eqnarray*}
We will see that the main contribution to $F_{xy}$ comes from the distances of the
order of typical wavelength of deformations far away from the core. Therefore we can
expand with respect to $\tilde{\phi}_{n}$ and can use linear and continuous
approximation
\[
F_{yz}\approx\frac{E_{0}}{2}\int\frac{d^{3}\bar{\mathbf{k}}}{\left(
2\pi\right)  ^{3}}\left(  \bar{k}_{y}^{2}+\bar{k}_{z}^{2}\right)
|\tilde {\phi}(\bar{\mathbf{k}})|^{2}.
\]
Substituting $\tilde{\phi}(\bar{\mathbf{k}})$ and computing integral with respect to
$\bar{k}_{y}$ and $\bar{k}_{z}$, which converges at $\bar{k}_{y}$,
$\bar{k}_{z}\sim\bar{k}_{x}$, we obtain
\begin{equation}
F_{yz}\approx\frac{\pi}{4}E_{0}\int\frac{d\bar{k}_{x}}{2\pi}\bar{k}_{x}
^{2}|\bar{u}(\bar{k}_{x})|^{2}.\label{eq:JVLineFyz}
\end{equation}
Finally, combining \eqref{eq:JVLineFx} and \eqref{eq:JVLineFyz}, we obtain the
line-tension energy of the Josephson vortex \eqref{eq:LineTensEnJV} which is
presented already in the real coordinates and the numerical constant is given by
$C_{t}=2\exp(-\Euler+C_{y})$.

\section{Discrete and nonlinear corrections to the
Josephson vortex phase and energy at large distances from the core\label{App:JVphaseAsymp}}

The phase distribution in the JV\ core $\phi_n (y)$ obeys equation
\eqref{eq:phase_no_screen}. We will measure the in-plane coordinate $y$ in units of
the Josephson length $\Lambda_{J}=\gamma d$ defining the dimensionless coordinate
$\scaledy=y/\gamma d$ and rewrite \eqref{eq:phase_no_screen} in the form
\[
\frac{\mathrm{d} ^{2}\phi_n  }{\mathrm{d} \scaledy^{2}}+ \sin \left[ \phi_{n+1}
\left( \scaledy \right) -\phi_n \left( \scaledy\right) \right] +\sin
\left[ \phi_{n-1} \left( \scaledy\right) -\phi_n \left(
\scaledy\right) \right] =0.
\]
At large distances from the core, $n^2+ \scaledy^2\gg 1$,this equation transforms
into the isotropic London equation $ \nabla^2 \phi =0$. In this region  $\phi_n
\left( \scaledy \right) $ can be approximated by a continuous function $\phi \left(
\scaledy ,\scaledz \right) $ with $ n\rightarrow \scaledz $. Using Taylor series for
the difference $\phi \left( \scaledy ,\scaledz +1\right) -\phi \left( \scaledy
,\scaledz \right)$,  we obtain,
\begin{eqnarray*}
&&\sin \left[ \phi \left( \scaledy ,\scaledz \!+\!1\right)
\!-\!\phi\left( \scaledy ,\scaledz \right) \right]
\!+\!\sin\left[ \phi \left( \scaledy ,\scaledz \!-\!1\right)
\!-\!\phi \left( \scaledy ,\scaledz \right) \right] \\
&&\hspace{1cm}\approx \frac{\partial ^{2}\phi }{\partial \scaledz
^{2}}+\frac{1}{12}\frac{
\partial ^{4}\phi }{\partial \scaledz ^{4}}-\frac{1}{2}\left( \frac{\partial \phi }{
\partial \scaledz }\right) ^{2}\frac{\partial ^{2}\phi }{\partial \scaledz ^{2}}
+\ldots .
\end{eqnarray*}
Therefore, the phase equation to 4-th order in the gradient (which is small at large
distances) is given by
\begin{equation}
\frac{\partial ^{2}\phi }{\partial \scaledy ^{2}}+\frac{\partial
^{2}\phi }{\partial \scaledz ^{2}}+\frac{1}{12}\frac{\partial
^{4}\phi }{\partial \scaledz ^{4}}-\frac{1}{2} \left( \frac{\partial
\phi }{\partial \scaledz }\right) ^{2}\frac{\partial ^{2}\phi
}{\partial \scaledz ^{2}}=0. \label{eq:JVphase_asypt}
\end{equation}
This equation can be solved iteratively. For the Josephson vortex located at
$\scaledy=0$ in between the layers $0$ and $1$ the zero-order solution $\phi^0$
(correct to second order in the gradients), is given by the angle $\phi
^{0}(\scaledy ,\scaledz )=-\tan^{-1} [(\scaledz-1/2) /\scaledy ]$ (note that for
$\scaledz = n$ we have $\phi^0(y/\gamma d,n) =\phi_n^{\rm Jv}(y)$ in
\eqref{eq:JVphase_outside}). The first-order correction $\delta\phi ^{1}(\scaledy
,\scaledz )$ obeys the equation
\begin{eqnarray*}
\nabla^2 \delta\phi ^{1} &= &-\frac{1}{12}\frac{\partial ^{4}\phi
^{0}}{\partial \scaledz ^{4}}+ \frac{1}{2}\left( \frac{\partial \phi
^{0}}{\partial \scaledz }\right) ^{2}\frac{
\partial ^{2}\phi ^{0}}{\partial \scaledz ^{2}} \\
&=&\frac{2\,\sin (2\,\phi^0 )+5\,\sin (4\,\phi^0 )}{8\,\scaledr^{4}}
\end{eqnarray*}
where $\scaledr^2=\scaledy^2+(\scaledz-1/2)^2$. Using the solutions of the
inhomogeneous Laplace equations
\begin{eqnarray*}
\nabla^2 \phi  &=&\frac{\sin 2\phi^0 }{\scaledr^{4}}\rightarrow \phi
=-\frac{\sin
2\phi^0 }{4\scaledr^{2}}\ln \scaledr \\
\nabla^2 \phi  &=&\frac{\sin 4\phi^0 }{\scaledr^{4}}\rightarrow \phi
=-\frac{\sin 4\phi^0 }{12\scaledr^{2}}
\end{eqnarray*}
we build the solution for $\delta\phi ^{1}\left( \scaledr,\phi^0 \right) $ and
arrive at the correction
\begin{equation}
\delta \phi \left( \scaledy,\scaledz\right) =\frac{\sin (2\,\phi^0
)}{16\,\scaledr^{2}} \left( \ln \scaledr+C_{\delta\phi}\right)
+\frac{5\sin (4\,\phi^0 )}{96\,\scaledr^{2}} +{\cal
O}(1/\scaledr^4). \label{eq:delta_phi}
\end{equation}
Here we have added the solution $\sin (2\,\phi^0 )/\scaledr^{2}$ of the homogeneous
Laplace equation with an unknown numerical constant $C_{\delta\phi}$. Comparison of
these asymptotics with the full numerical solution gives $C_{\delta\phi}\approx
4.362$. The result \eqref{eq:delta_phi} is given in unscaled coordinates in
\eqref{eq:JVcorr}.

In a similar way, we can derive a nonlinear/discrete correction to the energy far
away from the core. The reduced energy contribution to the Josephson vortex from the
region $\scaledr < \lambda_{\mathrm{ab}}/d$ is given by
\[
\varepsilon_{\mathrm{Jv}}=\int d\scaledy\sum_{n}\left[
\frac{1}{2}\left( \frac{d\phi_{n}}{d\scaledy}\right)
^{2}+1-\cos\left( \phi_{n+1}-\phi_{n}\right)\right].
\]
In the region $\scaledr\gg 1$ we can again apply expansion with respect to small
gradient along the z axis which leads us to the following result
\begin{eqnarray*}
\varepsilon_{\mathrm{Jv}}&\approx& \!\int\limits_{1\ll\scaledr\ll
\lambda_{\mathrm{ab}}/d} d^{2}
\scaledrvec\left[  \frac{1}{2}\left(  \frac{d\phi}%
{d\scaledy}\right)  ^{2}+\frac{1}{2}\left(
\frac{\partial\phi}{\partial \scaledz}\right)
^{2}\right.\\
&-&\left.\frac{1}{24}\left( \frac{\partial^{2}\phi}{\partial
\scaledz^{2}}\right) ^{2}-\frac{1}{24}\left(
\frac{\partial\phi}{\partial \scaledz}\right) ^{4}\right].
\end{eqnarray*}
In the lowest order with respect to small gradients, this gives us the correction to
the energy due to layered structure
\begin{equation}
\delta \varepsilon_{\mathrm{Jv}}\!=\!
-\frac{1}{24}\int\limits_{1\ll\scaledrvec\ll \lambda_{\mathrm{ab}}/d} d^{2}
\!\mathbf{\scaledr}\left[ \left( \frac{\partial^{2}\phi ^0}{\partial
\scaledz^{2}}\right) ^{2}\!+\!\left( \frac{\partial\phi ^0}{\partial
\scaledz}\right) ^{4}\right].%
\label{eq:nonlinJVcorr}
\end{equation}%
In the case of a single Josephson vortex this formula is not very useful because the
integral is formally diverging at small distances and is determined by the
small-distance cut off.  In the case of finite vortex density, however,
generalization of this equation will allow us to obtain a nontrivial correction to
the vortex-lattice energy.

In the vortex-lattice case at finite in-plane field following the same reasoning we
obtain the correction to the reduced energy per unit cell \eqref{eq:LDEner}
\begin{equation}
\hspace{-0.02in}\delta u\!=\!\frac{1}{\pi}\!\int\limits_{u.c.} d^{2}\scaledrvec\!\left[
\!-\!\frac{1}{24}\left( \frac{\partial^{2}\phi^0}{\partial
\scaledz^{2}}\right) ^{2}\!-\!\frac {1}{24}\left(
\frac{\partial\phi^0}{\partial \scaledz}\!+\!h\scaledy\right)
^{4}\right], \label{eq:nonlJVLcorr}
\end{equation}
where integration is performed over the unit cell and $\phi^0(\scaledrvec)$ is the
vortex-lattice phase within London approximation. To estimate the dominating
contribution, we use  the circular-cell approximation for the lattice phase. In this
approximation supercurrents flow radially within the cell $\scaledr<a_c =
\sqrt{2/h}$ and vanish at its boundary, so that the gauge-invariant phase gradient
is given by
\[
\frac{1}{\scaledr}\frac{\partial\phi_{\mathrm{cc}}}{\partial\alpha}
=-\frac{1} {\scaledr}+\frac{\scaledr}{a_{c}^{2}} \ \ \hbox{for
}0<r<a_c
\]
where $\alpha =\tan^{-1}(\scaledz/\scaledy)$ is the polar angle meaning that
$\partial\phi^0/\partial \scaledz+h\scaledy \rightarrow -\cos
(\alpha)(1/r-r/a_{c}^2)$. The integral is formally diverging at small distances.
This divergency, however, is due to the vortex-core energy. To find the nontrivial
correction to the lattice energy we subtract the diverging single-vortex term. The
dominating contribution to the rest part is coming from the second [nonlinear] term
\begin{equation}
\delta u\approx-\frac{1}{24\pi}\int_0^{2\pi}
d\alpha\int_0^{a_c}\scaledr d\scaledr \cos^4 (\alpha)\left[\left(
\frac{1} {\scaledr}-\frac{\scaledr}{a_{c}^{2}} \right) ^{4}-\frac{1}
{\scaledr^4}\right], \label{eq:nonlJVLcorrExp}
\end{equation}
and calculation gives the following result
\begin{equation}
\delta u=\frac{h}{32}\ln\frac{C_{h}}{h}.
\label{eq:nonlJVLcorrResult}
\end{equation}
From fit of the numerically computed energy to this formula we obtain numerical
constant $C_{h}\approx 110$. Note that this correction does not depend on lattice
orientation with respect to the layers. Interaction with the layers also eliminates
the ``elliptic'' rotation degeneracy of the lattice described in Section
\ref{sec:jvldilute}. The expansion \eqref{eq:nonlJVLcorrExp}, however, is not
sufficient to find the orientation-dependent correction to the energy. To obtain
such correction one has to obtain the next-order expansion with respect to the
gradients (6-th order terms).

\section{Calculation of the orientation-dependent fluctuation correction
to the free energy\label{App:EntCorrCalc}}

\begin{figure}[ptb]
\begin{center}
\includegraphics[width=3.4in]{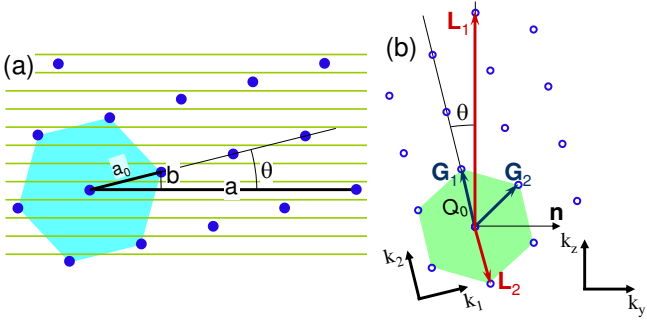}
\end{center}
\caption{\emph{Left part: }Josephson vortex lattice in reduced
coordinates rotated at finite angle $\theta$ with respect to the
layers in real space so that the layers align with the
crystallographic direction (3,1). \emph{Right part:} The
corresponding reciprocal lattice and illustration of two selections
for the basis used in the calculation of the entropy correction: the
basic wave vectors aligned with the lattice, $\mathbf{G}_{1,2}$, and
the basic wave vectors aligned with the layers, $\mathbf{L}_{1,2}$.}
\label{fig:AnEntrCorr}
\end{figure}
In this appendix we present calculation of the entropy correction to free energy
\eqref{eq:EnCorrDef} based on the planar elastic energy \eqref{eq:PlanElEner}. To
facilitate calculations we again introduce the reduced wave vectors, $\mathbf{\tilde
{k}}$, defined in Eq.\ \eqref{eq:JVL-red-k}
and the corresponding reciprocal-lattice vectors $\mathbf{\tilde{Q}}
=(\tilde{Q}_{y},\tilde{Q}_{z})$. In this presentation the reciprocal lattice becomes
a regular triangular lattice with the unit vector $Q_{0}=\sqrt {2\pi/\sqrt{3}}$ and
area of the Bravais cell equals to $\pi$. Using new variables, we represent
$\Phi_{\mathrm{JVL}}(\mathbf{k})$ in the compact reduced form as
\begin{eqnarray}
&&\Phi_{\mathrm{JVL}}(\mathbf{k})  \!=\!
\frac{B_{x}^{2}}{4\pi\lambda_{\mathrm{c}}^{2}}
\phi_{\mathrm{JVL}}(\mathbf{\tilde{k}})\hspace{1.7in}\label{eq:ElEnRed1}\\
&&\phi_{\mathrm{JVL}}(\mathbf{\tilde{k}})\!=\!
\sum_{\mathbf{\tilde{Q}}}\!\left[ \frac{\left(
\tilde{k}_{y}-\tilde{Q}_{y}\right)  ^{2}+\tilde{k}_{x}^{2}
}{b_{x}^{-\!1}\!+\!\left(
\mathbf{\tilde{k}}_{\mathrm{yz}}\!-\!\mathbf{\tilde{Q} }\right)
^{2}\!+\!\tilde{k}_{x}^{2}}\!-\!\frac{\tilde{Q}_{y}^{2}}{b_{x}^{-\!1}
\!+\!\tilde{Q}^{2}}\right]  \label{eq:ElEnRed2}
\end{eqnarray}
where $b_{x}=4\pi\lambda_{\mathrm{ab}}\lambda_{\mathrm{c}}B_{x}/\Phi_{0}=2\left(
\lambda_{\mathrm{ab}}/d\right) ^{2}h\gg1$ and
$\mathbf{\tilde{k}}_{\mathrm{yz}}=(0,\tilde{k}_{y,}\tilde {k}_{z})$.

We will assume that the lattice is rotated at a finite angle, $\theta$, with respect
to the layers selected in such a way that the layers are aligned with one of the
crystallographic directions as sketched in Fig.\ \ref{fig:AnEntrCorr}. This means
that the lattice, in general, has the form of a misaligned lattice sketched in Fig.
\ref{fig:JVLcommens}a and it is characterized by the aspect ratio, $r=\gamma b/a$,
and the shift parameter, $q$. For computing the sum over the reciprocal-lattice
vectors, we will use two equivalent parametrizations illustrated in Fig.\
\ref{fig:AnEntrCorr}. The first parametrization uses expansion over the two basic
vector of tilted lattice, $\mathbf{\tilde{Q}}=n\mathbf{G}_{1}+m\mathbf{G}_{2}$ with
$m,n=0,\pm 1,\pm2\ldots$. For such expansion we can simply represent the component
of $\mathbf{\tilde{Q}}$ along the two main directions of the tilted lattice,
($\mathbf{k}_{1}$, $\mathbf{k}_{2}$), shown in Fig.\ \ref{fig:AnEntrCorr},
\begin{equation}
\tilde{Q}_{1} =\frac{\sqrt{3}}{2}mQ_{0},\ \tilde{Q}_{2}=\left(
n+\frac{m} {2}\right)  Q_{0}. \label{eq:JVL-lattice-basis}
\end{equation}
This gives $\tilde{Q}^{2} =\left(  n^{2}+nm+m^{2}\right) Q_{0}^{2}$. The
($y,z$)-components of the wave vectors are related to the ($1,2$)-components by axis
rotation. For example, for the component, $\tilde {k}_{y}$, in Eq.\
\eqref{eq:ElEnRed2} we have $\tilde{k}_{y}=\cos\theta
\tilde{k}_{1}+\sin\theta\tilde{k}_{2}$. This parametrization allows us to trace
naturally the dependence on the rotation angle $\theta$. The second parametrization
utilizes the basic wave vectors aligned with the layers,
\begin{eqnarray}
\mathbf{\tilde{Q}}  &  =&n\mathbf{L}_{1}+m\mathbf{L}_{2},\label{eq:Q-rq}\\
\mathbf{L}_{1}  &  =&\left(  0,\sqrt{\frac{\pi}{r}}\right)  ;\
\mathbf{L} _{2}=\left(  \sqrt{\pi r},-q\sqrt{\frac{\pi}{r}}\right)
.\nonumber
\end{eqnarray}
This basis allows us to trace easily the dependence on the lattice-structure
parameters $r$ and $q$. It also allows us to reduce $\phi_{\mathrm{JVL}
}(\mathbf{\tilde{k}})$ to a simpler form. Substituting presentation \eqref{eq:Q-rq}
into Eq. \eqref{eq:ElEnRed2} and taking sum over $n$, we obtain
\begin{widetext}
\begin{eqnarray}\label{eq:JVL-el-matr-pres}
&&\phi_{\mathrm{JVL}}(\mathbf{\tilde{k}})\!
=\sqrt{\pi r}\sum_{m=-\infty }^{\infty}\left( \frac{\left(
\tilde{k}_{y}\!-\!m\sqrt{\pi r}\right)
^{2}+\tilde{k}_{x}^{2}}{\kappa(\tilde{k}_{y}\!-\!m\sqrt{\pi
r},\tilde{k}_{x} )}\frac{\sinh\left[  2\sqrt{\pi
r}\kappa(\tilde{k}_{y}\!-\!m\sqrt{\pi r},\tilde{k}_{x})\right]
}{\cosh\left[  2\sqrt{\pi r}\kappa(\tilde{k} _{y}\!-\!m\sqrt{\pi
r},\tilde{k}_{x})\right]  \!-\!\cos\left[ 2\pi\left(
qm\!+\!\tilde{k}_{z}\sqrt{\frac{r}{\pi}}\right)  \right]  }\right. \nonumber\\
&&-\left. \frac{\pi rm^{2}}{\kappa(m\sqrt{\pi
r},0)}\frac{\sinh\left[ 2\sqrt{\pi r}\kappa(m\sqrt{\pi r},0)\right]
}{\cosh\left[  2\sqrt{\pi r}\kappa(m\sqrt{\pi r},0)\right]
-\cos(2\pi qm)}\right)
\nonumber
\end{eqnarray}
\end{widetext}
with $\kappa(k_{y},k_{x})\equiv\sqrt{b_{x}^{-1}+k_{y}^{2}+k_{x}^{2}}$. This formula
contains only one summation which makes it convenient for numerical evaluations. On
the other hand, the dependence on the rotation angle here is not obvious and is
hidden in the dependence on the parameters $r$ and $q$.

The sums over the reciprocal-lattice vectors in Eqs.\ \eqref{eq:JVLel_matr} and
\eqref{eq:ElEnRed2} formally diverge logarithmically at large $\mathbf{Q}$
($\mathbf{\tilde{Q}}$). Correspondingly, the sum over $m$ in Eq.
\eqref{eq:JVL-el-matr-pres} also logarithmically diverges. This divergency is due to
the single-vortex tilt energy and it has to be cut at the core size,
$Q_{y}\sim1/\gamma d$. This energy has been considered in details in Sec.
\ref{sec:LineTensJV}. We will split the reduced elastic matrix,
$\phi_{\mathrm{JVL}}(\mathbf{\tilde {k}})$, into the single-vortex,
$\phi_{\mathrm{sv}}(\tilde{k}_{x})$, and interaction,
$\phi_{\mathrm{i}}(\mathbf{\tilde{k}})$, terms,
\[
\phi_{\mathrm{JVL}}(\mathbf{\tilde{k}})=\phi_{\mathrm{sv}}(\tilde{k}_{x}
)+\phi_{\mathrm{i}}(\mathbf{\tilde{k}}).
\]
The single-vortex term $\phi_{\mathrm{sv}}(\tilde{k}_{x})$ can be obtained from Eq.\
\eqref{eq:ElEnRed2} by replacing summation over $\mathbf{\tilde{Q}}$ with
integration
\[
\phi_{\mathrm{sv}}(\tilde{k}_{x})=\int\frac{d^{2}\mathbf{\tilde{Q}}}{\pi
}\left[
\frac{\tilde{Q}_{y}^{2}+\tilde{k}_{x}^{2}}{b_{x}^{-1}+\mathbf{\tilde
{Q}}^{2}+\tilde{k}_{x}^{2}}-\frac{\tilde{Q}_{y}^{2}}{b_{x}^{-1}+\mathbf{\tilde
{Q}}^{2}}\right]
\]
Using Eq.\ \eqref{eq:LineTensEnJV}, we obtain the line-tension term in real units,
$\Phi_{\mathrm{sv}}(k_{x})=\pi(B_{x}/\Phi_{0})\varepsilon_{\mathrm{J}
}k_{x}^{2}\ln(C_{t}/\gamma dk_{x})$ with $\varepsilon_{\mathrm{J}}\equiv
E_{0}/\gamma d$ and $C_{t}\approx2.86$. This corresponds to the following result for
the reduced line-tension term, $\phi_{\mathrm{sv}}(\tilde{k} _{x})=\left(
4\pi\lambda_{\mathrm{c}}^{2}/B_{x}^{2}\right) \Phi_{\mathrm{sv}}(k_{x} )$,
\begin{equation}
\phi_{\mathrm{sv}}(\tilde{k}_{x})\approx\frac{\tilde{k}_{x}^{2}}{2}\ln
\frac{4.09}{h\tilde{k}_{x}^{2}}. \label{eq:JV-sv-r}
\end{equation}
for $\tilde{k}_{x}^{2}\ll4/h$. In the interaction term, $\phi_{\mathrm{i}
}(\mathbf{\tilde{k}})=\phi_{\mathrm{JVL}}(\mathbf{\tilde{k}})-\phi
_{\mathrm{sv}}(\tilde{k}_{x})$, the logarithmic divergency is compensated and the
sum over $\mathbf{\tilde{Q}}$ converges roughly at $\tilde{Q}\sim1$. In particular,
using presentation given by Eq. \eqref{eq:JVL-el-matr-pres}, the interaction term
can be represented as a converging sum,
\begin{widetext}
\begin{eqnarray}\label{JVL-int-el-matr-pres}
&&\phi_{\mathrm{i}}(\mathbf{k}) =\sqrt{\pi r}
\sum_{m=-\infty}^{\infty} \left( \frac{\left(
\tilde{k}_{y}\!-\!m\sqrt{\pi r}\right)  ^{2}\!+\!\tilde{k}_{x}^{2}}
{\kappa(\tilde{k}_{y}\!-\!m\sqrt{\pi
r},\tilde{k}_{x})}\frac{\sinh\left[ 2\sqrt{\pi
r}\kappa(\tilde{k}_{y}\!-\!m\sqrt{\pi r},\tilde{k}_{x})\right]
}{\cosh\!\left[  2\sqrt{\pi r}\kappa(\tilde{k}_{y}\!-\!m\sqrt{\pi
r},\tilde {k}_{x})\right]  \!-\!\cos\!\left[  2\pi\left(
qm\!+\!\tilde{k}_{z}
\sqrt{\frac{r}{\pi}}\right)  \right]  }\right. \nonumber\\
&&  \left.  \!-\!\sum_{\delta=\pm1}\delta\ U\left[
\tilde{k}_{x},\left( m\!+\!\delta/2\right)  \sqrt{\pi
r}\!-\!\tilde{k}_{y}\right]  \right.
\nonumber\\
&&  \left.  -\frac{\pi rm^{2}}{\kappa(m\sqrt{\pi
r},0)}\frac{\sinh\left[ 2\sqrt{\pi r}\kappa(m\sqrt{\pi r},0)\right]
}{\cosh\!\left[  2\sqrt{\pi r}\kappa(m\sqrt{\pi r},0)\right]
\!-\!\cos(2\pi qm)}\!+\!\sum_{\delta=\pm1} \delta~U\!\left[
0,\left(  m\!+\!\delta/2\right)  \sqrt{\pi r}\right]  \right)
\nonumber
\end{eqnarray}
with
\[U\left[  k_{x},k_{y}\right]
\equiv\frac{1}{2}\left[
k_{y}\sqrt{b_{x}^{-1}+k_{x}^{2}+k_{y}^{2}}+\left(  -b_{x}^{-1}+k_{x}
^{2}\right)  \ln\left(
k_{y}+\sqrt{b_{x}^{-1}+k_{x}^{2}+k_{y}^{2}}\right) \right]
\]
\end{widetext}
Here the terms with $U\left[  \ldots,\ldots\right]  $ originate from the
single-vortex contribution $\phi_{\mathrm{sv}}(\tilde{k}_{x})$ which is properly
decomposed to compensate the summation divergency. In spite of its scary look, this
formula is the most suitable one for numerical calculations.

From Eq.\ \eqref{eq:EnCorrDef} we obtain the entropy correction to the free energy
in reduced form
\begin{equation}
\delta f_{T}  \!=\!-\frac{T}{2\sqrt{\gamma}}\left(  \frac{4\pi
B_{x}}{\Phi_{0}}\right)
^{3/2}\!\int\limits_{-\infty}^{\infty}\frac{d\tilde{k}_{x}}{2\pi}\int_{\mathrm{BZ}
}\frac{d\tilde{k}_{y}d\tilde{k}_{z}}{(2\pi)^{2}}\ln\frac{\tilde{C}}
{\phi_{\mathrm{JVL}}(\mathbf{\tilde{k}})} \label{eq:EnCorrRed}
\end{equation}
where $\int_{\mathrm{BZ}}\ldots$ notates the integral over the Brillouin zone and
$\tilde{C}$ is dimensionless constant. Integral over $k_{x}$ is formally diverging.
This divergency is due to short wavelength excitations in the vortex cores and it
does not contribute to the angular-dependent correction. To separate the regular
anisotropic correction, we subtract from the total free energy the isotropic
single-vortex contribution and represent the resulting anisotropic correction as
\begin{equation}
\delta f_{T,a}(\theta)=-\frac{T}{\sqrt{\pi\gamma}}\left(
\frac{B_{x}} {\Phi_{0}}\right)  ^{3/2}g_{a}
\label{eq:an-EntrCorrPres}
\end{equation}
with
\begin{equation}
g_{a}=\int_{\mathrm{BZ}}\frac{d^{2}\mathbf{\tilde{k}}_{\mathrm{yz}}}{\pi}
\int_{0}^{\infty}dk_{x}\ln\frac{\phi_{\mathrm{sv}}(\tilde{k}_{x})}
{\phi_{\mathrm{sv}}(\tilde{k}_{x})+\phi_{\mathrm{i}}(\mathbf{\tilde{k}})}.
\label{eq:EntrCorr-ga}
\end{equation}
This presentation is used in Eq. \eqref{eq:JVL-AnisEn-EntrCorr}.

\begin{figure}[ptb]
\centerline{\includegraphics[width=3.4in,clip]{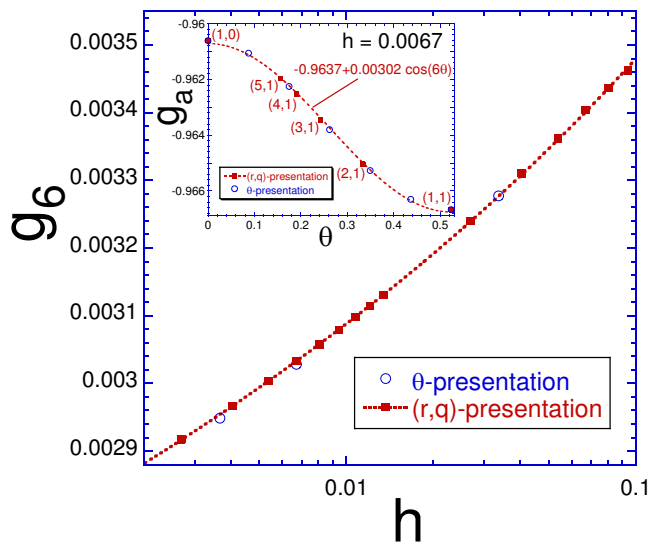}}
\caption{Inset shows the example of the numerically computed angular
dependence of the reduced entropy connection $g_a(\theta)$ defined
by Eq.\ \eqref{eq:EntrCorr-ga} for $h=0.0067$. Solid squares show
results obtained using representation for fixed lattice parameters
$r$ and $q$ given by Eq.\ \eqref{JVL-int-el-matr-pres}. This
computation is done for layers oriented along the crystal directions
$(m,n)$ which are also shown in the plot. Open symbols are obtained
using representation with the explicit dependence on the lattice
rotation angle $\theta$ using the expansion
\eqref{eq:JVL-lattice-basis}. Dashed line is the fit to the formula
$g_0+g_6\cos(6\theta)$. The main plot shows the field dependence of
the coefficient $g_6$ and the corresponding fit in Eq.\ \eqref{eq:ga-num}. } \label{fig:g6-h}
\end{figure}
The large logarithmic factor in $\phi_{\mathrm{sv}}(\tilde{k}_{x})$ in Eq.
\eqref{eq:JV-sv-r} allows us to obtain a useful approximate formula for $g_{a}$. As
$\phi_{\mathrm{i}}(\mathbf{\tilde{k}})\sim1$, integral over $\tilde{k}_{x}$
converges at $\tilde{k}_{x}\sim1/\sqrt{\ln(1/h)}\ll1$ meaning that for a
log-accuracy estimate we can neglect $\tilde{k}_{x}$-dependence of
$\phi_{\mathrm{i}}(\mathbf{\tilde{k}})$. Evaluating the integral over $k_{x}$, we
obtain
\begin{equation}
g_{a}\approx-\frac{\sqrt{2}}{\sqrt{\ln(A/h) }}\int_{\mathrm{BZ}}
d^{2}\mathbf{\tilde{k}}_{\mathrm{yz}}\sqrt{\phi_{\mathrm{i}}(\mathbf{\tilde
{k}}_{\mathrm{yz}})}.
\end{equation}
with $A\sim1$. If we neglect small parameter $b_{x}^{-1}$ in $\phi
_{\mathrm{i}}(\mathbf{\tilde{k}}_{\mathrm{yz}})$ then the integral in this formula
becomes field independent and the only field dependence of $g_{a}$ for
$h\rightarrow0$ is given by the factor $[\ln( A/h)]^{-1/2}$.

We numerically computed the reduced entropy correction $g_{a}$ for different lattice
orientations and reduced fields $h$. Example of the angular dependence of $g_{a}$
for $h=0.0067$ is shown in the inset of figure \ref{fig:g6-h}. We found that in the
range $0.001<h<0.1$ the orientation-dependent part of $g_{a}$ can be well fitted by
the formula \eqref{eq:ga-num}. The dependence $g_{6}(h)$ is plotted in figure
\ref{fig:g6-h}. Positive sign of $g_{6}(h)$ means that the fluctuations give the
largest negative contribution for $\theta=0$, i.e., they indeed favor the aligned
lattice (1,0). We also can see that the effect occurs to be quantitatively rather
small, at least in the considered Gaussian-fluctuations regime.

\newpage

\bibliography{JosVortLat}

\begin{thebibliography}{91}%
\makeatletter
\providecommand \@ifxundefined [1]{%
 \@ifx{#1\undefined}
}%
\providecommand \@ifnum [1]{%
 \ifnum #1\expandafter \@firstoftwo
 \else \expandafter \@secondoftwo
 \fi
}%
\providecommand \@ifx [1]{%
 \ifx #1\expandafter \@firstoftwo
 \else \expandafter \@secondoftwo
 \fi
}%
\providecommand \natexlab [1]{#1}%
\providecommand \enquote  [1]{``#1''}%
\providecommand \bibnamefont  [1]{#1}%
\providecommand \bibfnamefont [1]{#1}%
\providecommand \citenamefont [1]{#1}%
\providecommand \href@noop [0]{\@secondoftwo}%
\providecommand \href [0]{\begingroup \@sanitize@url \@href}%
\providecommand \@href[1]{\@@startlink{#1}\@@href}%
\providecommand \@@href[1]{\endgroup#1\@@endlink}%
\providecommand \@sanitize@url [0]{\catcode `\\12\catcode `\$12\catcode
  `\&12\catcode `\#12\catcode `\^12\catcode `\_12\catcode `\%12\relax}%
\providecommand \@@startlink[1]{}%
\providecommand \@@endlink[0]{}%
\providecommand \url  [0]{\begingroup\@sanitize@url \@url }%
\providecommand \@url [1]{\endgroup\@href {#1}{\urlprefix }}%
\providecommand \urlprefix  [0]{URL }%
\providecommand \Eprint [0]{\href }%
\providecommand \doibase [0]{http://dx.doi.org/}%
\providecommand \selectlanguage [0]{\@gobble}%
\providecommand \bibinfo  [0]{\@secondoftwo}%
\providecommand \bibfield  [0]{\@secondoftwo}%
\providecommand \translation [1]{[#1]}%
\providecommand \BibitemOpen [0]{}%
\providecommand \bibitemStop [0]{}%
\providecommand \bibitemNoStop [0]{.\EOS\space}%
\providecommand \EOS [0]{\spacefactor3000\relax}%
\providecommand \BibitemShut  [1]{\csname bibitem#1\endcsname}%
\let\auto@bib@innerbib\@empty
\bibitem [{\citenamefont {Josephson}(1962)}]{Josephson1962}%
  \BibitemOpen
  \bibfield  {author} {\bibinfo {author} {\bibfnamefont {B.~D.}\ \bibnamefont
  {Josephson}},\ }\href@noop {} {\bibfield  {journal} {\bibinfo  {journal}
  {Phys. Letters}\ }\textbf {\bibinfo {volume} {1}},\ \bibinfo {pages} {251}
  (\bibinfo {year} {1962})}\BibitemShut {NoStop}%
\bibitem [{\citenamefont {Bednorz}\ and\ \citenamefont
  {Muller}(1986)}]{Bednorz1986}%
  \BibitemOpen
  \bibfield  {author} {\bibinfo {author} {\bibfnamefont {J.~G.}\ \bibnamefont
  {Bednorz}}\ and\ \bibinfo {author} {\bibfnamefont {K.~A.}\ \bibnamefont
  {Muller}},\ }\href@noop {} {\bibfield  {journal} {\bibinfo  {journal} {Z.
  Phys. B}\ }\textbf {\bibinfo {volume} {64}},\ \bibinfo {pages} {189}
  (\bibinfo {year} {1986})}\BibitemShut {NoStop}%
\bibitem [{\citenamefont {Wu}\ \emph {et~al.}(1987)\citenamefont {Wu},
  \citenamefont {Ashburn}, \citenamefont {Torng}, \citenamefont {Hor},
  \citenamefont {Meng}, \citenamefont {Gao}, \citenamefont {Huang},
  \citenamefont {Wang},\ and\ \citenamefont {Chu}}]{WuATHMGHWC1987}%
  \BibitemOpen
  \bibfield  {author} {\bibinfo {author} {\bibfnamefont {M.~K.}\ \bibnamefont
  {Wu}}, \bibinfo {author} {\bibfnamefont {J.~R.}\ \bibnamefont {Ashburn}},
  \bibinfo {author} {\bibfnamefont {C.~J.}\ \bibnamefont {Torng}}, \bibinfo
  {author} {\bibfnamefont {P.~H.}\ \bibnamefont {Hor}}, \bibinfo {author}
  {\bibfnamefont {R.~L.}\ \bibnamefont {Meng}}, \bibinfo {author}
  {\bibfnamefont {L.}~\bibnamefont {Gao}}, \bibinfo {author} {\bibfnamefont
  {Z.~J.}\ \bibnamefont {Huang}}, \bibinfo {author} {\bibfnamefont {Y.~Q.}\
  \bibnamefont {Wang}}, \ and\ \bibinfo {author} {\bibfnamefont {C.~W.}\
  \bibnamefont {Chu}},\ }\href@noop {} {\bibfield  {journal} {\bibinfo
  {journal} {Phys. Rev. Lett.}\ }\textbf {\bibinfo {volume} {58}},\ \bibinfo
  {pages} {908} (\bibinfo {year} {1987})}\BibitemShut {NoStop}%
\bibitem [{\citenamefont {Maeda}\ \emph {et~al.}(1988)\citenamefont {Maeda},
  \citenamefont {Tanaka}, \citenamefont {Fukutomi},\ and\ \citenamefont
  {Asano}}]{MaedaTFA1988}%
  \BibitemOpen
  \bibfield  {author} {\bibinfo {author} {\bibfnamefont {H.}~\bibnamefont
  {Maeda}}, \bibinfo {author} {\bibfnamefont {Y.}~\bibnamefont {Tanaka}},
  \bibinfo {author} {\bibfnamefont {M.}~\bibnamefont {Fukutomi}}, \ and\
  \bibinfo {author} {\bibfnamefont {T.}~\bibnamefont {Asano}},\ }\href@noop {}
  {\bibfield  {journal} {\bibinfo  {journal} {Japan. J. Appl. Phys.}\ }\textbf
  {\bibinfo {volume} {27}},\ \bibinfo {pages} {L209} (\bibinfo {year}
  {1988})}\BibitemShut {NoStop}%
\bibitem [{\citenamefont {Chu}\ \emph {et~al.}(1988)\citenamefont {Chu},
  \citenamefont {Bechtold}, \citenamefont {Gao}, \citenamefont {Hor},
  \citenamefont {Huang}, \citenamefont {Meng}, \citenamefont {Sun},
  \citenamefont {Wang},\ and\ \citenamefont {Xue}}]{ChuBGHHMSWX1988}%
  \BibitemOpen
  \bibfield  {author} {\bibinfo {author} {\bibfnamefont {C.~W.}\ \bibnamefont
  {Chu}}, \bibinfo {author} {\bibfnamefont {J.}~\bibnamefont {Bechtold}},
  \bibinfo {author} {\bibfnamefont {L.}~\bibnamefont {Gao}}, \bibinfo {author}
  {\bibfnamefont {P.~H.}\ \bibnamefont {Hor}}, \bibinfo {author} {\bibfnamefont
  {Z.~J.}\ \bibnamefont {Huang}}, \bibinfo {author} {\bibfnamefont {R.~L.}\
  \bibnamefont {Meng}}, \bibinfo {author} {\bibfnamefont {Y.~Y.}\ \bibnamefont
  {Sun}}, \bibinfo {author} {\bibfnamefont {Y.~Q.}\ \bibnamefont {Wang}}, \
  and\ \bibinfo {author} {\bibfnamefont {Y.~Y.}\ \bibnamefont {Xue}},\
  }\href@noop {} {\bibfield  {journal} {\bibinfo  {journal} {Phys. Rev. Lett.}\
  }\textbf {\bibinfo {volume} {60}},\ \bibinfo {pages} {941} (\bibinfo {year}
  {1988})}\BibitemShut {NoStop}%
\bibitem [{\citenamefont {Wilson}\ and\ \citenamefont
  {Yoffe}(1969)}]{WilsonY69}%
  \BibitemOpen
  \bibfield  {author} {\bibinfo {author} {\bibfnamefont {J.}~\bibnamefont
  {Wilson}}\ and\ \bibinfo {author} {\bibfnamefont {A.}~\bibnamefont {Yoffe}},\
  }\href {\doibase 10.1080/00018736900101307} {\bibfield  {journal} {\bibinfo
  {journal} {Adv. in Phys.}\ }\textbf {\bibinfo {volume} {18}},\ \bibinfo
  {pages} {193} (\bibinfo {year} {1969})}\BibitemShut {NoStop}%
\bibitem [{\citenamefont {Bulaevskii}(1975)}]{Bulaevskii1975}%
  \BibitemOpen
  \bibfield  {author} {\bibinfo {author} {\bibfnamefont {L.~N.}\ \bibnamefont
  {Bulaevskii}},\ }\href@noop {} {\bibfield  {journal} {\bibinfo  {journal}
  {Usp. Fiz. Nauk}\ }\textbf {\bibinfo {volume} {116}},\ \bibinfo {pages} {449}
  (\bibinfo {year} {1975})},\ \bibinfo {note} {[{\em Sov. Phys.--Usp.} {\bf
  18}, 514 (1976)]}\BibitemShut {NoStop}%
\bibitem [{\citenamefont {Ishiguro}\ and\ \citenamefont
  {Yamaji}(1990)}]{IshiguroY1990}%
  \BibitemOpen
  \bibfield  {author} {\bibinfo {author} {\bibfnamefont {T.}~\bibnamefont
  {Ishiguro}}\ and\ \bibinfo {author} {\bibfnamefont {K.}~\bibnamefont
  {Yamaji}},\ }\href@noop {} {\emph {\bibinfo {title} {Organic
  Superconductors}}}\ (\bibinfo  {publisher} {Springer},\ \bibinfo {address}
  {Berlin},\ \bibinfo {year} {1990})\ \bibinfo {note} {chapter 5}\BibitemShut
  {NoStop}%
\bibitem [{\citenamefont {Singleton}(2000)}]{Singleton2000}%
  \BibitemOpen
  \bibfield  {author} {\bibinfo {author} {\bibfnamefont {J.}~\bibnamefont
  {Singleton}},\ }\href@noop {} {\bibfield  {journal} {\bibinfo  {journal}
  {Rep. Prog. Phys.}\ }\textbf {\bibinfo {volume} {63}},\ \bibinfo {pages}
  {1111} (\bibinfo {year} {2000})}\BibitemShut {NoStop}%
\bibitem [{\citenamefont {Kamihara}\ \emph {et~al.}(2008)\citenamefont
  {Kamihara}, \citenamefont {Watanabe}, \citenamefont {Hirano},\ and\
  \citenamefont {Hosono}}]{KamiharaWHH2008}%
  \BibitemOpen
  \bibfield  {author} {\bibinfo {author} {\bibfnamefont {Y.}~\bibnamefont
  {Kamihara}}, \bibinfo {author} {\bibfnamefont {T.}~\bibnamefont {Watanabe}},
  \bibinfo {author} {\bibfnamefont {M.}~\bibnamefont {Hirano}}, \ and\ \bibinfo
  {author} {\bibfnamefont {H.}~\bibnamefont {Hosono}},\ }\href {\doibase
  10.1021/ja800073m} {\bibfield  {journal} {\bibinfo  {journal} {J. Am. Chem.
  Soc.}\ }\textbf {\bibinfo {volume} {130}},\ \bibinfo {pages} {3296} (\bibinfo
  {year} {2008})}\BibitemShut {NoStop}%
\bibitem [{\citenamefont {Johnston}(2010)}]{Johnston2010}%
  \BibitemOpen
  \bibfield  {author} {\bibinfo {author} {\bibfnamefont {D.~C.}\ \bibnamefont
  {Johnston}},\ }\href {\doibase 10.1080/00018732.2010.513480} {\bibfield
  {journal} {\bibinfo  {journal} {Adv. in Phys.}\ }\textbf {\bibinfo {volume}
  {59}},\ \bibinfo {pages} {803} (\bibinfo {year} {2010})}\BibitemShut
  {NoStop}%
\bibitem [{\citenamefont {Canfield}\ and\ \citenamefont
  {Bud'ko}(2010)}]{CanfieldB2010}%
  \BibitemOpen
  \bibfield  {author} {\bibinfo {author} {\bibfnamefont {P.~C.}\ \bibnamefont
  {Canfield}}\ and\ \bibinfo {author} {\bibfnamefont {S.~L.}\ \bibnamefont
  {Bud'ko}},\ }\href {\doibase 10.1146/annurev-conmatphys-070909-104041}
  {\bibfield  {journal} {\bibinfo  {journal} {Ann. Rev. of Cond. Mat. Phys.}\
  }\textbf {\bibinfo {volume} {1}},\ \bibinfo {pages} {27} (\bibinfo {year}
  {2010})}\BibitemShut {NoStop}%
\bibitem [{\citenamefont {Stewart}(2011)}]{Stewart2011}%
  \BibitemOpen
  \bibfield  {author} {\bibinfo {author} {\bibfnamefont {G.~R.}\ \bibnamefont
  {Stewart}},\ }\href {\doibase 10.1103/RevModPhys.83.1589} {\bibfield
  {journal} {\bibinfo  {journal} {Rev. Mod. Phys.}\ }\textbf {\bibinfo {volume}
  {83}},\ \bibinfo {pages} {1589} (\bibinfo {year} {2011})}\BibitemShut
  {NoStop}%
\bibitem [{\citenamefont {Chen}\ \emph {et~al.}(2008)\citenamefont {Chen},
  \citenamefont {Wu}, \citenamefont {Wu}, \citenamefont {Liu}, \citenamefont
  {Chen},\ and\ \citenamefont {Fang}}]{ChenWWLCF2008}%
  \BibitemOpen
  \bibfield  {author} {\bibinfo {author} {\bibfnamefont {X.~H.}\ \bibnamefont
  {Chen}}, \bibinfo {author} {\bibfnamefont {T.}~\bibnamefont {Wu}}, \bibinfo
  {author} {\bibfnamefont {G.}~\bibnamefont {Wu}}, \bibinfo {author}
  {\bibfnamefont {R.~H.}\ \bibnamefont {Liu}}, \bibinfo {author} {\bibfnamefont
  {H.}~\bibnamefont {Chen}}, \ and\ \bibinfo {author} {\bibfnamefont {D.~F.}\
  \bibnamefont {Fang}},\ }\href {\doibase {10.1038/nature07045}} {\bibfield
  {journal} {\bibinfo  {journal} {Nature}\ }\textbf {\bibinfo {volume} {453}},\
  \bibinfo {pages} {761} (\bibinfo {year} {2008})}\BibitemShut {NoStop}%
\bibitem [{\citenamefont {Moll}\ \emph {et~al.}(2013)\citenamefont {Moll},
  \citenamefont {Balicas}, \citenamefont {Geshkenbein}, \citenamefont
  {Blatter}, \citenamefont {Karpinski}, \citenamefont {Zhigadlo},\ and\
  \citenamefont {Batlogg}}]{MollBGB2013}%
  \BibitemOpen
  \bibfield  {author} {\bibinfo {author} {\bibfnamefont {P.~J.~W.}\
  \bibnamefont {Moll}}, \bibinfo {author} {\bibfnamefont {L.}~\bibnamefont
  {Balicas}}, \bibinfo {author} {\bibfnamefont {V.}~\bibnamefont
  {Geshkenbein}}, \bibinfo {author} {\bibfnamefont {G.}~\bibnamefont
  {Blatter}}, \bibinfo {author} {\bibfnamefont {J.}~\bibnamefont {Karpinski}},
  \bibinfo {author} {\bibfnamefont {N.~D.}\ \bibnamefont {Zhigadlo}}, \ and\
  \bibinfo {author} {\bibfnamefont {B.}~\bibnamefont {Batlogg}},\ }\href
  {\doibase 10.1038/NMAT3489} {\bibfield  {journal} {\bibinfo  {journal}
  {Nature Materials}\ }\textbf {\bibinfo {volume} {12}},\ \bibinfo {pages}
  {134} (\bibinfo {year} {2013})}\BibitemShut {NoStop}%
\bibitem [{\citenamefont {Zhu}\ \emph {et~al.}(2009)\citenamefont {Zhu},
  \citenamefont {Han}, \citenamefont {Mu}, \citenamefont {Cheng}, \citenamefont
  {Shen}, \citenamefont {Zeng},\ and\ \citenamefont {Wen}}]{ZhuHM2009}%
  \BibitemOpen
  \bibfield  {author} {\bibinfo {author} {\bibfnamefont {X.}~\bibnamefont
  {Zhu}}, \bibinfo {author} {\bibfnamefont {F.}~\bibnamefont {Han}}, \bibinfo
  {author} {\bibfnamefont {G.}~\bibnamefont {Mu}}, \bibinfo {author}
  {\bibfnamefont {P.}~\bibnamefont {Cheng}}, \bibinfo {author} {\bibfnamefont
  {B.}~\bibnamefont {Shen}}, \bibinfo {author} {\bibfnamefont {B.}~\bibnamefont
  {Zeng}}, \ and\ \bibinfo {author} {\bibfnamefont {H.-H.}\ \bibnamefont
  {Wen}},\ }\href {\doibase 10.1103/PhysRevB.79.220512} {\bibfield  {journal}
  {\bibinfo  {journal} {Phys. Rev. B}\ }\textbf {\bibinfo {volume} {79}},\
  \bibinfo {pages} {220512} (\bibinfo {year} {2009})}\BibitemShut {NoStop}%
\bibitem [{\citenamefont {Ogino}\ \emph {et~al.}(2009)\citenamefont {Ogino},
  \citenamefont {Matsumura}, \citenamefont {Katsura}, \citenamefont {Ushiyama},
  \citenamefont {Horii}, \citenamefont {Kishio},\ and\ \citenamefont {ichi
  Shimoyama}}]{OginoMK2009}%
  \BibitemOpen
  \bibfield  {author} {\bibinfo {author} {\bibfnamefont {H.}~\bibnamefont
  {Ogino}}, \bibinfo {author} {\bibfnamefont {Y.}~\bibnamefont {Matsumura}},
  \bibinfo {author} {\bibfnamefont {Y.}~\bibnamefont {Katsura}}, \bibinfo
  {author} {\bibfnamefont {K.}~\bibnamefont {Ushiyama}}, \bibinfo {author}
  {\bibfnamefont {S.}~\bibnamefont {Horii}}, \bibinfo {author} {\bibfnamefont
  {K.}~\bibnamefont {Kishio}}, \ and\ \bibinfo {author} {\bibfnamefont
  {J.}~\bibnamefont {ichi Shimoyama}},\ }\href
  {http://stacks.iop.org/0953-2048/22/i=7/a=075008} {\bibfield  {journal}
  {\bibinfo  {journal} {Supercond. Sci. Technol.}\ }\textbf {\bibinfo {volume}
  {22}},\ \bibinfo {pages} {075008} (\bibinfo {year} {2009})}\BibitemShut
  {NoStop}%
\bibitem [{\citenamefont {Sato}\ \emph {et~al.}(2010)\citenamefont {Sato},
  \citenamefont {Ogino}, \citenamefont {Kawaguchi}, \citenamefont {Katsura},
  \citenamefont {Kishio}, \citenamefont {ichi Shimoyama}, \citenamefont
  {Kotegawa},\ and\ \citenamefont {Tou}}]{SatoOK2010}%
  \BibitemOpen
  \bibfield  {author} {\bibinfo {author} {\bibfnamefont {S.}~\bibnamefont
  {Sato}}, \bibinfo {author} {\bibfnamefont {H.}~\bibnamefont {Ogino}},
  \bibinfo {author} {\bibfnamefont {N.}~\bibnamefont {Kawaguchi}}, \bibinfo
  {author} {\bibfnamefont {Y.}~\bibnamefont {Katsura}}, \bibinfo {author}
  {\bibfnamefont {K.}~\bibnamefont {Kishio}}, \bibinfo {author} {\bibfnamefont
  {J.}~\bibnamefont {ichi Shimoyama}}, \bibinfo {author} {\bibfnamefont
  {H.}~\bibnamefont {Kotegawa}}, \ and\ \bibinfo {author} {\bibfnamefont
  {H.}~\bibnamefont {Tou}},\ }\href
  {http://stacks.iop.org/0953-2048/23/i=4/a=045001} {\bibfield  {journal}
  {\bibinfo  {journal} {Supercond. Sci. Technol.}\ }\textbf {\bibinfo {volume}
  {23}},\ \bibinfo {pages} {045001} (\bibinfo {year} {2010})}\BibitemShut
  {NoStop}%
\bibitem [{\citenamefont {Blatter}\ \emph {et~al.}(1994)\citenamefont
  {Blatter}, \citenamefont {Feigelman}, \citenamefont {Geshkenbein},
  \citenamefont {Larkin},\ and\ \citenamefont {Vinokur}}]{BlatterFGLV1994}%
  \BibitemOpen
  \bibfield  {author} {\bibinfo {author} {\bibfnamefont {G.}~\bibnamefont
  {Blatter}}, \bibinfo {author} {\bibfnamefont {M.~V.}\ \bibnamefont
  {Feigelman}}, \bibinfo {author} {\bibfnamefont {V.~B.}\ \bibnamefont
  {Geshkenbein}}, \bibinfo {author} {\bibfnamefont {A.~I.}\ \bibnamefont
  {Larkin}}, \ and\ \bibinfo {author} {\bibfnamefont {V.~M.}\ \bibnamefont
  {Vinokur}},\ }\href@noop {} {\bibfield  {journal} {\bibinfo  {journal} {Rev.
  Mod. Phys.}\ }\textbf {\bibinfo {volume} {66}},\ \bibinfo {pages} {1125}
  (\bibinfo {year} {1994})}\BibitemShut {NoStop}%
\bibitem [{\citenamefont {Feinberg}(1994)}]{Feinberg1994}%
  \BibitemOpen
  \bibfield  {author} {\bibinfo {author} {\bibfnamefont {D.}~\bibnamefont
  {Feinberg}},\ }\href@noop {} {\bibfield  {journal} {\bibinfo  {journal} {J.
  Physique III}\ }\textbf {\bibinfo {volume} {4}},\ \bibinfo {pages} {169}
  (\bibinfo {year} {1994})}\BibitemShut {NoStop}%
\bibitem [{\citenamefont {Brandt}(1995)}]{Brandt1995}%
  \BibitemOpen
  \bibfield  {author} {\bibinfo {author} {\bibfnamefont {E.~H.}\ \bibnamefont
  {Brandt}},\ }\href@noop {} {\bibfield  {journal} {\bibinfo  {journal} {Rep.
  Prog. Phys.}\ }\textbf {\bibinfo {volume} {58}},\ \bibinfo {pages} {1465}
  (\bibinfo {year} {1995})}\BibitemShut {NoStop}%
\bibitem [{\citenamefont {Nattermann}\ and\ \citenamefont
  {Scheidl}(2000)}]{NattermannS2000}%
  \BibitemOpen
  \bibfield  {author} {\bibinfo {author} {\bibfnamefont {T.}~\bibnamefont
  {Nattermann}}\ and\ \bibinfo {author} {\bibfnamefont {S.}~\bibnamefont
  {Scheidl}},\ }\href@noop {} {\bibfield  {journal} {\bibinfo  {journal} {Adv.
  Phys.}\ }\textbf {\bibinfo {volume} {49}},\ \bibinfo {pages} {607} (\bibinfo
  {year} {2000})}\BibitemShut {NoStop}%
\bibitem [{\citenamefont {Blatter}\ and\ \citenamefont
  {Geshkenbein}(2003)}]{BlatterG2003}%
  \BibitemOpen
  \bibfield  {author} {\bibinfo {author} {\bibfnamefont {G.}~\bibnamefont
  {Blatter}}\ and\ \bibinfo {author} {\bibfnamefont {V.~B.}\ \bibnamefont
  {Geshkenbein}},\ }\href@noop {} {\emph {\bibinfo {title} {in ``The Physics of
  Superconductors'', Vol 1: Conventional and High-$T_c$ Superconductors}}},\
  edited by\ \bibinfo {editor} {\bibfnamefont {K.}~\bibnamefont {Bennemann}}\
  and\ \bibinfo {editor} {\bibfnamefont {J.}~\bibnamefont {Ketterson}}\
  (\bibinfo  {publisher} {Springer},\ \bibinfo {address} {Berlin},\ \bibinfo
  {year} {2003})\ p.\ \bibinfo {pages} {726},\ \bibinfo {note} {ed Bennemann K
  H and Ketterson J B}\BibitemShut {NoStop}%
\bibitem [{\citenamefont {Campbell}\ \emph {et~al.}(1988)\citenamefont
  {Campbell}, \citenamefont {Doria},\ and\ \citenamefont
  {Kogan}}]{CampbellDK1988}%
  \BibitemOpen
  \bibfield  {author} {\bibinfo {author} {\bibfnamefont {L.~J.}\ \bibnamefont
  {Campbell}}, \bibinfo {author} {\bibfnamefont {M.~M.}\ \bibnamefont {Doria}},
  \ and\ \bibinfo {author} {\bibfnamefont {V.~G.}\ \bibnamefont {Kogan}},\
  }\href@noop {} {\bibfield  {journal} {\bibinfo  {journal} {Phys. Rev. B}\
  }\textbf {\bibinfo {volume} {38}},\ \bibinfo {pages} {2439} (\bibinfo {year}
  {1988})}\BibitemShut {NoStop}%
\bibitem [{\citenamefont {Bulaevskii}\ and\ \citenamefont
  {Clem}(1991)}]{BulaevskiiC1991}%
  \BibitemOpen
  \bibfield  {author} {\bibinfo {author} {\bibfnamefont {L.}~\bibnamefont
  {Bulaevskii}}\ and\ \bibinfo {author} {\bibfnamefont {J.~R.}\ \bibnamefont
  {Clem}},\ }\href@noop {} {\bibfield  {journal} {\bibinfo  {journal} {Phys.
  Rev. B}\ }\textbf {\bibinfo {volume} {44}},\ \bibinfo {pages} {10234}
  (\bibinfo {year} {1991})}\BibitemShut {NoStop}%
\bibitem [{Note1()}]{Note1}%
  \BibitemOpen
  \bibinfo {note} {In the literature the layer plane and the axis perpendicular
  to the layers are frequently called ``ab plane'' and ``c axis''.}\BibitemShut
  {Stop}%
\bibitem [{\citenamefont {Clem}(2004)}]{Clem04}%
  \BibitemOpen
  \bibfield  {author} {\bibinfo {author} {\bibfnamefont {J.~R.}\ \bibnamefont
  {Clem}},\ }\href {\doibase 10.1007/s10948-004-0774-z} {\bibfield  {journal}
  {\bibinfo  {journal} {Journal of Superconductivity}\ }\textbf {\bibinfo
  {volume} {17}},\ \bibinfo {pages} {613} (\bibinfo {year} {2004})}\BibitemShut
  {NoStop}%
\bibitem [{\citenamefont {Dolan}\ \emph {et~al.}(1989)\citenamefont {Dolan},
  \citenamefont {Holtzberg}, \citenamefont {Feild},\ and\ \citenamefont
  {Dinger}}]{DolanHFD1989}%
  \BibitemOpen
  \bibfield  {author} {\bibinfo {author} {\bibfnamefont {G.~J.}\ \bibnamefont
  {Dolan}}, \bibinfo {author} {\bibfnamefont {F.}~\bibnamefont {Holtzberg}},
  \bibinfo {author} {\bibfnamefont {C.}~\bibnamefont {Feild}}, \ and\ \bibinfo
  {author} {\bibfnamefont {T.~R.}\ \bibnamefont {Dinger}},\ }\href@noop {}
  {\bibfield  {journal} {\bibinfo  {journal} {Phys. Rev. Lett.}\ }\textbf
  {\bibinfo {volume} {62}},\ \bibinfo {pages} {2184} (\bibinfo {year}
  {1989})}\BibitemShut {NoStop}%
\bibitem [{\citenamefont {Oussena}\ \emph {et~al.}(1994)\citenamefont
  {Oussena}, \citenamefont {de~Groot}, \citenamefont {Gagnon},\ and\
  \citenamefont {Taillefer}}]{OussenaGGT94}%
  \BibitemOpen
  \bibfield  {author} {\bibinfo {author} {\bibfnamefont {M.}~\bibnamefont
  {Oussena}}, \bibinfo {author} {\bibfnamefont {P.~A.~J.}\ \bibnamefont
  {de~Groot}}, \bibinfo {author} {\bibfnamefont {R.}~\bibnamefont {Gagnon}}, \
  and\ \bibinfo {author} {\bibfnamefont {L.}~\bibnamefont {Taillefer}},\ }\href
  {\doibase 10.1103/PhysRevLett.72.3606} {\bibfield  {journal} {\bibinfo
  {journal} {Phys. Rev. Lett.}\ }\textbf {\bibinfo {volume} {72}},\ \bibinfo
  {pages} {3606} (\bibinfo {year} {1994})}\BibitemShut {NoStop}%
\bibitem [{\citenamefont {Zhukov}\ \emph {et~al.}(1999)\citenamefont {Zhukov},
  \citenamefont {K\"upfer}, \citenamefont {Perkins}, \citenamefont {Caplin},
  \citenamefont {Wolf}, \citenamefont {Kugel}, \citenamefont {Rakhmanov},
  \citenamefont {Mikheev}, \citenamefont {Voronkova}, \citenamefont
  {Kl\"aser},\ and\ \citenamefont {W\"uhl}}]{ZhukovKPC1999}%
  \BibitemOpen
  \bibfield  {author} {\bibinfo {author} {\bibfnamefont {A.~A.}\ \bibnamefont
  {Zhukov}}, \bibinfo {author} {\bibfnamefont {H.}~\bibnamefont {K\"upfer}},
  \bibinfo {author} {\bibfnamefont {G.~K.}\ \bibnamefont {Perkins}}, \bibinfo
  {author} {\bibfnamefont {A.~D.}\ \bibnamefont {Caplin}}, \bibinfo {author}
  {\bibfnamefont {T.}~\bibnamefont {Wolf}}, \bibinfo {author} {\bibfnamefont
  {K.~I.}\ \bibnamefont {Kugel}}, \bibinfo {author} {\bibfnamefont {A.~L.}\
  \bibnamefont {Rakhmanov}}, \bibinfo {author} {\bibfnamefont {M.~G.}\
  \bibnamefont {Mikheev}}, \bibinfo {author} {\bibfnamefont {V.~I.}\
  \bibnamefont {Voronkova}}, \bibinfo {author} {\bibfnamefont {M.}~\bibnamefont
  {Kl\"aser}}, \ and\ \bibinfo {author} {\bibfnamefont {H.}~\bibnamefont
  {W\"uhl}},\ }\href {\doibase 10.1103/PhysRevB.59.11213} {\bibfield  {journal}
  {\bibinfo  {journal} {Phys. Rev. B}\ }\textbf {\bibinfo {volume} {59}},\
  \bibinfo {pages} {11213} (\bibinfo {year} {1999})}\BibitemShut {NoStop}%
\bibitem [{\citenamefont {Gordeev}\ \emph {et~al.}(2000)\citenamefont
  {Gordeev}, \citenamefont {Zhukov}, \citenamefont {de~Groot}, \citenamefont
  {Jansen}, \citenamefont {Gagnon},\ and\ \citenamefont
  {Taillefer}}]{GordeevZdG2000}%
  \BibitemOpen
  \bibfield  {author} {\bibinfo {author} {\bibfnamefont {S.~N.}\ \bibnamefont
  {Gordeev}}, \bibinfo {author} {\bibfnamefont {A.~A.}\ \bibnamefont {Zhukov}},
  \bibinfo {author} {\bibfnamefont {P.~A.~J.}\ \bibnamefont {de~Groot}},
  \bibinfo {author} {\bibfnamefont {A.~G.~M.}\ \bibnamefont {Jansen}}, \bibinfo
  {author} {\bibfnamefont {R.}~\bibnamefont {Gagnon}}, \ and\ \bibinfo {author}
  {\bibfnamefont {L.}~\bibnamefont {Taillefer}},\ }\href {\doibase
  10.1103/PhysRevLett.85.4594} {\bibfield  {journal} {\bibinfo  {journal}
  {Phys. Rev. Lett.}\ }\textbf {\bibinfo {volume} {85}},\ \bibinfo {pages}
  {4594} (\bibinfo {year} {2000})}\BibitemShut {NoStop}%
\bibitem [{\citenamefont {Bolle}\ \emph {et~al.}(1991)\citenamefont {Bolle},
  \citenamefont {Gammel}, \citenamefont {Grier}, \citenamefont {Murray},
  \citenamefont {Bishop}, \citenamefont {Mitzi},\ and\ \citenamefont
  {Kapitulnik}}]{BolleGGMBMK1991}%
  \BibitemOpen
  \bibfield  {author} {\bibinfo {author} {\bibfnamefont {C.~A.}\ \bibnamefont
  {Bolle}}, \bibinfo {author} {\bibfnamefont {P.~L.}\ \bibnamefont {Gammel}},
  \bibinfo {author} {\bibfnamefont {D.~G.}\ \bibnamefont {Grier}}, \bibinfo
  {author} {\bibfnamefont {C.~A.}\ \bibnamefont {Murray}}, \bibinfo {author}
  {\bibfnamefont {D.~J.}\ \bibnamefont {Bishop}}, \bibinfo {author}
  {\bibfnamefont {D.~B.}\ \bibnamefont {Mitzi}}, \ and\ \bibinfo {author}
  {\bibfnamefont {A.}~\bibnamefont {Kapitulnik}},\ }\href@noop {} {\bibfield
  {journal} {\bibinfo  {journal} {Phys. Rev. Lett.}\ }\textbf {\bibinfo
  {volume} {66}},\ \bibinfo {pages} {112} (\bibinfo {year} {1991})}\BibitemShut
  {NoStop}%
\bibitem [{\citenamefont {Tokunaga}\ \emph {et~al.}(2003)\citenamefont
  {Tokunaga}, \citenamefont {Tamegai}, \citenamefont {Fasano},\ and\
  \citenamefont {de~la Cruz}}]{TokunagaTFC2003}%
  \BibitemOpen
  \bibfield  {author} {\bibinfo {author} {\bibfnamefont {M.}~\bibnamefont
  {Tokunaga}}, \bibinfo {author} {\bibfnamefont {T.}~\bibnamefont {Tamegai}},
  \bibinfo {author} {\bibfnamefont {Y.}~\bibnamefont {Fasano}}, \ and\ \bibinfo
  {author} {\bibfnamefont {F.}~\bibnamefont {de~la Cruz}},\ }\href@noop {}
  {\bibfield  {journal} {\bibinfo  {journal} {Phys. Rev. B}\ }\textbf {\bibinfo
  {volume} {67}},\ \bibinfo {pages} {134501} (\bibinfo {year}
  {2003})}\BibitemShut {NoStop}%
\bibitem [{\citenamefont {Grigorenko}\ \emph {et~al.}(2001)\citenamefont
  {Grigorenko}, \citenamefont {Bending}, \citenamefont {Tamegai}, \citenamefont
  {Ooi},\ and\ \citenamefont {Henini}}]{GrigorenkoBTOH2001}%
  \BibitemOpen
  \bibfield  {author} {\bibinfo {author} {\bibfnamefont {A.}~\bibnamefont
  {Grigorenko}}, \bibinfo {author} {\bibfnamefont {S.}~\bibnamefont {Bending}},
  \bibinfo {author} {\bibfnamefont {T.}~\bibnamefont {Tamegai}}, \bibinfo
  {author} {\bibfnamefont {S.}~\bibnamefont {Ooi}}, \ and\ \bibinfo {author}
  {\bibfnamefont {M.}~\bibnamefont {Henini}},\ }\href@noop {} {\bibfield
  {journal} {\bibinfo  {journal} {Nature}\ }\textbf {\bibinfo {volume} {414}},\
  \bibinfo {pages} {728} (\bibinfo {year} {2001})}\BibitemShut {NoStop}%
\bibitem [{\citenamefont {Matsuda}\ \emph {et~al.}(2001)\citenamefont
  {Matsuda}, \citenamefont {Kamimura}, \citenamefont {Kasai}, \citenamefont
  {Harada}, \citenamefont {Yoshida}, \citenamefont {Akashi}, \citenamefont
  {Tonomura}, \citenamefont {Nakayama}, \citenamefont {Shimoyama},
  \citenamefont {Kishio}, \citenamefont {Hanaguri},\ and\ \citenamefont
  {Kitazawa}}]{MatsudaKKHYATNSKHK2001}%
  \BibitemOpen
  \bibfield  {author} {\bibinfo {author} {\bibfnamefont {T.}~\bibnamefont
  {Matsuda}}, \bibinfo {author} {\bibfnamefont {O.}~\bibnamefont {Kamimura}},
  \bibinfo {author} {\bibfnamefont {H.}~\bibnamefont {Kasai}}, \bibinfo
  {author} {\bibfnamefont {K.}~\bibnamefont {Harada}}, \bibinfo {author}
  {\bibfnamefont {T.}~\bibnamefont {Yoshida}}, \bibinfo {author} {\bibfnamefont
  {T.}~\bibnamefont {Akashi}}, \bibinfo {author} {\bibfnamefont
  {A.}~\bibnamefont {Tonomura}}, \bibinfo {author} {\bibfnamefont
  {Y.}~\bibnamefont {Nakayama}}, \bibinfo {author} {\bibfnamefont
  {J.}~\bibnamefont {Shimoyama}}, \bibinfo {author} {\bibfnamefont
  {K.}~\bibnamefont {Kishio}}, \bibinfo {author} {\bibfnamefont
  {T.}~\bibnamefont {Hanaguri}}, \ and\ \bibinfo {author} {\bibfnamefont
  {K.}~\bibnamefont {Kitazawa}},\ }\href@noop {} {\bibfield  {journal}
  {\bibinfo  {journal} {Science}\ }\textbf {\bibinfo {volume} {294}},\ \bibinfo
  {pages} {2136} (\bibinfo {year} {2001})}\BibitemShut {NoStop}%
\bibitem [{\citenamefont {Tonomura}\ \emph {et~al.}(2002)\citenamefont
  {Tonomura}, \citenamefont {Kasai}, \citenamefont {Kamimura}, \citenamefont
  {Matsuda}, \citenamefont {Harada}, \citenamefont {Yoshida}, \citenamefont
  {Akashi}, \citenamefont {Shimoyama}, \citenamefont {Kishio}, \citenamefont
  {Hanaguri}, \citenamefont {Kitazawa}, \citenamefont {Masui}, \citenamefont
  {Tajima}, \citenamefont {Koshizuka}, \citenamefont {Gammel}, \citenamefont
  {Bishop}, \citenamefont {Sasase},\ and\ \citenamefont
  {Okayasu}}]{TonomuraKKMHYASKHKMTKGBSO2002}%
  \BibitemOpen
  \bibfield  {author} {\bibinfo {author} {\bibfnamefont {A.}~\bibnamefont
  {Tonomura}}, \bibinfo {author} {\bibfnamefont {H.}~\bibnamefont {Kasai}},
  \bibinfo {author} {\bibfnamefont {O.}~\bibnamefont {Kamimura}}, \bibinfo
  {author} {\bibfnamefont {T.}~\bibnamefont {Matsuda}}, \bibinfo {author}
  {\bibfnamefont {K.}~\bibnamefont {Harada}}, \bibinfo {author} {\bibfnamefont
  {T.}~\bibnamefont {Yoshida}}, \bibinfo {author} {\bibfnamefont
  {T.}~\bibnamefont {Akashi}}, \bibinfo {author} {\bibfnamefont
  {J.}~\bibnamefont {Shimoyama}}, \bibinfo {author} {\bibfnamefont
  {K.}~\bibnamefont {Kishio}}, \bibinfo {author} {\bibfnamefont
  {T.}~\bibnamefont {Hanaguri}}, \bibinfo {author} {\bibfnamefont
  {K.}~\bibnamefont {Kitazawa}}, \bibinfo {author} {\bibfnamefont
  {T.}~\bibnamefont {Masui}}, \bibinfo {author} {\bibfnamefont
  {S.}~\bibnamefont {Tajima}}, \bibinfo {author} {\bibfnamefont
  {N.}~\bibnamefont {Koshizuka}}, \bibinfo {author} {\bibfnamefont {P.~L.}\
  \bibnamefont {Gammel}}, \bibinfo {author} {\bibfnamefont {D.}~\bibnamefont
  {Bishop}}, \bibinfo {author} {\bibfnamefont {M.}~\bibnamefont {Sasase}}, \
  and\ \bibinfo {author} {\bibfnamefont {S.}~\bibnamefont {Okayasu}},\
  }\href@noop {} {\bibfield  {journal} {\bibinfo  {journal} {Phys. Rev. Lett.}\
  }\textbf {\bibinfo {volume} {88}},\ \bibinfo {pages} {237001} (\bibinfo
  {year} {2002})}\BibitemShut {NoStop}%
\bibitem [{\citenamefont {Vlasko-Vlasov}\ \emph
  {et~al.}(2002{\natexlab{a}})\citenamefont {Vlasko-Vlasov}, \citenamefont
  {Koshelev}, \citenamefont {Welp}, \citenamefont {Crabtree},\ and\
  \citenamefont {Kadowaki}}]{VlaskoVlasovKWCK2002}%
  \BibitemOpen
  \bibfield  {author} {\bibinfo {author} {\bibfnamefont {V.~K.}\ \bibnamefont
  {Vlasko-Vlasov}}, \bibinfo {author} {\bibfnamefont {A.}~\bibnamefont
  {Koshelev}}, \bibinfo {author} {\bibfnamefont {U.}~\bibnamefont {Welp}},
  \bibinfo {author} {\bibfnamefont {G.~W.}\ \bibnamefont {Crabtree}}, \ and\
  \bibinfo {author} {\bibfnamefont {K.}~\bibnamefont {Kadowaki}},\ }\href@noop
  {} {\bibfield  {journal} {\bibinfo  {journal} {Phys. Rev. B}\ }\textbf
  {\bibinfo {volume} {66}},\ \bibinfo {pages} {014523} (\bibinfo {year}
  {2002}{\natexlab{a}})}\BibitemShut {NoStop}%
\bibitem [{\citenamefont {Vlasko-Vlasov}\ \emph
  {et~al.}(2002{\natexlab{b}})\citenamefont {Vlasko-Vlasov}, \citenamefont
  {Koshelev}, \citenamefont {Welp}, \citenamefont {Crabtree},\ and\
  \citenamefont {Kadowaki}}]{VlaskoVlasovKWCK2003}%
  \BibitemOpen
  \bibfield  {author} {\bibinfo {author} {\bibfnamefont {V.~K.}\ \bibnamefont
  {Vlasko-Vlasov}}, \bibinfo {author} {\bibfnamefont {A.}~\bibnamefont
  {Koshelev}}, \bibinfo {author} {\bibfnamefont {U.}~\bibnamefont {Welp}},
  \bibinfo {author} {\bibfnamefont {G.~W.}\ \bibnamefont {Crabtree}}, \ and\
  \bibinfo {author} {\bibfnamefont {K.}~\bibnamefont {Kadowaki}},\ }\href@noop
  {} {\emph {\bibinfo {title} {in ``Proc. NATO Advanced Research Workshop on
  Magneto-Optical Imaging (Oystese, Norway, Aug. 2003)" vol. 142}}}\ (\bibinfo
  {publisher} {Kluwer Academic},\ \bibinfo {address} {Dordrecht},\ \bibinfo
  {year} {2002})\ p.~\bibinfo {pages} {39},\ \bibinfo {note} {ed Johansen T H
  and Shantsev D V}\BibitemShut {NoStop}%
\bibitem [{\citenamefont {Tokunaga}\ \emph {et~al.}(2002)\citenamefont
  {Tokunaga}, \citenamefont {Kobayashi}, \citenamefont {Tokunaga},\ and\
  \citenamefont {Tamegai}}]{TokunagaKTT2002}%
  \BibitemOpen
  \bibfield  {author} {\bibinfo {author} {\bibfnamefont {M.}~\bibnamefont
  {Tokunaga}}, \bibinfo {author} {\bibfnamefont {M.}~\bibnamefont {Kobayashi}},
  \bibinfo {author} {\bibfnamefont {Y.}~\bibnamefont {Tokunaga}}, \ and\
  \bibinfo {author} {\bibfnamefont {T.}~\bibnamefont {Tamegai}},\ }\href@noop
  {} {\bibfield  {journal} {\bibinfo  {journal} {Phys. Rev. B}\ }\textbf
  {\bibinfo {volume} {66}},\ \bibinfo {pages} {060507} (\bibinfo {year}
  {2002})}\BibitemShut {NoStop}%
\bibitem [{\citenamefont {Bending}\ and\ \citenamefont
  {Dodgson}(2005)}]{BendingD2005}%
  \BibitemOpen
  \bibfield  {author} {\bibinfo {author} {\bibfnamefont {S.~J.}\ \bibnamefont
  {Bending}}\ and\ \bibinfo {author} {\bibfnamefont {M.~J.~W.}\ \bibnamefont
  {Dodgson}},\ }\href@noop {} {\bibfield  {journal} {\bibinfo  {journal} {J.
  Phys.: Condens. Matt.}\ }\textbf {\bibinfo {volume} {17}},\ \bibinfo {pages}
  {R955} (\bibinfo {year} {2005})}\BibitemShut {NoStop}%
\bibitem [{\citenamefont {Lee}\ \emph {et~al.}(1995)\citenamefont {Lee},
  \citenamefont {Nordman},\ and\ \citenamefont {Hohenwarter}}]{LeeNH95}%
  \BibitemOpen
  \bibfield  {author} {\bibinfo {author} {\bibfnamefont {J.~U.}\ \bibnamefont
  {Lee}}, \bibinfo {author} {\bibfnamefont {J.~E.}\ \bibnamefont {Nordman}}, \
  and\ \bibinfo {author} {\bibfnamefont {G.}~\bibnamefont {Hohenwarter}},\
  }\href {\doibase 10.1063/1.114498} {\bibfield  {journal} {\bibinfo  {journal}
  {Appl. Phys. Lett.}\ }\textbf {\bibinfo {volume} {67}},\ \bibinfo {pages}
  {1471} (\bibinfo {year} {1995})}\BibitemShut {NoStop}%
\bibitem [{\citenamefont {Lee}\ \emph {et~al.}(1997)\citenamefont {Lee},
  \citenamefont {Guptasarma}, \citenamefont {Hornbaker}, \citenamefont
  {El-Kortas}, \citenamefont {Hinks},\ and\ \citenamefont {Gray}}]{LeeGHE97}%
  \BibitemOpen
  \bibfield  {author} {\bibinfo {author} {\bibfnamefont {J.~U.}\ \bibnamefont
  {Lee}}, \bibinfo {author} {\bibfnamefont {P.}~\bibnamefont {Guptasarma}},
  \bibinfo {author} {\bibfnamefont {D.}~\bibnamefont {Hornbaker}}, \bibinfo
  {author} {\bibfnamefont {A.}~\bibnamefont {El-Kortas}}, \bibinfo {author}
  {\bibfnamefont {D.}~\bibnamefont {Hinks}}, \ and\ \bibinfo {author}
  {\bibfnamefont {K.~E.}\ \bibnamefont {Gray}},\ }\href {\doibase
  10.1063/1.119909} {\bibfield  {journal} {\bibinfo  {journal} {Appl. Phys.
  Lett.}\ }\textbf {\bibinfo {volume} {71}},\ \bibinfo {pages} {1412} (\bibinfo
  {year} {1997})}\BibitemShut {NoStop}%
\bibitem [{\citenamefont {Hechtfischer}\ \emph
  {et~al.}(1997{\natexlab{a}})\citenamefont {Hechtfischer}, \citenamefont
  {Kleiner}, \citenamefont {Ustinov},\ and\ \citenamefont
  {M\"uller}}]{HechtfischerKUM97}%
  \BibitemOpen
  \bibfield  {author} {\bibinfo {author} {\bibfnamefont {G.}~\bibnamefont
  {Hechtfischer}}, \bibinfo {author} {\bibfnamefont {R.}~\bibnamefont
  {Kleiner}}, \bibinfo {author} {\bibfnamefont {A.~V.}\ \bibnamefont
  {Ustinov}}, \ and\ \bibinfo {author} {\bibfnamefont {P.}~\bibnamefont
  {M\"uller}},\ }\href {\doibase 10.1103/PhysRevLett.79.1365} {\bibfield
  {journal} {\bibinfo  {journal} {Phys. Rev. Lett.}\ }\textbf {\bibinfo
  {volume} {79}},\ \bibinfo {pages} {1365} (\bibinfo {year}
  {1997}{\natexlab{a}})}\BibitemShut {NoStop}%
\bibitem [{\citenamefont {Hechtfischer}\ \emph
  {et~al.}(1997{\natexlab{b}})\citenamefont {Hechtfischer}, \citenamefont
  {Kleiner}, \citenamefont {Schlenga}, \citenamefont {Walkenhorst},
  \citenamefont {M\"uller},\ and\ \citenamefont {Johnson}}]{HechtfischerKSW97}%
  \BibitemOpen
  \bibfield  {author} {\bibinfo {author} {\bibfnamefont {G.}~\bibnamefont
  {Hechtfischer}}, \bibinfo {author} {\bibfnamefont {R.}~\bibnamefont
  {Kleiner}}, \bibinfo {author} {\bibfnamefont {K.}~\bibnamefont {Schlenga}},
  \bibinfo {author} {\bibfnamefont {W.}~\bibnamefont {Walkenhorst}}, \bibinfo
  {author} {\bibfnamefont {P.}~\bibnamefont {M\"uller}}, \ and\ \bibinfo
  {author} {\bibfnamefont {H.~L.}\ \bibnamefont {Johnson}},\ }\href {\doibase
  10.1103/PhysRevB.55.14638} {\bibfield  {journal} {\bibinfo  {journal} {Phys.
  Rev. B}\ }\textbf {\bibinfo {volume} {55}},\ \bibinfo {pages} {14638}
  (\bibinfo {year} {1997}{\natexlab{b}})}\BibitemShut {NoStop}%
\bibitem [{\citenamefont {Latyshev}\ \emph {et~al.}(2001)\citenamefont
  {Latyshev}, \citenamefont {Gaifullin}, \citenamefont {Yamashita},
  \citenamefont {Machida},\ and\ \citenamefont {Matsuda}}]{LatyshevGYMM01}%
  \BibitemOpen
  \bibfield  {author} {\bibinfo {author} {\bibfnamefont {Y.~I.}\ \bibnamefont
  {Latyshev}}, \bibinfo {author} {\bibfnamefont {M.~B.}\ \bibnamefont
  {Gaifullin}}, \bibinfo {author} {\bibfnamefont {T.}~\bibnamefont
  {Yamashita}}, \bibinfo {author} {\bibfnamefont {M.}~\bibnamefont {Machida}},
  \ and\ \bibinfo {author} {\bibfnamefont {Y.}~\bibnamefont {Matsuda}},\ }\href
  {\doibase 10.1103/PhysRevLett.87.247007} {\bibfield  {journal} {\bibinfo
  {journal} {Phys. Rev. Lett.}\ }\textbf {\bibinfo {volume} {87}},\ \bibinfo
  {pages} {247007} (\bibinfo {year} {2001})}\BibitemShut {NoStop}%
\bibitem [{\citenamefont {Latyshev}\ \emph {et~al.}(2003)\citenamefont
  {Latyshev}, \citenamefont {Koshelev},\ and\ \citenamefont
  {Bulaevskii}}]{LatyshevKB2003}%
  \BibitemOpen
  \bibfield  {author} {\bibinfo {author} {\bibfnamefont {Y.~I.}\ \bibnamefont
  {Latyshev}}, \bibinfo {author} {\bibfnamefont {A.~E.}\ \bibnamefont
  {Koshelev}}, \ and\ \bibinfo {author} {\bibfnamefont {L.~N.}\ \bibnamefont
  {Bulaevskii}},\ }\href {\doibase 10.1103/PhysRevB.68.134504} {\bibfield
  {journal} {\bibinfo  {journal} {Phys. Rev. B}\ }\textbf {\bibinfo {volume}
  {68}},\ \bibinfo {pages} {134504} (\bibinfo {year} {2003})}\BibitemShut
  {NoStop}%
\bibitem [{\citenamefont {Kim}\ \emph {et~al.}(2005)\citenamefont {Kim},
  \citenamefont {Wang}, \citenamefont {Hatano}, \citenamefont {Urayama},
  \citenamefont {Kawakami}, \citenamefont {Nagao}, \citenamefont {Takano},
  \citenamefont {Yamashita},\ and\ \citenamefont {Lee}}]{KimWH2005}%
  \BibitemOpen
  \bibfield  {author} {\bibinfo {author} {\bibfnamefont {S.~M.}\ \bibnamefont
  {Kim}}, \bibinfo {author} {\bibfnamefont {H.~B.}\ \bibnamefont {Wang}},
  \bibinfo {author} {\bibfnamefont {T.}~\bibnamefont {Hatano}}, \bibinfo
  {author} {\bibfnamefont {S.}~\bibnamefont {Urayama}}, \bibinfo {author}
  {\bibfnamefont {S.}~\bibnamefont {Kawakami}}, \bibinfo {author}
  {\bibfnamefont {M.}~\bibnamefont {Nagao}}, \bibinfo {author} {\bibfnamefont
  {Y.}~\bibnamefont {Takano}}, \bibinfo {author} {\bibfnamefont
  {T.}~\bibnamefont {Yamashita}}, \ and\ \bibinfo {author} {\bibfnamefont
  {K.}~\bibnamefont {Lee}},\ }\href {\doibase 10.1103/PhysRevB.72.140504}
  {\bibfield  {journal} {\bibinfo  {journal} {Phys. Rev. B}\ }\textbf {\bibinfo
  {volume} {72}},\ \bibinfo {pages} {140504} (\bibinfo {year}
  {2005})}\BibitemShut {NoStop}%
\bibitem [{\citenamefont {Ooi}\ \emph {et~al.}(2002)\citenamefont {Ooi},
  \citenamefont {Mochiku},\ and\ \citenamefont {Hirata}}]{OoiMK02}%
  \BibitemOpen
  \bibfield  {author} {\bibinfo {author} {\bibfnamefont {S.}~\bibnamefont
  {Ooi}}, \bibinfo {author} {\bibfnamefont {T.}~\bibnamefont {Mochiku}}, \ and\
  \bibinfo {author} {\bibfnamefont {K.}~\bibnamefont {Hirata}},\ }\href
  {\doibase 10.1103/PhysRevLett.89.247002} {\bibfield  {journal} {\bibinfo
  {journal} {Phys. Rev. Lett.}\ }\textbf {\bibinfo {volume} {89}},\ \bibinfo
  {pages} {247002} (\bibinfo {year} {2002})}\BibitemShut {NoStop}%
\bibitem [{\citenamefont {Zhu}\ \emph {et~al.}(2005)\citenamefont {Zhu},
  \citenamefont {Wang}, \citenamefont {Kim}, \citenamefont {Urayama},
  \citenamefont {Hatano},\ and\ \citenamefont {Hu}}]{ZhuWK2005}%
  \BibitemOpen
  \bibfield  {author} {\bibinfo {author} {\bibfnamefont {B.~Y.}\ \bibnamefont
  {Zhu}}, \bibinfo {author} {\bibfnamefont {H.~B.}\ \bibnamefont {Wang}},
  \bibinfo {author} {\bibfnamefont {S.~M.}\ \bibnamefont {Kim}}, \bibinfo
  {author} {\bibfnamefont {S.}~\bibnamefont {Urayama}}, \bibinfo {author}
  {\bibfnamefont {T.}~\bibnamefont {Hatano}}, \ and\ \bibinfo {author}
  {\bibfnamefont {X.}~\bibnamefont {Hu}},\ }\href {\doibase
  10.1103/PhysRevB.72.174514} {\bibfield  {journal} {\bibinfo  {journal} {Phys.
  Rev. B}\ }\textbf {\bibinfo {volume} {72}},\ \bibinfo {pages} {174514}
  (\bibinfo {year} {2005})}\BibitemShut {NoStop}%
\bibitem [{\citenamefont {Latyshev}\ \emph {et~al.}(2005)\citenamefont
  {Latyshev}, \citenamefont {Pavlenko}, \citenamefont {Orlov},\ and\
  \citenamefont {Hu}}]{LatyshevPOH2005}%
  \BibitemOpen
  \bibfield  {author} {\bibinfo {author} {\bibfnamefont {Y.}~\bibnamefont
  {Latyshev}}, \bibinfo {author} {\bibfnamefont {V.}~\bibnamefont {Pavlenko}},
  \bibinfo {author} {\bibfnamefont {A.~P.}\ \bibnamefont {Orlov}}, \ and\
  \bibinfo {author} {\bibfnamefont {X.}~\bibnamefont {Hu}},\ }\href@noop {}
  {\bibfield  {journal} {\bibinfo  {journal} {Pis'ma v ZhETF}\ }\textbf
  {\bibinfo {volume} {82}},\ \bibinfo {pages} {251} (\bibinfo {year} {2005})},\
  \bibinfo {note} {[{\em JETP Lett.} {\bf 82}, 232 (2005)]}\BibitemShut
  {NoStop}%
\bibitem [{\citenamefont {Katterwe}\ and\ \citenamefont
  {Krasnov}(2009)}]{KatterweKrasnov2009}%
  \BibitemOpen
  \bibfield  {author} {\bibinfo {author} {\bibfnamefont {S.~O.}\ \bibnamefont
  {Katterwe}}\ and\ \bibinfo {author} {\bibfnamefont {V.~M.}\ \bibnamefont
  {Krasnov}},\ }\href {\doibase 10.1103/PhysRevB.80.020502} {\bibfield
  {journal} {\bibinfo  {journal} {Phys. Rev. B}\ }\textbf {\bibinfo {volume}
  {80}},\ \bibinfo {pages} {020502} (\bibinfo {year} {2009})}\BibitemShut
  {NoStop}%
\bibitem [{\citenamefont {Kakeya}\ \emph {et~al.}(2009)\citenamefont {Kakeya},
  \citenamefont {Kubo}, \citenamefont {Kohri}, \citenamefont {Iwase},
  \citenamefont {Yamamoto},\ and\ \citenamefont {Kadowaki}}]{KakeyaKK2009}%
  \BibitemOpen
  \bibfield  {author} {\bibinfo {author} {\bibfnamefont {I.}~\bibnamefont
  {Kakeya}}, \bibinfo {author} {\bibfnamefont {Y.}~\bibnamefont {Kubo}},
  \bibinfo {author} {\bibfnamefont {M.}~\bibnamefont {Kohri}}, \bibinfo
  {author} {\bibfnamefont {M.}~\bibnamefont {Iwase}}, \bibinfo {author}
  {\bibfnamefont {T.}~\bibnamefont {Yamamoto}}, \ and\ \bibinfo {author}
  {\bibfnamefont {K.}~\bibnamefont {Kadowaki}},\ }\href {\doibase
  10.1103/PhysRevB.79.212503} {\bibfield  {journal} {\bibinfo  {journal} {Phys.
  Rev. B}\ }\textbf {\bibinfo {volume} {79}},\ \bibinfo {pages} {212503}
  (\bibinfo {year} {2009})}\BibitemShut {NoStop}%
\bibitem [{Note4()}]{Note4}%
  \BibitemOpen
  \bibinfo {note} {Here $e$ is chosen to be positive, $e>0$, i.e., the charge
  of an electron is $-e$.}\BibitemShut {Stop}%
\bibitem [{\citenamefont {Lawrence}\ and\ \citenamefont
  {Doniach}(1970)}]{Lawrence1970}%
  \BibitemOpen
  \bibfield  {author} {\bibinfo {author} {\bibfnamefont {W.~E.}\ \bibnamefont
  {Lawrence}}\ and\ \bibinfo {author} {\bibfnamefont {S.}~\bibnamefont
  {Doniach}},\ }in\ \href@noop {} {\emph {\bibinfo {booktitle} {Proc. 12th Int.
  Conf. Low Temp. Phys., Kyoto, Japan}}},\ \bibinfo {editor} {edited by\
  \bibinfo {editor} {\bibfnamefont {E.}~\bibnamefont {Kanda}}}\ (\bibinfo
  {year} {1970})\ p.\ \bibinfo {pages} {361}\BibitemShut {NoStop}%
\bibitem [{\citenamefont {Bulaevskii}(1973)}]{Bulaevskii1973}%
  \BibitemOpen
  \bibfield  {author} {\bibinfo {author} {\bibfnamefont {L.~N.}\ \bibnamefont
  {Bulaevskii}},\ }\href@noop {} {\bibfield  {journal} {\bibinfo  {journal}
  {Zh. Eksp. Teor. Fiz.}\ }\textbf {\bibinfo {volume} {64}},\ \bibinfo {pages}
  {2241} (\bibinfo {year} {1973})},\ \bibinfo {note} {[{\em Sov. Phys.--JETP}
  {\bf 37}, 1133 (1973)]}\BibitemShut {NoStop}%
\bibitem [{\citenamefont {Josephson}(1965)}]{Josephson1965}%
  \BibitemOpen
  \bibfield  {author} {\bibinfo {author} {\bibfnamefont {B.~D.}\ \bibnamefont
  {Josephson}},\ }\href@noop {} {\bibfield  {journal} {\bibinfo  {journal}
  {Adv. Phys.}\ }\textbf {\bibinfo {volume} {14}},\ \bibinfo {pages} {419}
  (\bibinfo {year} {1965})}\BibitemShut {NoStop}%
\bibitem [{Note3()}]{Note3}%
  \BibitemOpen
  \bibinfo {note} {This characteristic length was noted soon after the
  discovery of the Josephson effect \cite {Ferrell1963}.}\BibitemShut {Stop}%
\bibitem [{\citenamefont {Clem}\ and\ \citenamefont
  {Coffey}(1990)}]{ClemCoffey1990}%
  \BibitemOpen
  \bibfield  {author} {\bibinfo {author} {\bibfnamefont {J.~R.}\ \bibnamefont
  {Clem}}\ and\ \bibinfo {author} {\bibfnamefont {M.}~\bibnamefont {Coffey}},\
  }\href@noop {} {\bibfield  {journal} {\bibinfo  {journal} {Phys. Rev. B}\
  }\textbf {\bibinfo {volume} {42}},\ \bibinfo {pages} {6209} (\bibinfo {year}
  {1990})}\BibitemShut {NoStop}%
\bibitem [{\citenamefont {Koshelev}(1993)}]{Koshelev1993}%
  \BibitemOpen
  \bibfield  {author} {\bibinfo {author} {\bibfnamefont {A.~E.}\ \bibnamefont
  {Koshelev}},\ }\href@noop {} {\bibfield  {journal} {\bibinfo  {journal}
  {Phys. Rev. B}\ }\textbf {\bibinfo {volume} {48}},\ \bibinfo {pages} {1180}
  (\bibinfo {year} {1993})}\BibitemShut {NoStop}%
\bibitem [{\citenamefont {Clem}\ \emph {et~al.}(1991)\citenamefont {Clem},
  \citenamefont {Coffey},\ and\ \citenamefont {Hao}}]{ClemCH1991}%
  \BibitemOpen
  \bibfield  {author} {\bibinfo {author} {\bibfnamefont {J.~R.}\ \bibnamefont
  {Clem}}, \bibinfo {author} {\bibfnamefont {M.~W.}\ \bibnamefont {Coffey}}, \
  and\ \bibinfo {author} {\bibfnamefont {Z.}~\bibnamefont {Hao}},\ }\href@noop
  {} {\bibfield  {journal} {\bibinfo  {journal} {Phys. Rev. B}\ }\textbf
  {\bibinfo {volume} {44}},\ \bibinfo {pages} {2732} (\bibinfo {year}
  {1991})}\BibitemShut {NoStop}%
\bibitem [{\citenamefont {Klemm}\ and\ \citenamefont
  {Clem}(1980)}]{KlemmClem1980}%
  \BibitemOpen
  \bibfield  {author} {\bibinfo {author} {\bibfnamefont {R.~A.}\ \bibnamefont
  {Klemm}}\ and\ \bibinfo {author} {\bibfnamefont {J.~R.}\ \bibnamefont
  {Clem}},\ }\href@noop {} {\bibfield  {journal} {\bibinfo  {journal} {Phys.
  Rev. B}\ }\textbf {\bibinfo {volume} {21}},\ \bibinfo {pages} {1868}
  (\bibinfo {year} {1980})}\BibitemShut {NoStop}%
\bibitem [{\citenamefont {Blatter}\ \emph {et~al.}(1992)\citenamefont
  {Blatter}, \citenamefont {Geshkenbein},\ and\ \citenamefont
  {Larkin}}]{BlatterGL1992}%
  \BibitemOpen
  \bibfield  {author} {\bibinfo {author} {\bibfnamefont {G.}~\bibnamefont
  {Blatter}}, \bibinfo {author} {\bibfnamefont {V.~B.}\ \bibnamefont
  {Geshkenbein}}, \ and\ \bibinfo {author} {\bibfnamefont {A.~I.}\ \bibnamefont
  {Larkin}},\ }\href@noop {} {\bibfield  {journal} {\bibinfo  {journal} {Phys.
  Rev. Lett.}\ }\textbf {\bibinfo {volume} {68}},\ \bibinfo {pages} {875}
  (\bibinfo {year} {1992})}\BibitemShut {NoStop}%
\bibitem [{\citenamefont {Ivlev}\ \emph {et~al.}(1990)\citenamefont {Ivlev},
  \citenamefont {Kopnin},\ and\ \citenamefont {Pokrovskii}}]{IvlevKP1990}%
  \BibitemOpen
  \bibfield  {author} {\bibinfo {author} {\bibfnamefont {B.~I.}\ \bibnamefont
  {Ivlev}}, \bibinfo {author} {\bibfnamefont {N.~B.}\ \bibnamefont {Kopnin}}, \
  and\ \bibinfo {author} {\bibfnamefont {V.~L.}\ \bibnamefont {Pokrovskii}},\
  }\href@noop {} {\bibfield  {journal} {\bibinfo  {journal} {J. Low Temp.
  Phys.}\ }\textbf {\bibinfo {volume} {80}},\ \bibinfo {pages} {187} (\bibinfo
  {year} {1990})}\BibitemShut {NoStop}%
\bibitem [{\citenamefont {Levitov}(1992)}]{Levitov1991}%
  \BibitemOpen
  \bibfield  {author} {\bibinfo {author} {\bibfnamefont {L.~S.}\ \bibnamefont
  {Levitov}},\ }\href@noop {} {\bibfield  {journal} {\bibinfo  {journal} {Phys.
  Rev. Lett.}\ }\textbf {\bibinfo {volume} {66}},\ \bibinfo {pages} {224}
  (\bibinfo {year} {1992})}\BibitemShut {NoStop}%
\bibitem [{\citenamefont {Laguna}\ \emph {et~al.}(2000)\citenamefont {Laguna},
  \citenamefont {Dominguez},\ and\ \citenamefont {Balseiro}}]{LagunaDB2000}%
  \BibitemOpen
  \bibfield  {author} {\bibinfo {author} {\bibfnamefont {M.~F.}\ \bibnamefont
  {Laguna}}, \bibinfo {author} {\bibfnamefont {D.}~\bibnamefont {Dominguez}}, \
  and\ \bibinfo {author} {\bibfnamefont {C.~A.}\ \bibnamefont {Balseiro}},\
  }\href@noop {} {\bibfield  {journal} {\bibinfo  {journal} {Phys. Rev. B}\
  }\textbf {\bibinfo {volume} {62}},\ \bibinfo {pages} {6692} (\bibinfo {year}
  {2000})}\BibitemShut {NoStop}%
\bibitem [{\citenamefont {Stewart}(1964)}]{Stewart64}%
  \BibitemOpen
  \bibfield  {author} {\bibinfo {author} {\bibfnamefont {B.~M.}\ \bibnamefont
  {Stewart}},\ }\href@noop {} {\emph {\bibinfo {title} {Theory of numbers}}}\
  (\bibinfo  {publisher} {The Macmillan Company},\ \bibinfo {address} {New
  York},\ \bibinfo {year} {1964})\BibitemShut {NoStop}%
\bibitem [{\citenamefont {Ichioka}(1995)}]{Ichioka95}%
  \BibitemOpen
  \bibfield  {author} {\bibinfo {author} {\bibfnamefont {M.}~\bibnamefont
  {Ichioka}},\ }\href@noop {} {\bibfield  {journal} {\bibinfo  {journal} {Phys.
  Rev. B}\ }\textbf {\bibinfo {volume} {51}},\ \bibinfo {pages} {9423}
  (\bibinfo {year} {1995})}\BibitemShut {NoStop}%
\bibitem [{Note2()}]{Note2}%
  \BibitemOpen
  \bibinfo {note} {A. E. Koshelev, Proceedings of FIMS/ITS-NS/CTC/PLASMA 2004,
  Tsukuba, Japan, Nov. 24–28, 2004; arXiv:cond-mat/0602341.}\BibitemShut
  {Stop}%
\bibitem [{\citenamefont {Nonomura}\ and\ \citenamefont
  {Hu}(2006)}]{NonomuraHu2006}%
  \BibitemOpen
  \bibfield  {author} {\bibinfo {author} {\bibfnamefont {Y.}~\bibnamefont
  {Nonomura}}\ and\ \bibinfo {author} {\bibfnamefont {X.}~\bibnamefont {Hu}},\
  }\href {\doibase 10.1103/PhysRevB.74.024504} {\bibfield  {journal} {\bibinfo
  {journal} {Phys. Rev. B}\ }\textbf {\bibinfo {volume} {74}},\ \bibinfo
  {pages} {024504} (\bibinfo {year} {2006})}\BibitemShut {NoStop}%
\bibitem [{\citenamefont {Ikeda}\ and\ \citenamefont
  {Isotani}(1999)}]{IkedaI1999}%
  \BibitemOpen
  \bibfield  {author} {\bibinfo {author} {\bibfnamefont {R.}~\bibnamefont
  {Ikeda}}\ and\ \bibinfo {author} {\bibfnamefont {K.}~\bibnamefont
  {Isotani}},\ }\href@noop {} {\bibfield  {journal} {\bibinfo  {journal}
  {Journ. Phys. Soc. japan}\ }\textbf {\bibinfo {volume} {68}},\ \bibinfo
  {pages} {599} (\bibinfo {year} {1999})}\BibitemShut {NoStop}%
\bibitem [{\citenamefont {Korshunov}(1990)}]{Korshunov1991}%
  \BibitemOpen
  \bibfield  {author} {\bibinfo {author} {\bibfnamefont {S.~E.}\ \bibnamefont
  {Korshunov}},\ }\href@noop {} {\bibfield  {journal} {\bibinfo  {journal}
  {Europhys. Lett.}\ }\textbf {\bibinfo {volume} {15}},\ \bibinfo {pages} {771}
  (\bibinfo {year} {1990})}\BibitemShut {NoStop}%
\bibitem [{\citenamefont {Korshunov}\ and\ \citenamefont
  {Larkin}(1992)}]{KorshunovL1992}%
  \BibitemOpen
  \bibfield  {author} {\bibinfo {author} {\bibfnamefont {S.~E.}\ \bibnamefont
  {Korshunov}}\ and\ \bibinfo {author} {\bibfnamefont {A.~I.}\ \bibnamefont
  {Larkin}},\ }\href@noop {} {\bibfield  {journal} {\bibinfo  {journal} {Phys.
  Rev. B}\ }\textbf {\bibinfo {volume} {46}},\ \bibinfo {pages} {6395}
  (\bibinfo {year} {1992})}\BibitemShut {NoStop}%
\bibitem [{\citenamefont {Koshelev}(2002)}]{Koshelev2002}%
  \BibitemOpen
  \bibfield  {author} {\bibinfo {author} {\bibfnamefont {A.~E.}\ \bibnamefont
  {Koshelev}},\ }\href {\doibase 10.1103/PhysRevB.66.224514} {\bibfield
  {journal} {\bibinfo  {journal} {Phys. Rev. B}\ }\textbf {\bibinfo {volume}
  {66}},\ \bibinfo {pages} {224514} (\bibinfo {year} {2002})}\BibitemShut
  {NoStop}%
\bibitem [{\citenamefont {Koshelev}(2007)}]{Koshelev2007}%
  \BibitemOpen
  \bibfield  {author} {\bibinfo {author} {\bibfnamefont {A.~E.}\ \bibnamefont
  {Koshelev}},\ }\href {\doibase 10.1103/PhysRevB.75.214513} {\bibfield
  {journal} {\bibinfo  {journal} {Phys. Rev. B}\ }\textbf {\bibinfo {volume}
  {75}},\ \bibinfo {pages} {214513} (\bibinfo {year} {2007})}\BibitemShut
  {NoStop}%
\bibitem [{\citenamefont {Machida}(2003)}]{Machida2003}%
  \BibitemOpen
  \bibfield  {author} {\bibinfo {author} {\bibfnamefont {M.}~\bibnamefont
  {Machida}},\ }\href {\doibase 10.1103/PhysRevLett.90.037001} {\bibfield
  {journal} {\bibinfo  {journal} {Phys. Rev. Lett.}\ }\textbf {\bibinfo
  {volume} {90}},\ \bibinfo {pages} {037001} (\bibinfo {year}
  {2003})}\BibitemShut {NoStop}%
\bibitem [{\citenamefont {Machida}(2006)}]{Machida2006}%
  \BibitemOpen
  \bibfield  {author} {\bibinfo {author} {\bibfnamefont {M.}~\bibnamefont
  {Machida}},\ }\href {\doibase 10.1103/PhysRevLett.96.097002} {\bibfield
  {journal} {\bibinfo  {journal} {Phys. Rev. Lett.}\ }\textbf {\bibinfo
  {volume} {96}},\ \bibinfo {pages} {097002} (\bibinfo {year}
  {2006})}\BibitemShut {NoStop}%
\bibitem [{\citenamefont {Irie}\ and\ \citenamefont {Oya}(2007)}]{IrieOya2007}%
  \BibitemOpen
  \bibfield  {author} {\bibinfo {author} {\bibfnamefont {A.}~\bibnamefont
  {Irie}}\ and\ \bibinfo {author} {\bibfnamefont {G.}~\bibnamefont {Oya}},\
  }\href {http://stacks.iop.org/0953-2048/20/i=2/a=S05} {\bibfield  {journal}
  {\bibinfo  {journal} {Supercond. Sci. Technol.}\ }\textbf {\bibinfo {volume}
  {20}},\ \bibinfo {pages} {S18} (\bibinfo {year} {2007})}\BibitemShut
  {NoStop}%
\bibitem [{\citenamefont {Efetov}(1979)}]{Efetov1979}%
  \BibitemOpen
  \bibfield  {author} {\bibinfo {author} {\bibfnamefont {K.~B.}\ \bibnamefont
  {Efetov}},\ }\href@noop {} {\bibfield  {journal} {\bibinfo  {journal} {Zh.
  Eksp. Teor. Fiz.}\ }\textbf {\bibinfo {volume} {76}},\ \bibinfo {pages}
  {1781} (\bibinfo {year} {1979})},\ \bibinfo {note} {[{\em Sov. Phys.--JETP}
  {\bf 49}, 905 (1979 )]}\BibitemShut {NoStop}%
\bibitem [{\citenamefont {Horovitz}(1993)}]{Horovitz1993II}%
  \BibitemOpen
  \bibfield  {author} {\bibinfo {author} {\bibfnamefont {B.}~\bibnamefont
  {Horovitz}},\ }\href {\doibase 10.1103/PhysRevB.47.5964} {\bibfield
  {journal} {\bibinfo  {journal} {Phys. Rev. B}\ }\textbf {\bibinfo {volume}
  {47}},\ \bibinfo {pages} {5964} (\bibinfo {year} {1993})}\BibitemShut
  {NoStop}%
\bibitem [{\citenamefont {Balents}\ and\ \citenamefont
  {Nelson}(1995)}]{BalentsN1995}%
  \BibitemOpen
  \bibfield  {author} {\bibinfo {author} {\bibfnamefont {L.}~\bibnamefont
  {Balents}}\ and\ \bibinfo {author} {\bibfnamefont {D.~R.}\ \bibnamefont
  {Nelson}},\ }\href {\doibase 10.1103/PhysRevB.52.12951} {\bibfield  {journal}
  {\bibinfo  {journal} {Phys. Rev. B}\ }\textbf {\bibinfo {volume} {52}},\
  \bibinfo {pages} {12951} (\bibinfo {year} {1995})}\BibitemShut {NoStop}%
\bibitem [{\citenamefont {Balents}\ and\ \citenamefont
  {Radzihovsky}(1996)}]{BalentsR1996}%
  \BibitemOpen
  \bibfield  {author} {\bibinfo {author} {\bibfnamefont {L.}~\bibnamefont
  {Balents}}\ and\ \bibinfo {author} {\bibfnamefont {L.}~\bibnamefont
  {Radzihovsky}},\ }\href {\doibase 10.1103/PhysRevLett.76.3416} {\bibfield
  {journal} {\bibinfo  {journal} {Phys. Rev. Lett.}\ }\textbf {\bibinfo
  {volume} {76}},\ \bibinfo {pages} {3416} (\bibinfo {year}
  {1996})}\BibitemShut {NoStop}%
\bibitem [{\citenamefont {Hu}\ \emph {et~al.}(2005)\citenamefont {Hu},
  \citenamefont {Luo},\ and\ \citenamefont {Ma}}]{HuLM2005}%
  \BibitemOpen
  \bibfield  {author} {\bibinfo {author} {\bibfnamefont {X.}~\bibnamefont
  {Hu}}, \bibinfo {author} {\bibfnamefont {M.}~\bibnamefont {Luo}}, \ and\
  \bibinfo {author} {\bibfnamefont {Y.}~\bibnamefont {Ma}},\ }\href
  {http://link.aps.org/abstract/PRB/v72/e174503} {\bibfield  {journal}
  {\bibinfo  {journal} {Phys. Rev. B}\ }\textbf {\bibinfo {volume} {72}},\
  \bibinfo {eid} {174503} (\bibinfo {year} {2005})}\BibitemShut {NoStop}%
\bibitem [{\citenamefont {Hu}\ and\ \citenamefont {Tachiki}(2000)}]{HuT2000}%
  \BibitemOpen
  \bibfield  {author} {\bibinfo {author} {\bibfnamefont {X.}~\bibnamefont
  {Hu}}\ and\ \bibinfo {author} {\bibfnamefont {M.}~\bibnamefont {Tachiki}},\
  }\href {\doibase 10.1103/PhysRevLett.85.2577} {\bibfield  {journal} {\bibinfo
   {journal} {Phys. Rev. Lett.}\ }\textbf {\bibinfo {volume} {85}},\ \bibinfo
  {pages} {2577} (\bibinfo {year} {2000})}\BibitemShut {NoStop}%
\bibitem [{\citenamefont {Hu}\ and\ \citenamefont {Tachiki}(2004)}]{HuT2004}%
  \BibitemOpen
  \bibfield  {author} {\bibinfo {author} {\bibfnamefont {X.}~\bibnamefont
  {Hu}}\ and\ \bibinfo {author} {\bibfnamefont {M.}~\bibnamefont {Tachiki}},\
  }\href {http://link.aps.org/abstract/PRB/v70/e064506} {\bibfield  {journal}
  {\bibinfo  {journal} {Phys. Rev. B}\ }\textbf {\bibinfo {volume} {70}},\
  \bibinfo {eid} {064506} (\bibinfo {year} {2004})}\BibitemShut {NoStop}%
\bibitem [{Note5()}]{Note5}%
  \BibitemOpen
  \bibinfo {note} {As in most theoretical papers, the temperature is measured
  in energy units.}\BibitemShut {Stop}%
\bibitem [{\citenamefont {Koshelev}(1994)}]{Koshelev1994}%
  \BibitemOpen
  \bibfield  {author} {\bibinfo {author} {\bibfnamefont {A.~E.}\ \bibnamefont
  {Koshelev}},\ }\href@noop {} {\bibfield  {journal} {\bibinfo  {journal}
  {Phys. Rev. B}\ }\textbf {\bibinfo {volume} {50}},\ \bibinfo {pages} {506}
  (\bibinfo {year} {1994})}\BibitemShut {NoStop}%
\bibitem [{\citenamefont {Kwok}\ \emph {et~al.}(1994)\citenamefont {Kwok},
  \citenamefont {Fendrich}, \citenamefont {Welp}, \citenamefont {Fleshler},
  \citenamefont {Downey},\ and\ \citenamefont {Crabtree}}]{KwokFWFDC1994}%
  \BibitemOpen
  \bibfield  {author} {\bibinfo {author} {\bibfnamefont {W.~K.}\ \bibnamefont
  {Kwok}}, \bibinfo {author} {\bibfnamefont {J.}~\bibnamefont {Fendrich}},
  \bibinfo {author} {\bibfnamefont {U.}~\bibnamefont {Welp}}, \bibinfo {author}
  {\bibfnamefont {S.}~\bibnamefont {Fleshler}}, \bibinfo {author}
  {\bibfnamefont {J.}~\bibnamefont {Downey}}, \ and\ \bibinfo {author}
  {\bibfnamefont {G.~W.}\ \bibnamefont {Crabtree}},\ }\href {\doibase
  10.1103/PhysRevLett.72.1088} {\bibfield  {journal} {\bibinfo  {journal}
  {Phys. Rev. Lett.}\ }\textbf {\bibinfo {volume} {72}},\ \bibinfo {pages}
  {1088} (\bibinfo {year} {1994})}\BibitemShut {NoStop}%
\bibitem [{\citenamefont {Villain}\ \emph {et~al.}(1980)\citenamefont
  {Villain}, \citenamefont {Bidaux}, \citenamefont {Carton},\ and\
  \citenamefont {Conte}}]{VillainBCC1980}%
  \BibitemOpen
  \bibfield  {author} {\bibinfo {author} {\bibfnamefont {J.}~\bibnamefont
  {Villain}}, \bibinfo {author} {\bibfnamefont {R.}~\bibnamefont {Bidaux}},
  \bibinfo {author} {\bibfnamefont {J.}~\bibnamefont {Carton}}, \ and\ \bibinfo
  {author} {\bibfnamefont {R.}~\bibnamefont {Conte}},\ }\href@noop {}
  {\bibfield  {journal} {\bibinfo  {journal} {J. Phys. (France)}\ }\textbf
  {\bibinfo {volume} {41}},\ \bibinfo {pages} {1263} (\bibinfo {year}
  {1980})}\BibitemShut {NoStop}%
\bibitem [{\citenamefont {Pokrovskii}\ and\ \citenamefont
  {Uimin}(1973)}]{PokrovskiiU1973}%
  \BibitemOpen
  \bibfield  {author} {\bibinfo {author} {\bibfnamefont {V.~L.}\ \bibnamefont
  {Pokrovskii}}\ and\ \bibinfo {author} {\bibfnamefont {G.~V.}\ \bibnamefont
  {Uimin}},\ }\href@noop {} {\bibfield  {journal} {\bibinfo  {journal} {Zh.
  Eksp. Teor. Fiz.}\ }\textbf {\bibinfo {volume} {65}},\ \bibinfo {pages}
  {1961} (\bibinfo {year} {1973})},\ \bibinfo {note} {[{\em Sov. Phys.--JETP}
  {\bf 38}, 847 (1974)]}\BibitemShut {NoStop}%
\bibitem [{\citenamefont {Glazman}\ and\ \citenamefont
  {Koshelev}(1990)}]{GlazmanKoshelev1990}%
  \BibitemOpen
  \bibfield  {author} {\bibinfo {author} {\bibfnamefont {L.~I.}\ \bibnamefont
  {Glazman}}\ and\ \bibinfo {author} {\bibfnamefont {A.~E.}\ \bibnamefont
  {Koshelev}},\ }\href@noop {} {\bibfield  {journal} {\bibinfo  {journal} {Zh.
  Eksp. Teor. Fiz}\ }\textbf {\bibinfo {volume} {97}},\ \bibinfo {pages} {1371}
  (\bibinfo {year} {1990})},\ \bibinfo {note} {[{\em Sov. Phys.--JETP} {\bf
  70}, 774 (1990)]}\BibitemShut {NoStop}%
\bibitem [{\citenamefont {Ferrell}\ and\ \citenamefont
  {Prange}(1963)}]{Ferrell1963}%
  \BibitemOpen
  \bibfield  {author} {\bibinfo {author} {\bibfnamefont {R.~A.}\ \bibnamefont
  {Ferrell}}\ and\ \bibinfo {author} {\bibfnamefont {R.~E.}\ \bibnamefont
  {Prange}},\ }\href@noop {} {\bibfield  {journal} {\bibinfo  {journal} {Phys.
  Rev. Lett.}\ }\textbf {\bibinfo {volume} {10}},\ \bibinfo {pages} {479}
  (\bibinfo {year} {1963})}\BibitemShut {NoStop}%
\end{thebibliography}%

\end{document}